\documentclass[preprint, showpacs]{revtex4}
\usepackage{graphicx}
\usepackage{psfrag}
\usepackage{color}
\usepackage{amssymb}
\usepackage{amsmath}
\usepackage{epstopdf}

\usepackage{multirow}
\usepackage{booktabs}
\usepackage{float}

\newcommand{\be}{\begin{eqnarray}}
\newcommand{\ee}{\end{eqnarray}}

\newcommand{\bfk}{{\bf k}_{\perp}}

\newcommand{\bfb}{{\bf b}_{\perp}}

\newcommand{\tvec}[1]{\mbox{\boldmath{$#1$}}}
\newcommand{\svec}[1]{\mbox{\boldmath{$\scriptstyle #1$}}}

\begin{document}

\title{Leading twist generalized parton distributions and spin densities in a proton }
\author{Tanmay Maji$^1$, Chandan Mondal$^{1,2}$, Dipankar Chakrabarti$^1$}
\affiliation{ $^1$Department of Physics, Indian Institute of Technology Kanpur, Kanpur 208016, India}
\affiliation{$^2$Institute of Modern Physics, Chinese Academy of Sciences, Lanzhou 730000, China}

\date{\today}

\begin{abstract}
We evaluate both chirally even and odd generalized parton distributions(GPDs) in the leading twist in a  recently proposed quark-diquark model for the proton where the light front wavefunctions  are constructed from the soft-wall AdS/QCD prediction.  The GPDs in transverse impact parameter space give the spin densities for different quark and proton polarizations. For longitudinally polarized proton only chiral even GPDs contribute but for transversely polarized proton both chiral even and chiral odd GPDs contribute to the spin densities.  We present a detail study of the spin densities in this model. 

\end{abstract}

\maketitle

\section{introduction}
Generalized parton distributions(GPDs)\cite{rev} encode spatial as well as partonic spin structure in a proton.  GPDs are functions of three variables: longitudinal momentum fraction $x$ of the parton, longitudinal momentum fraction transferred in the process which is given by skewness $\xi$  and square of the momentum transferred  $t^2$. GPDs give a unified picture of the nucleon, in the sense
that the $x$ moments of them give the form factors accessible in exclusive processes like deeply virtual Compton scattering
(DVCS) or vector-meson productions whereas in
the forward limit they reduce to parton distributions, accessible in inclusive processes. Being off-forward matrix elements, GPDs have no probabilistic interpretation, but at zero skewness, the Fourier transforms of the GPDs 
with respect to the transverse momentum transfer $\Delta_\perp$ give the impact parameter dependent parton distributions \cite{burkardt} which tells us how  the partons of
a given longitudinal momentum are distributed in transverse position space.  The GPDs not only provide the spatial structure of the proton  but also  encode the partonic spin information.  Ji sum rule\cite{Ji} relates the GPDs with the angular momentum  of the proton. For different polarizations of the proton, spin densities can be expressed in terms of the impact parameter dependent GPDs.

For many years, DVCS data have been collceted in different experimental labs. Recently JLab has also started DVCS experiment, 
COMPASS at CERN will start to have more DVCS data and future Electron-Ion Collider 
is planned to explore the GPDs through DVCS.
But   experimental extractions of GPDs are not straight forward as fitting of DVCS data does not give direct information about the GPDs
but some weighted integrals of GPDs. Since nonperturbative QCD prediction is not yet possible, different model predictions of 
GPDs are very important to constrain the GPDs and  data fitting to extract GPDs from  DVCS data.
 
In this paper, we consider a  light front quark-diquark model of proton \cite{MC}. Both scalar and vector diquarks are considered in this model where the light front wave functions are constructed from soft-wall AdS/QCD prediction. The wave functions predicted in AdS/QCD \cite{BT} can not be derived in perturbation theory with few Fock states,  and hence contains nonperturbative structure of the proton. We present results for both chiral-even and chiral-odd GPDs. At leading twist, there are eight GPDs, four of them are chiral even and four are chiral-odd.  Similar to the definition of PDFs, we can define three different types of correlators  of  vector, axial vector and tensor quark currents.  The off-forward matrix elements of the first two currents involve four chiral-even GPDs, $H^q,~E^q, \tilde{H}^q,~ \tilde{E}^q$. The off-forward matrix elements of  the third one are chirally odd and involves four GPDs, namely, $H^q_T,~E^q_T, \tilde{H}^q_T,~ \tilde{E}^q_T$.  Chiral even GPDs are studied in different models. e.g., bag model\cite{bag},  constituent quark model\cite{cqm}, soliton model\cite{soliton}, dressed electron\cite{qed},  AdS/QCD models\cite{ads}, scalar diquark model\cite{sdq}, in basis light front quantization\cite{blfq} etc.
 The chiral-odd GPDs in a constituent quark model have been studied for nonzero skewness using the overlap representation in terms of light-front wave functions (LFWFs) in~\cite{Pasquini1}. The general properties of the chiral-odd GPDs in a QED 
model have been investigated in both momentum and transverse position as well as longitudinal position spaces in~\cite{CMM1}; the impact parameter representation of the GPDs have been studied in a QED model of a dressed electron for $\xi=0$\cite{Dahiya07} and  in quark-diquark models for nucleon \cite{kumar,chandan}. Recently, it has been demostrated that the transversity quark GPDs can be measured from neutrino-production of a charmed meson \cite{pire17}. The Mellin moments of the transverse GPDs have been evaluated on lattice \cite{gock,Hagler:2009,Hagler:2004,Bratt:2010}. 

The chiral even GPD $E^q$ is responsible for the distortion in the  unpolarized quark distribution in a transversely polarized proton. Similarly,  chirally odd GPDs affect the transversely polarized quark distributions in a unpolarized or transversely polarized  proton in different ways. Chiral even GPDs are accessible in exclusive processes like deeply virtual Compton scattering or  deeply virtual meson production. But,  the chiral-odd GPDs are not easy to measure as they require another chiral-odd object in the amplitude to combine.   The chiral-odd GPDs can be measured in the diffractive electroproduction of two vector mesons with large rapidity gap \cite{chi1} or exclusive $\pi^0$ electroproduction \cite{chi2}.
For longitudinally polarized proton, the spin density involves only chirally even GPDs while the spin density for a transversely polarized proton involve both chiral even and odd GPDs.  In this work, we evaluate all the leading twist GPDs in an AdS/QCD inspired light front quark diquark model. 
Then  we present a detail study of the spin densities for both longitudinally and transversely polarized proton and show how different GPDs contribute to different spin densities.

\section{Generalized parton distributions}
At leading twist, one can define three generalized distributions in parallel to three PDFs, namely, the unpolarized, helicity, and transversity distributions. The GPDs are defined as off-forward matrix elements of the bilocal operator of light-front correlation functions of vector, axial vector, and tensor current
\be
& &\frac{1}{2}\int\frac{dz^-}{2\pi}e^{i \bar xP^+z^-}\langle
p', \lambda'|\bar{\psi}_q(-{z}/2)\gamma^+
\psi_q({z}/2)|p, \lambda\rangle_{|_{z^+ = 0, \vec z_\perp=
0}}
\nonumber\\
& &=\frac{1}{2P^+}\bar{u}(p',\lambda')\bigg[H^q\gamma^+ + E^q\frac{i}{2M}\sigma^{+\alpha}\Delta_{\alpha}\bigg]u(p,\lambda),\label{gpd_unpol}
\\
& &\frac{1}{2}\int\frac{dz^-}{2\pi}e^{i \bar xP^+z^-}\langle
p', \lambda'|\bar{\psi}_q(-{z}/2)\gamma^+\gamma_5
\psi_q({z}/2)|p, \lambda\rangle_{|_{z^+ = 0, \vec z_\perp=
0}}
\nonumber\\
& &= \frac{1}{2P^+} \bar{u}(p',\lambda') \left[
  \widetilde{H}^q\, \gamma^+ \gamma_5 +
  \widetilde{E}^q\, \frac{\gamma_5 \Delta^+}{2M}
  \right] u(p,\lambda),\label{gpd_helicity}
\\
& &\frac{1}{2}\int\frac{dz^-}{2\pi}e^{i \bar xP^+z^-}\langle
p', \lambda'|\bar{\psi}_q(-{z}/2)\sigma^{+j}\gamma_5
\psi_q({z}/2)|p, \lambda\rangle_{|_{z^+ = 0, \vec z_\perp=
0}}
\nonumber\\
& &= \frac{1}{2P^+}\bar{u}(p',
\lambda')\bigg[H^q_T\sigma^{+j}\gamma_5 +
\widetilde{H}^q_T\frac{\epsilon^{+j\alpha\beta}\Delta_\alpha
P_\beta}{M^2} +
E^q_T\frac{\epsilon^{+j\alpha\beta}\Delta_\alpha\gamma_\beta}{2M}
+
\widetilde{E}^q_T\frac{\epsilon^{+j\alpha\beta}P_\alpha\gamma_\beta}{M}\bigg]u(p,
\lambda), \label{gpd_transverse}
\ee
where $p$ $(p')$ and $\lambda$ $(\lambda')$ denote the proton momenta and the helicity of the initial (final) state of proton, respectively. 
$M$ denotes the mass of proton and $j=1,2$ is a transverse index. 
The $H$ and $E$, so-called unpolarized GPDs and the helicity dependent GPDs, $\widetilde H$ and $\widetilde E$ are chiral-even, while
$H_T$ , $\widetilde{H}_T$ , $E_T$ , and $\widetilde{E}_T$ are chiral-odd. 
In the symmetric frame, the kinematical variables are 
$
P^\mu=\frac{(p+p')^\mu}{2}, \quad\quad \Delta^\mu=p'^\mu-p^\mu, \quad\quad \xi=-\Delta^+/2P^+,
$
and $t=\Delta^2$. We choose the light-front gauge $A^+=0$, so that the gauge link appearing in between the quark fields in Eqs.(\ref{gpd_unpol}-\ref{gpd_transverse}) becomes unity. All the GPDs can be related to the following matrix elements depending on various helicity configurations of proton and quark \cite{diehl01,Boffi:2007yc}
\be
  \label{amplitude}
A_{\lambda'\mu', \lambda\mu} &=&
\int \frac{d z^-}{2\pi}\, e^{ix P^+ z^-}
  \langle p',\lambda'|\, {\cal O}_{\mu',\mu}(z)
  \,|p,\lambda \rangle \Big|_{z^+=0,\, \vec z_\perp=0} ,
\ee
with definite quark helicities $\mu$ and $\mu'$ and the operators ${\cal O}_{\mu,\mu'}$ occurring in the
definitions of the quark distributions are given by
\be
{\cal O}_{+,+} 
&=& \frac{1}{4}\, 
  \bar{\psi}\, \gamma^+ (1+\gamma_5)\, \psi ,
\nonumber \quad\quad
{\cal O}_{-,-}
= \frac{1}{4}\, \bar{\psi}\, \gamma^+ (1-\gamma_5)\, \psi ,
\nonumber \\
{\cal O}_{-,+} 
&=& - \frac{i}{4}\, \bar{\psi}\, \sigma^{+1} (1+\gamma_5)\, \psi \quad~
{\cal O}_{+,-}
= \frac{i}{4}\, 
  \bar{\psi}\, \sigma^{+1} (1-\gamma_5)\, \psi \phantom{-}
\hspace{2em}
\ee
Due to parity invariance one has the relation 
$A_{-\lambda'-,-\lambda +}=(-1)^{\lambda'-\lambda}A_{\lambda'+,\lambda -}$. Using  the reference frame where  the momenta $\vec p$ and $\vec p\,'$ lie in the $x-z$ plane, one can explicitly derive the following relations for the chiral-even GPDs~\cite{diehl01,Boffi:2007yc}
\be
H^q&=&\frac{1}{\sqrt{1-\xi^2}}T^q_1 - \frac{2M\xi^2} {\sqrt{t_0 - t}(1 - \xi^2)}T^q_3,\label{H}\\
E^q&=&-\frac{2M} {\epsilon \sqrt{t_0 - t}}T^q_3,\label{E}\\
\widetilde{H}^q&=&\frac{1}{\sqrt{1-\xi^2}}T^q_2+\frac{2M\xi} {\sqrt{t_0 - t}(1 - \xi^2)}T^q_4,\label{Ht}\\
\widetilde{E}^q&=&\frac{2M} {\epsilon \xi\sqrt{t_0 - t}}T^q_4,\label{Et}
\ee
and for the chiral-odd GPDs, the relations are given by
\be
H^{q}_T &=& \frac{1}{\sqrt{1 - \xi^2}}\widetilde{T}^q_{1} -
\frac{2M\xi} {\epsilon\sqrt{t_0 - t}(1 - \xi^2)}\widetilde{T}^q_{3},\label{HT}
\\
E_T^{q} &=& \frac{2M}{\epsilon\sqrt{t_0 - t}(1 -\xi^2)}
\Big(\xi \widetilde{T}^q_{3}
+ \widetilde{T}^q_{4}\Big)
- \frac{4M^2}{(t_0 - t)\sqrt{1 - \xi^2}(1 - \xi^2)}
\Big(\widetilde{T}^q_{2} - \widetilde{T}^q_{1}\Big).\label{ET}
\\
\widetilde{H}_T^{q} &=& \frac{2M^2}{(t_0 - t)\sqrt{1 -\xi^2}}
\Big(\widetilde{T}^q_{2} - \widetilde{T}^q_{1}\Big),\label{HtT}
\\
\widetilde{E}^{q}_T &=& \frac{2M}{\epsilon\sqrt{t_0 - t}(1 -\xi^2)}
\Big(\widetilde{T}^q_{3}
+ \xi\widetilde{T}^q_{4}\Big)
- \frac{4M^2\xi}{(t_0 - t)\sqrt{1 - \xi^2}(1 - \xi^2)}
\Big(\widetilde{T}^q_{2} - \widetilde{T}^q_{1}\Big).\label{EtT}
\ee
where the matrix elements $T^q_i$ and $\widetilde{T}^q_i$, in terms of the quark helicity basis are given by
\be
T^q_{1} =A_{++,++} + A_{-+,-+}, &\quad&
T^q_{2} =A_{++,++} - A_{-+,-+},\nonumber\\
T^q_{3} =A_{++,-+} - A_{-+,++}. &\quad&
T^q_{4} =A_{++,-+} + A_{-+,++},
\ee
and 
\be
\widetilde{T}^q_{1} =A_{++,--} + A_{-+,+-}, &\quad&
\widetilde{T}^q_{2} =A_{++,--} - A_{-+,+-},\nonumber\\
\widetilde{T}^q_{3} = A_{++,+-} - A_{-+,--}, &\quad&
\widetilde{T}^q_{4} = A_{++,+-} + A_{-+,--}.
\ee
For a given $\xi$, The minimum value of $-t$ is $- t_0 = 4 M^2 \xi^2/(1-\xi^2)$ , and $\epsilon =
\mathrm{sgn}(D^1)$, where $D^1$ is the $x$-component of $D^\alpha =
P^+ \Delta^\alpha - \Delta^+ P^\alpha$ and $D^1=0$ corresponds to $t=t_0$.
\subsection{Overlap formalism}
We evaluate all the GPDs in light front quark-diquark model\cite{MC} using the overlap representation of light front wave functions. Both the scalar and axial vector diquark are considered in this quark-diquark model and the nucleon wave functions are constructed from the framework of soft-wall AdS/QCD correspondence.   In this model, the proton is written 
as a sum of isoscalar-scalar diquark singlet $|u~ S^0\rangle$, isoscalar-vector diquark $|u~ A^0\rangle$ and isovector-vector diquark $|d~ A^1\rangle$ states having a spin-flavor $SU(4)$ structure 
\be 
|P; \pm\rangle = C_S|u~ S^0\rangle^\pm + C_V|u~ A^0\rangle^\pm + C_{VV}|d~ A^1\rangle^\pm. \label{PS_state}
\ee
Where $S$ and $A$ represent the scalar and axial-vector diquark and their superscripts  represent the isospin of that diquark.  
The two particle Fock-state expansion for $J^z =\pm1/2$  can be written for scalar diquark  as
\be
|u~ S\rangle^\pm & =& \int \frac{dx~ d^2\bfk}{2(2\pi)^3\sqrt{x(1-x)}} \bigg[ \psi^{\pm(u)}_{+}(x,\bfk)|+\frac{1}{2}~s; xP^+,\bfk\rangle \nonumber \\
 &+& \psi^{\pm(u)}_{-}(x,\bfk)|-\frac{1}{2}~s; xP^+,\bfk\rangle\bigg].\label{fock_SD}
\ee
 The corresponding  light front wave functions are given by
\be 
\psi^{+(u)}_+(x,\bfk)&=& N_S~ \varphi^{(u)}_{1}(x,\bfk),\nonumber \\
\psi^{+(u)}_-(x,\bfk)&=& N_S\bigg(- \frac{k^1+ik^2}{xM} \bigg)\varphi^{(u)}_{2}(x,\bfk),\label{LFWF_S}\\
\psi^{-(u)}_+(x,\bfk)&=& N_S \bigg(\frac{k^1-ik^2}{xM}\bigg) \varphi^{(u)}_{2}(x,\bfk),\nonumber \\
\psi^{-(u)}_-(x,\bfk)&=&  N_S~ \varphi^{(u)}_{1}(x,\bfk),\nonumber
\ee
where $|\lambda_q~\lambda_S; xP^+,\bfk\rangle$ represents the two particle state having a quark of helicity $\lambda_q$ and a scalar diquark having helicity $\lambda_S$ ($\lambda_S=s$ in the state indicates scalar).
Similarly the two particle fock-state expansion for vector diquark is written as
\be
|\nu~ A \rangle^\pm & =& \int \frac{dx~ d^2\bfk}{2(2\pi)^3\sqrt{x(1-x)}} \bigg[ \psi^{\pm(\nu)}_{++}(x,\bfk)|+\frac{1}{2}~+1; xP^+,\bfk\rangle \nonumber\\
 &+& \psi^{\pm(\nu)}_{-+}(x,\bfk)|-\frac{1}{2}~+1; xP^+,\bfk\rangle +\psi^{\pm(\nu)}_{+0}(x,\bfk)|+\frac{1}{2}~0; xP^+,\bfk\rangle \nonumber \\
 &+& \psi^{\pm(\nu)}_{-0}(x,\bfk)|-\frac{1}{2}~0; xP^+,\bfk\rangle + \psi^{\pm(\nu)}_{+-}(x,\bfk)|+\frac{1}{2}~-1; xP^+,\bfk\rangle \nonumber\\
 &+& \psi^{\pm(\nu)}_{--}(x,\bfk)|-\frac{1}{2}~-1; xP^+,\bfk\rangle  \bigg].\label{fock_VD}
\ee
Where $|\lambda_q~\lambda_D; xP^+,\bfk\rangle$ is the two-particle state with a quark of helicity $\lambda_q=\pm\frac{1}{2}$ and a vector diquark of helicity $\lambda_D=\pm 1,0(triplet)$.
The light front wave functions for the axial-vector diquark are given as, for $J^z=+1/2$
\be 
\psi^{+(\nu)}_{+~+}(x,\bfk)&=& N^{(\nu)}_1 \sqrt{\frac{2}{3}} \bigg(\frac{k^1-ik^2}{xM}\bigg) \varphi^{(\nu)}_{2}(x,\bfk),\nonumber \\
\psi^{+(\nu)}_{-~+}(x,\bfk)&=& N^{(\nu)}_1 \sqrt{\frac{2}{3}} \varphi^{(\nu)}_{1}(x,\bfk),\nonumber \\
\psi^{+(\nu)}_{+~0}(x,\bfk)&=& - N^{(\nu)}_0 \sqrt{\frac{1}{3}} \varphi^{(\nu)}_{1}(x,\bfk),\label{LFWF_Vp}\\
\psi^{+(\nu)}_{-~0}(x,\bfk)&=& N^{(\nu)}_0 \sqrt{\frac{1}{3}} \bigg(\frac{k^1+ik^2}{xM} \bigg)\varphi^{(\nu)}_{2}(x,\bfk),\nonumber \\
\psi^{+(\nu)}_{+~-}(x,\bfk)&=& 0,~~~~
\psi^{+(\nu)}_{-~-}(x,\bfk)=  0, \nonumber 
\ee
and for $J^z=-1/2$
\be 
\psi^{-(\nu)}_{+~+}(x,\bfk)&=& 0,~~~~
\psi^{-(\nu)}_{-~+}(x,\bfk)= 0,\nonumber \\
\psi^{-(\nu)}_{+~0}(x,\bfk)&=& N^{(\nu)}_0 \sqrt{\frac{1}{3}} \bigg( \frac{k^1-ik^2}{xM} \bigg) \varphi^{(\nu)}_{2}(x,\bfk),\label{LFWF_Vm}\\
\psi^{-(\nu)}_{-~0}(x,\bfk)&=& N^{(\nu)}_0\sqrt{\frac{1}{3}} \varphi^{(\nu)}_{1}(x,\bfk),\nonumber \\
\psi^{-(\nu)}_{+~-}(x,\bfk)&=& - N^{(\nu)}_1 \sqrt{\frac{2}{3}} \varphi^{(\nu)}_{1}(x,\bfk),\nonumber \\
\psi^{-(\nu)}_{-~-}(x,\bfk)&=& N^{(\nu)}_1 \sqrt{\frac{2}{3}} \bigg(\frac{k^1+ik^2}{xM}\bigg) \varphi^{(\nu)}_{2}(x,\bfk),\nonumber
\ee
having flavour index $\nu=u,d$.
We adopt a generic anstz of LFWFs $\varphi^{(\nu)}_i(x,\bfk)$ from the soft-wall AdS/QCD prediction\cite{BT} and introduce 
the parameters $a^\nu_i,~b^\nu_i$ and $\delta^\nu$ as
\be
\varphi_i^{(\nu)}(x,\bfk)=\frac{4\pi}{\kappa}\sqrt{\frac{\log(1/x)}{1-x}}x^{a_i^\nu}(1-x)^{b_i^\nu}\exp\bigg[-\delta^\nu\frac{\bfk^2}{2\kappa^2}\frac{\log(1/x)}{(1-x)^2}\bigg].
\label{LFWF_phi}
\ee
The values of the parameters   at the initial scale $\mu_0=0.313$ GeV are fitted to the nucleon and formfactor data and taken from Ref.\cite{MC}.

This kinematical domain i.e., $0<x<1$ where $x$ is the light front longitudinal momentum fraction carried by the struck quark corresponds to the situation where one removes a quark from the initial proton  with light-front longitudinal momentum $xP^+$ and re-inserts it into the final proton with the same longitudinal momentum. Thus the change in momentum occurs only in the transverse momentum. The particle number remain conserved in this kinematical region which describes the diagonal $n\rightarrow n$ overlaps. The matrix elements $T^{q}_{i}$ and $\widetilde{T}^{q}_{i}$ in the diagonal $2\rightarrow 2$ overlap representation in terms of light-front wave functions for scalar diquark are given by
\be
T^{qS}_{1(2)}&=& \int \frac{d^2\bfk}{16\pi^3}~\Big[\psi_{+q}^{+*}(x',\bfk')\psi_{+q}^+(x'',\bfk'') 
\pm \psi_{+q}^{-*}(x',\bfk')\psi_{+q}^-(x'',\bfk'')\Big],\label{T1}
\\
T^{qS}_{3(4)}&=& \int \frac{d^2\bfk}{16\pi^3}~\Big[\psi_{+q}^{+*}(x',\bfk')\psi_{+q}^-(x'',\bfk'') 
\mp \psi_{+q}^{-*}(x',\bfk')\psi_{+q}^+(x'',\bfk'')\Big],\label{T2}
\\
\widetilde{T}^{qS}_{1(2)}&=& \int \frac{d^2\bfk}{16\pi^3}~\Big[\psi_{+q}^{+*}(x',\bfk')\psi_{-q}^-(x'',\bfk'') 
\pm \psi_{+q}^{-*}(x',\bfk')\psi_{-q}^+(x'',\bfk'')\Big],\label{Tt1}
\\
\widetilde{T}^{qS}_{3(4)}&=& \int \frac{d^2\bfk}{16\pi^3}~\Big[\psi_{+q}^{+*}(x',\bfk')\psi_{-q}^+(x'',\bfk'') 
\mp \psi_{+q}^{-*}(x',\bfk')\psi_{-q}^-(x'',\bfk'')\Big],\label{Tt2}
\ee
and for the vector diquark the matrix elements are 
\be
T^{qA}_{1(2)}&=& \int \frac{d^2\bfk}{16\pi^3}~\Big[\Big\{\psi_{++q}^{+*}(x',\bfk')\psi_{++q}^+(x'',\bfk'') 
+\psi_{+0q}^{+*}(x',\bfk')\psi_{+0q}^+(x'',\bfk'')\nonumber\\
&&+\psi_{+-q}^{+*}(x',\bfk')\psi_{+-q}^+(x'',\bfk'')\Big\} \pm \Big\{\psi_{++q}^{-*}(x',\bfk')\psi_{++q}^-(x'',\bfk'')\nonumber\\
&&+\psi_{+0q}^{-*}(x',\bfk')\psi_{+0q}^-(x'',\bfk'')+\psi_{+-q}^{-*}(x',\bfk')\psi_{+-q}^-(x'',\bfk'')\Big\}\Big],\label{T1_v}
\ee
\be
T^{qA}_{3(4)}&=& \int \frac{d^2\bfk}{16\pi^3}~\Big[\Big\{\psi_{++q}^{+*}(x',\bfk')\psi_{++q}^-(x'',\bfk'')+\psi_{+0q}^{+*}(x',\bfk')\psi_{+0q}^-(x'',\bfk'')\nonumber\\
&&+\psi_{+-q}^{+*}(x',\bfk')\psi_{+-q}^-(x'',\bfk'') \Big\}
\mp \Big\{\psi_{++q}^{-*}(x',\bfk')\psi_{++q}^+(x'',\bfk'')\nonumber\\
&&+\psi_{+0q}^{-*}(x',\bfk')\psi_{+0q}^+(x'',\bfk'')+\psi_{+-q}^{-*}(x',\bfk')\psi_{+-q}^+(x'',\bfk'')\Big\}\Big],\label{T2_v}
\ee
\be
\widetilde{T}^{qA}_{1(2)}&=& \int \frac{d^2\bfk}{16\pi^3}~\Big[\Big\{\psi_{++q}^{+*}(x',\bfk')\psi_{-+q}^-(x'',\bfk'')+\psi_{+0q}^{+*}(x',\bfk')\psi_{-0q}^-(x'',\bfk'')\nonumber\\
&&+\psi_{+-q}^{+*}(x',\bfk')\psi_{--q}^-(x'',\bfk'')\Big\} 
\pm \Big\{\psi_{++q}^{-*}(x',\bfk')\psi_{-+q}^+(x'',\bfk'')\nonumber\\
&&+\psi_{+0q}^{-*}(x',\bfk')\psi_{-0q}^+(x'',\bfk'')+\psi_{+-q}^{-*}(x',\bfk')\psi_{--q}^+(x'',\bfk'')\Big\}\Big],\label{Tt1_v}
\ee
\be
\widetilde{T}^{qA}_{3(4)}&=& \int \frac{d^2\bfk}{16\pi^3}~\Big[\Big\{\psi_{++q}^{+*}(x',\bfk')\psi_{-+q}^+(x'',\bfk'')+\psi_{+0q}^{+*}(x',\bfk')\psi_{-0q}^+(x'',\bfk'')\nonumber\\
&&+\psi_{+-q}^{+*}(x',\bfk')\psi_{--q}^+(x'',\bfk'')\Big\} 
\mp \Big\{\psi_{++q}^{-*}(x',\bfk')\psi_{-+q}^-(x'',\bfk'')\nonumber\\
&&+\psi_{+0q}^{-*}(x',\bfk')\psi_{-0q}^-(x'',\bfk'')+\psi_{+-q}^{-*}(x',\bfk')\psi_{--q}^-(x'',\bfk'')\Big\}\Big],\label{Tt2_v}
\ee
where, for the final struck quark
\be
x'=\frac{x-\xi}{1-\xi}, \quad\quad\quad \bfk'=\bfk+(1-x')\frac{\bf{\Delta}_{\perp}}{2},
\ee
and for the initial struck quark
\be
x''=\frac{x+\xi}{1+\xi}, \quad\quad\quad \bfk''=\bfk-(1-x'')\frac{\bf{\Delta}_{\perp}}{2}.
\ee
The label $S$ represents the scalar and $A$ denotes the isoscalar-vector(V) diquark corresponding to u quark and isovector-vector(VV) diquark corresponding to d quark. The explicit calculations of the matrix elements $T^{q(S/A)}_{i}$ and $\widetilde{T}^{q(S/A)}_{i}$ using the light front wave functions of the quark-diquark model given in Eqs.(\ref{LFWF_S},\ref{LFWF_Vp},\ref{LFWF_Vm}) give
\be 
T^{qS}_{1(2)}&=&N_S^2 \mathcal{G}^q_{1(2)}, \quad\quad T^{qA}_{1(2)}=\Big(\frac{1}{3}N^{(\nu)2}_0\pm\frac{2}{3}N^{(\nu)2}_1\Big) \mathcal{G}^{q}_{1(2)},
\\
T^{qS}_{3(4)}&=&N_S^2 \mathcal{G}^q_{3(4)}, \quad\quad T^{qA}_{3(4)}=-\frac{1}{3}N^{(\nu)2}_0 \mathcal{G}^{q}_{3(4)},
\\
\widetilde{T}^{qS}_{1(2)}&=&N_S^2 \widetilde{\mathcal{G}}^q_{1(2)}, \quad\quad \widetilde{T}^{qA}_{1(2)}=-\frac{1}{3}N^{(\nu)2}_0\widetilde{\mathcal{G}}^{q}_{1(2)},
\\
\widetilde{T}^{qS}_{3(4)}&=&N_S^2 \widetilde{\mathcal{G}}^q_{3(4)}, \quad\quad \widetilde{T}^{qA}_{3(4)}=\Big(\frac{1}{3}N^{(\nu)2}_0\mp\frac{2}{3}N^{(\nu)2}_1\Big)\widetilde{\mathcal{G}}^{q}_{3(4)},
\ee
where
\be
\mathcal{G}^q_{1(2)}&=&\Big[F_1(x',x'')\frac{1}{A} \pm F_2(x',x'')
\Big\{\frac{1}{A^2}+\Big(\frac{B^2}{4A^2}\nonumber\\
&&-\frac{1}{4}(1-x')(1-x'')+\frac{B}{4A}(x''-x')\Big)\frac{Q^2}{A}\Big\}\Big]
\exp\Big[Q^2\Big(C+\frac{B^2}{4A}\Big)\Big] ,\label{T12F}\\
\mathcal{G}^q_{3(4)}&=&\Big[F_3(x',x'')\Big\{\frac{BQ}{2A^2}-\frac{Q}{2A}(1-x'')\Big\}\nonumber\\
&&\mp F_4(x',x'')
\Big\{\frac{BQ}{2A^2}+\frac{Q}{2A}(1-x'')\Big\}\Big]
\exp\Big[Q^2\Big(C+\frac{B^2}{4A}\Big)\Big] \label{T34F},\\
\widetilde{\mathcal{G}}^q_{1(2)}&=&\Big[F_1(x',x'')\frac{1}{A} \mp F_2(x',x'')
\Big\{\frac{B^2}{4A^2}\nonumber\\
&&-\frac{1}{4}(1-x')(1-x'')+\frac{B}{4A}(x''-x')\Big\}\frac{Q^2}{A}\Big]
\exp\Big[Q^2\Big(C+\frac{B^2}{4A}\Big)\Big] ,\label{Tt12F}\\
\widetilde{\mathcal{G}}^q_{3(4)}&=&-\Big[F_3(x',x'')\Big\{\frac{BQ}{2A^2}-\frac{Q}{2A}(1-x'')\Big\}\nonumber\\
&&\pm F_4(x',x'')
\Big\{\frac{BQ}{2A^2}+\frac{Q}{2A}(1-x'')\Big\}\Big]
\exp\Big[Q^2\Big(C+\frac{B^2}{4A}\Big)\Big] \label{Tt34F},
\ee
with 
\be
F_1(x',x'')&=&\frac{1}{\kappa^2}\Big[{\frac{\log x'\log x''}{(1-x')(1-x'')}}\Big]^{1/2}\Big[(x'x'')^{a_q^{(1)}}\{(1-x')(1-x'')\}^{b_q^{(1)}}\Big],\nonumber\\
F_2(x',x'')&=&\frac{1}{\kappa^2}\Big[{\frac{\log x'\log x''}{(1-x')(1-x'')}}\Big]^{1/2}\Big[\frac{1}{M^2}(x'x'')^{a_q^{(2)}-1}\{(1-x')(1-x'')\}^{b_q^{(2)}}\Big],\nonumber\\
F_3(x',x'')&=&\frac{1}{\kappa^2}\Big[{\frac{\log x'\log x''}{(1-x')(1-x'')}}\Big]^{1/2}\Big[\frac{1}{M}(x')^{a_q^{(1)}}(1-x')^{b_q^{(1)}}(x'')^{a_q^{(2)}-1}(1-x'')^{b_q^{(2)}}\Big],\nonumber\\
F_4(x',x'')&=&\frac{1}{\kappa^2}\Big[{\frac{\log x'\log x''}{(1-x')(1-x'')}}\Big]^{1/2}\Big[\frac{1}{M}(x')^{a_q^{(2)}-1}(1-x')^{b_q^{(2)}}(x'')^{a_q^{(1)}}(1-x'')^{b_q^{(1)}}\Big],\nonumber
\ee
and $A$, $B$ and $C$ are functions of $x'$ and $x''$
\be
A&=&A^q(x',x'')=-\frac{\delta^q\log x'}{2\kappa^2(1-x')^2}-\frac{\delta^q\log x''}{2\kappa^2(1-x'')^2},\nonumber\\
B&=&B^q(x',x'')=\frac{\delta^q\log x'}{2\kappa^2(1-x')}-\frac{\delta^q\log x''}{2\kappa^2(1-x'')},\nonumber\\
C&=&C^q(x',x'')=\frac{1}{4}\Big[\frac{\delta^q\log x'}{2\kappa^2}+\frac{\delta^q\log x''}{2\kappa^2}\Big].\nonumber
\ee
Combining the contributions from scalar and vector parts, one can write the matrix elements $T^{q}_{i}$ for $u$ and $d$ as 
\be 
T^{u}_{i}&=&C_S^2 T^{uS}_{i}+C_V^2 T^{uA}_{i},\label{Tu}\\
T^{d}_{i}&=&C_{VV}^2 T^{dA}_{i},\label{Td}
\ee
$\widetilde{T}^{q}_{i}$ also follows a similar expression as Eqs.(\ref{Tu}-\ref{Td}). Using the matrix elements $T^q_i~(\widetilde{T}^{q}_{i})$ calculated in Eqs.(\ref{Tu}-\ref{Td}), we evaluate all the GPDs in Eqs.(\ref{H}-\ref{EtT}).

\begin{figure}[H]
\begin{minipage}[c]{0.98\textwidth}
\small{(a)}
\includegraphics[width=7cm,clip]{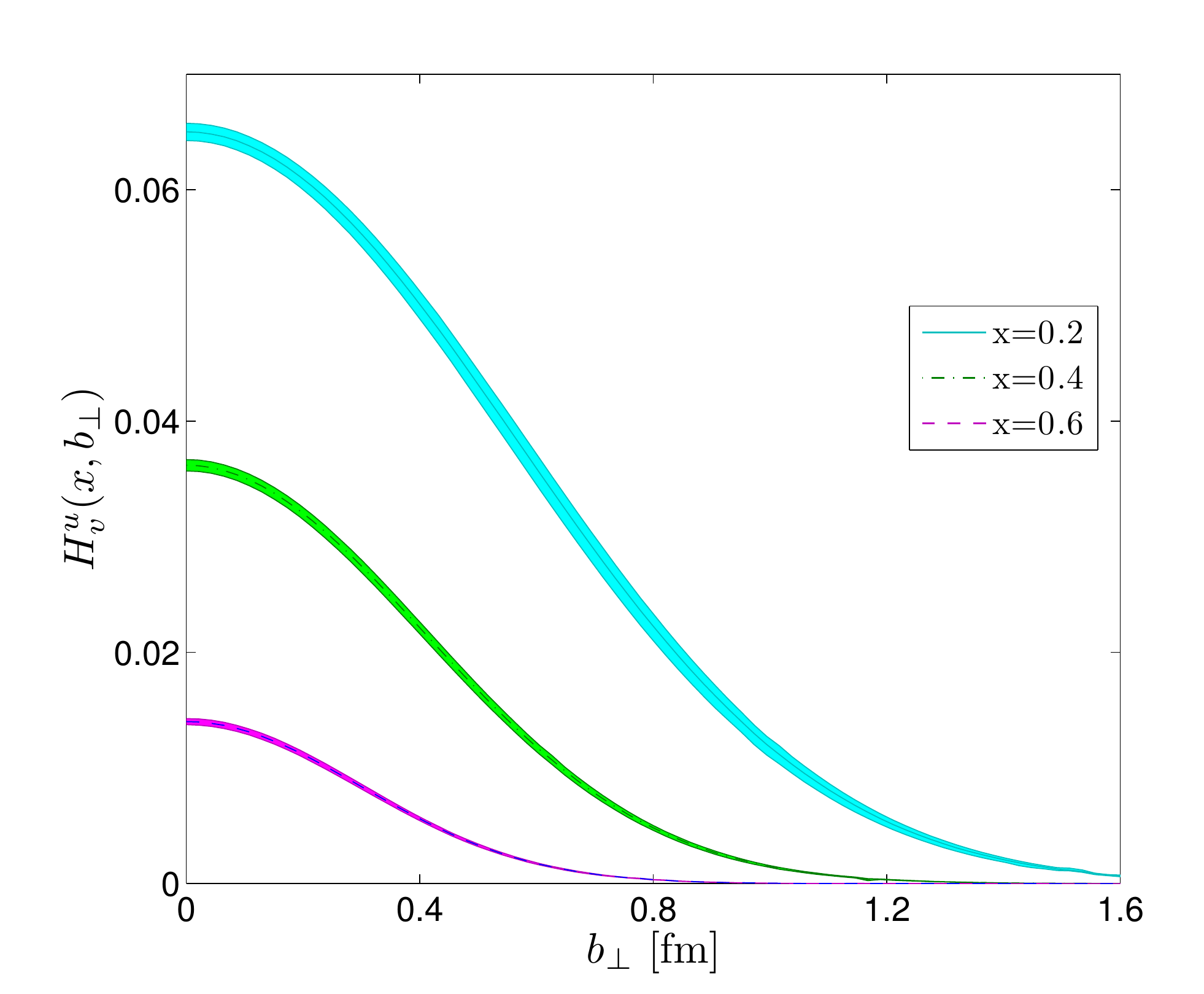}
\hspace{0.1cm}%
\small{(b)}\includegraphics[width=7cm,clip]{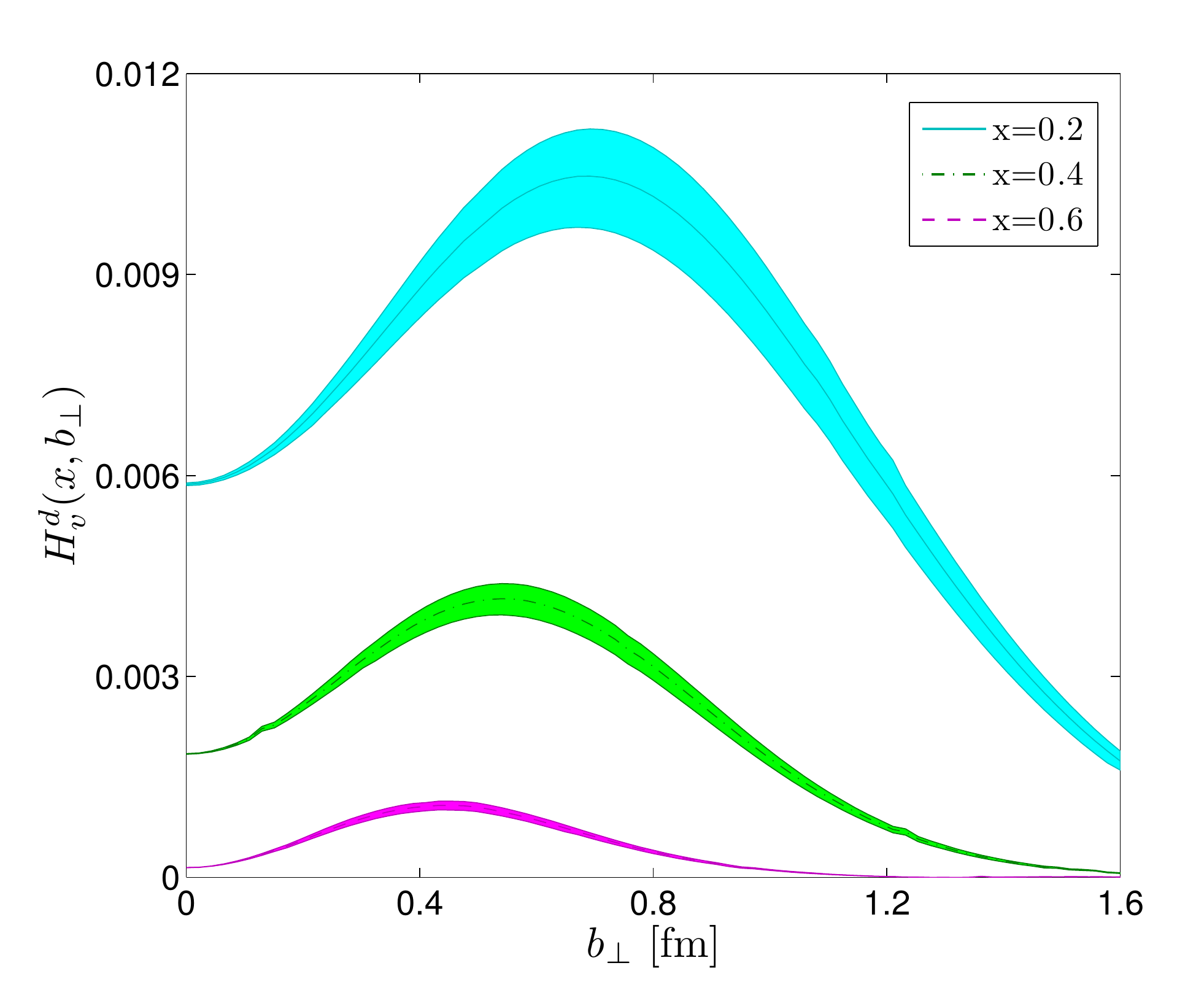}
\end{minipage}
\begin{minipage}[c]{0.98\textwidth}
\small{(c)}\includegraphics[width=7cm,clip]{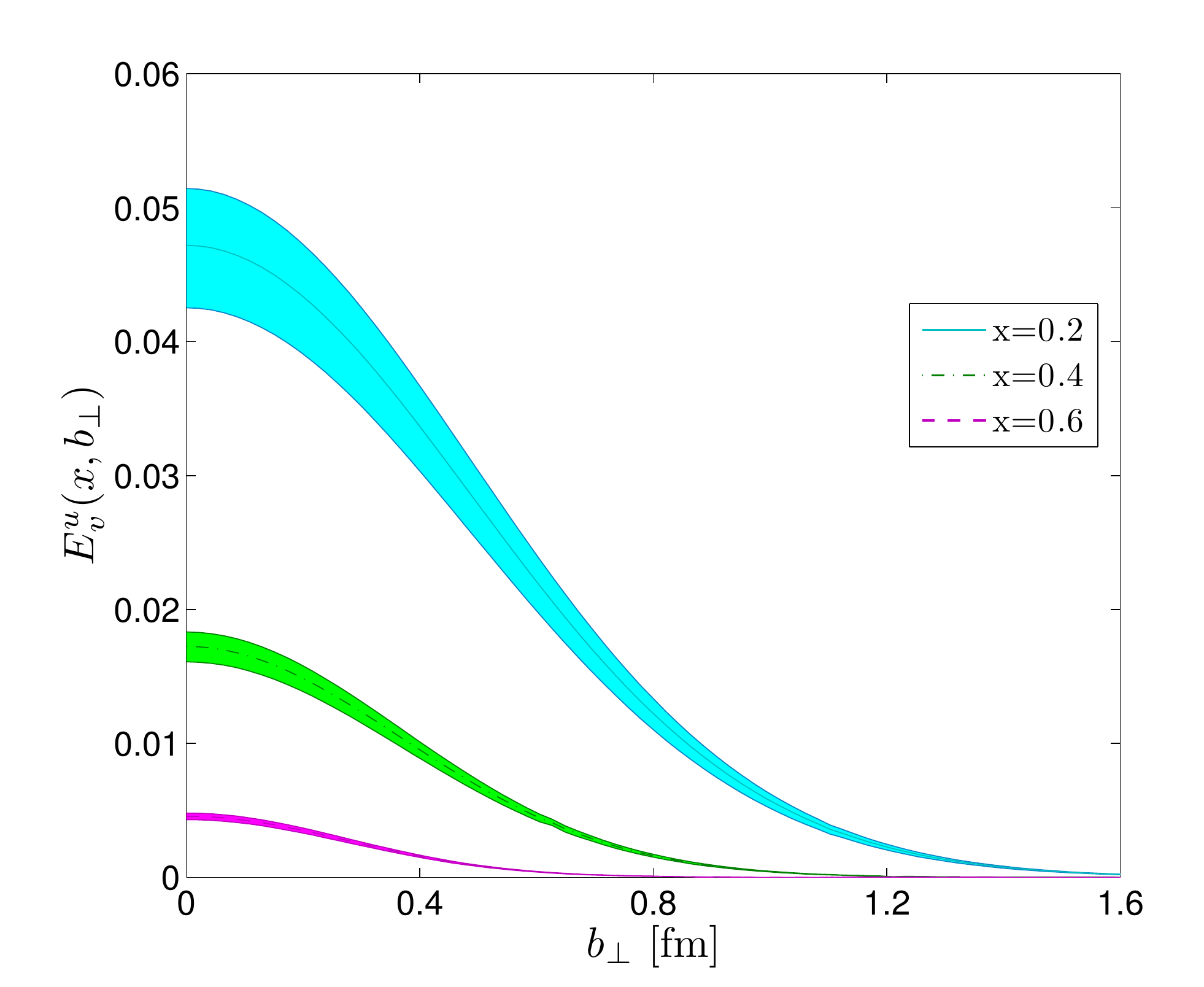}
\hspace{0.1cm}%
\small{(d)}\includegraphics[width=7cm,clip]{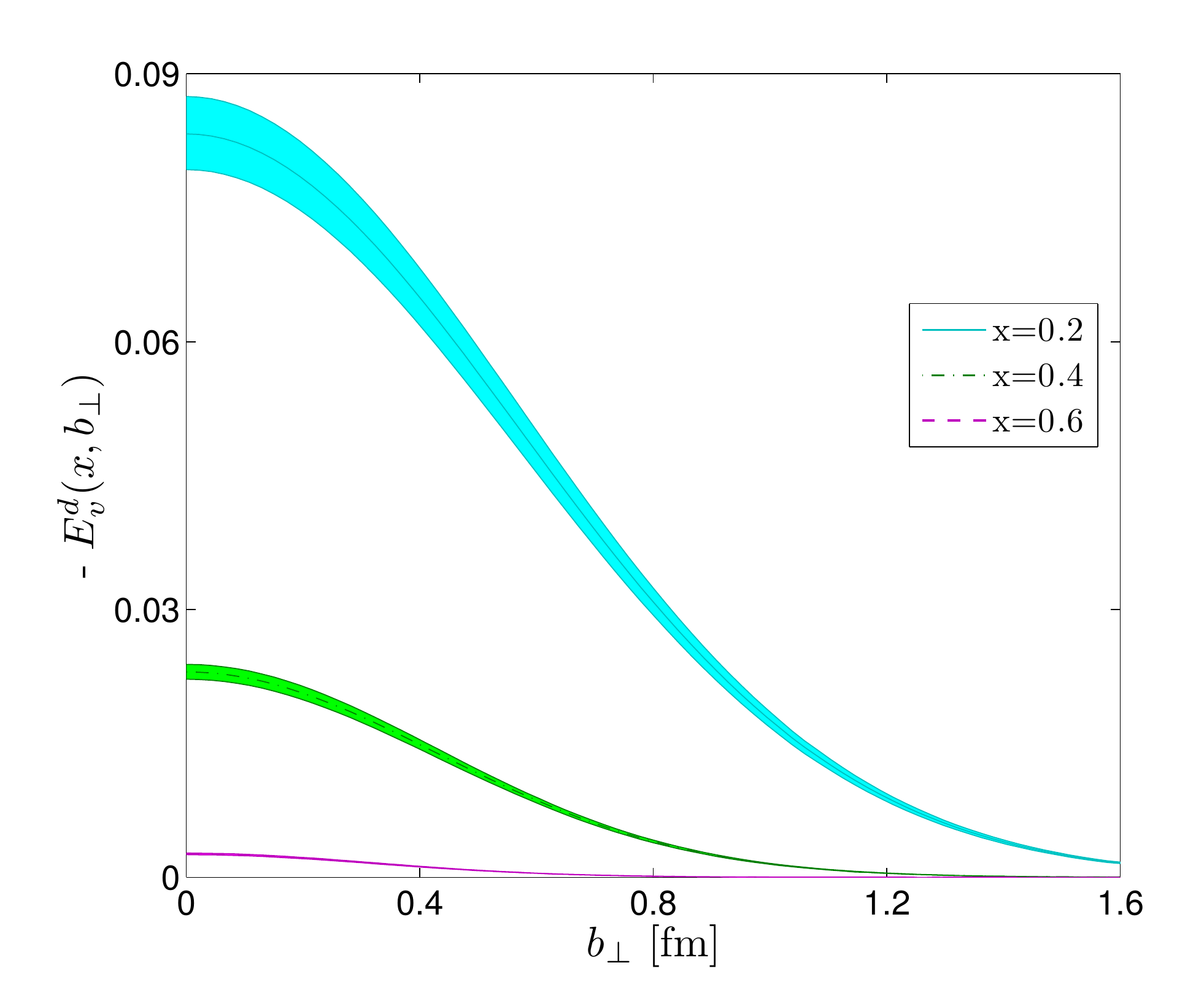}
\end{minipage}
\begin{minipage}[c]{0.98\textwidth}
\small{(e)}
\includegraphics[width=7cm,clip]{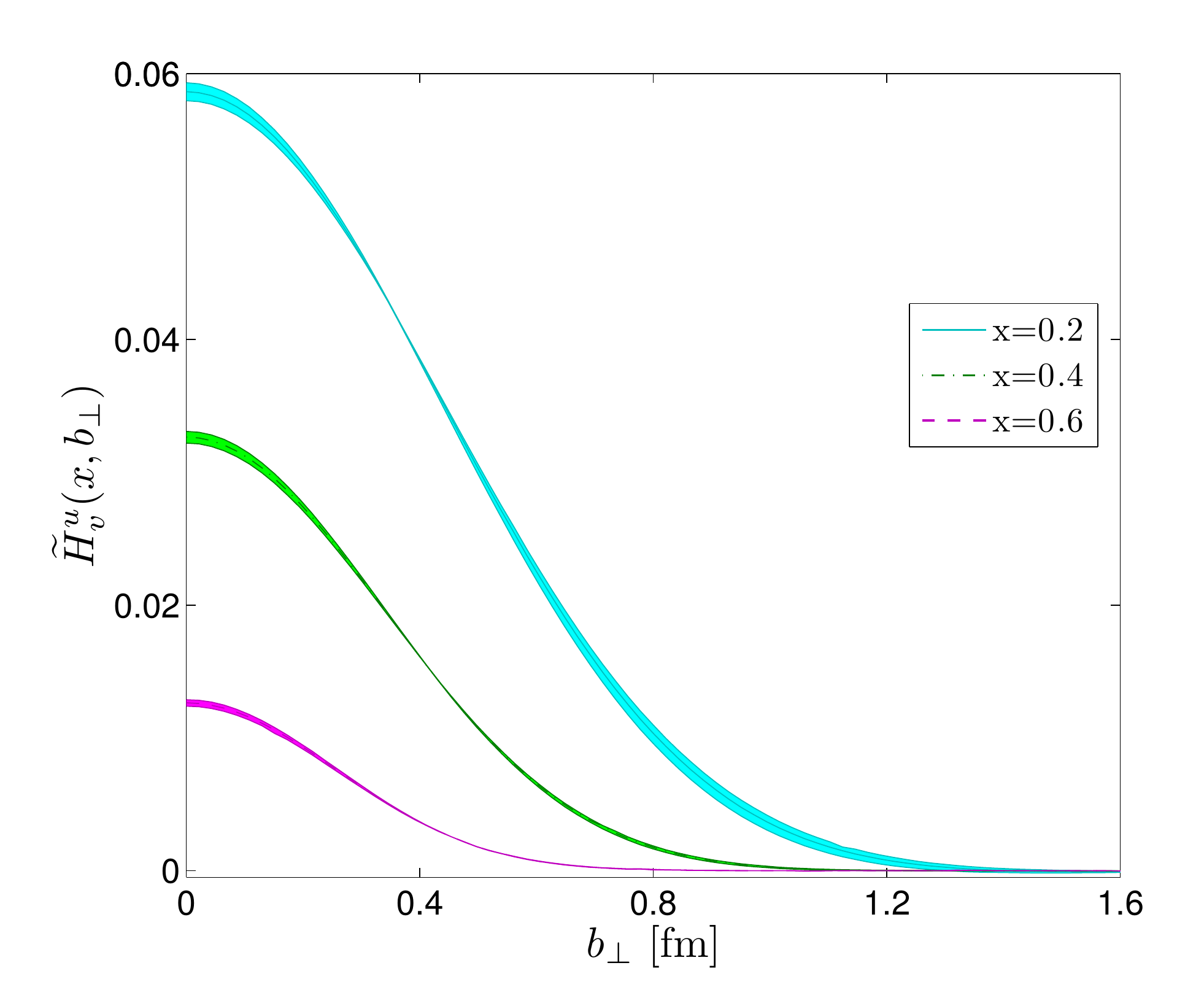}
\hspace{0.1cm}%
\small{(f)}\includegraphics[width=7cm,clip]{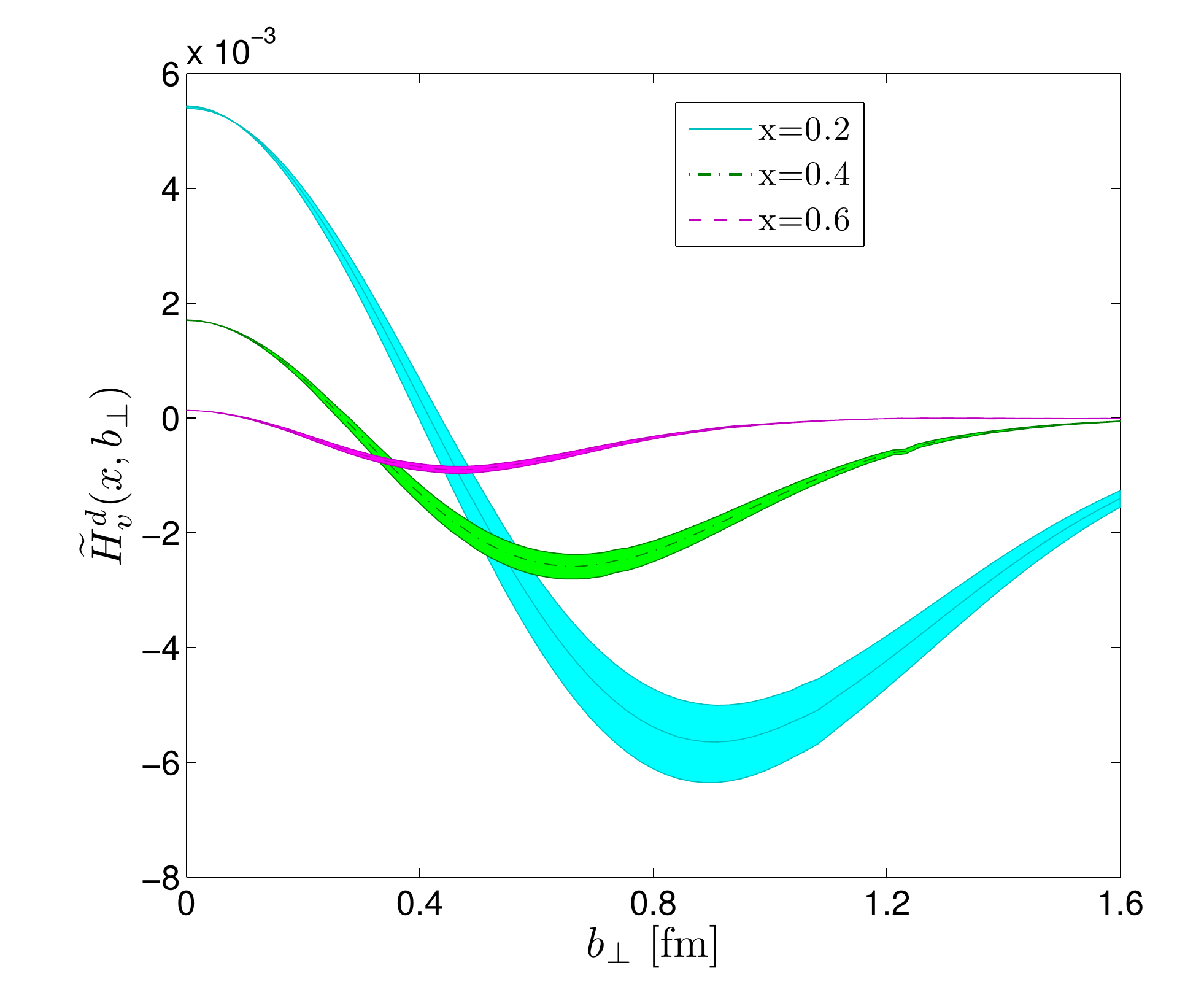}
\end{minipage}
\begin{minipage}[c]{0.98\textwidth}
\small{(g)}\includegraphics[width=7cm,clip]{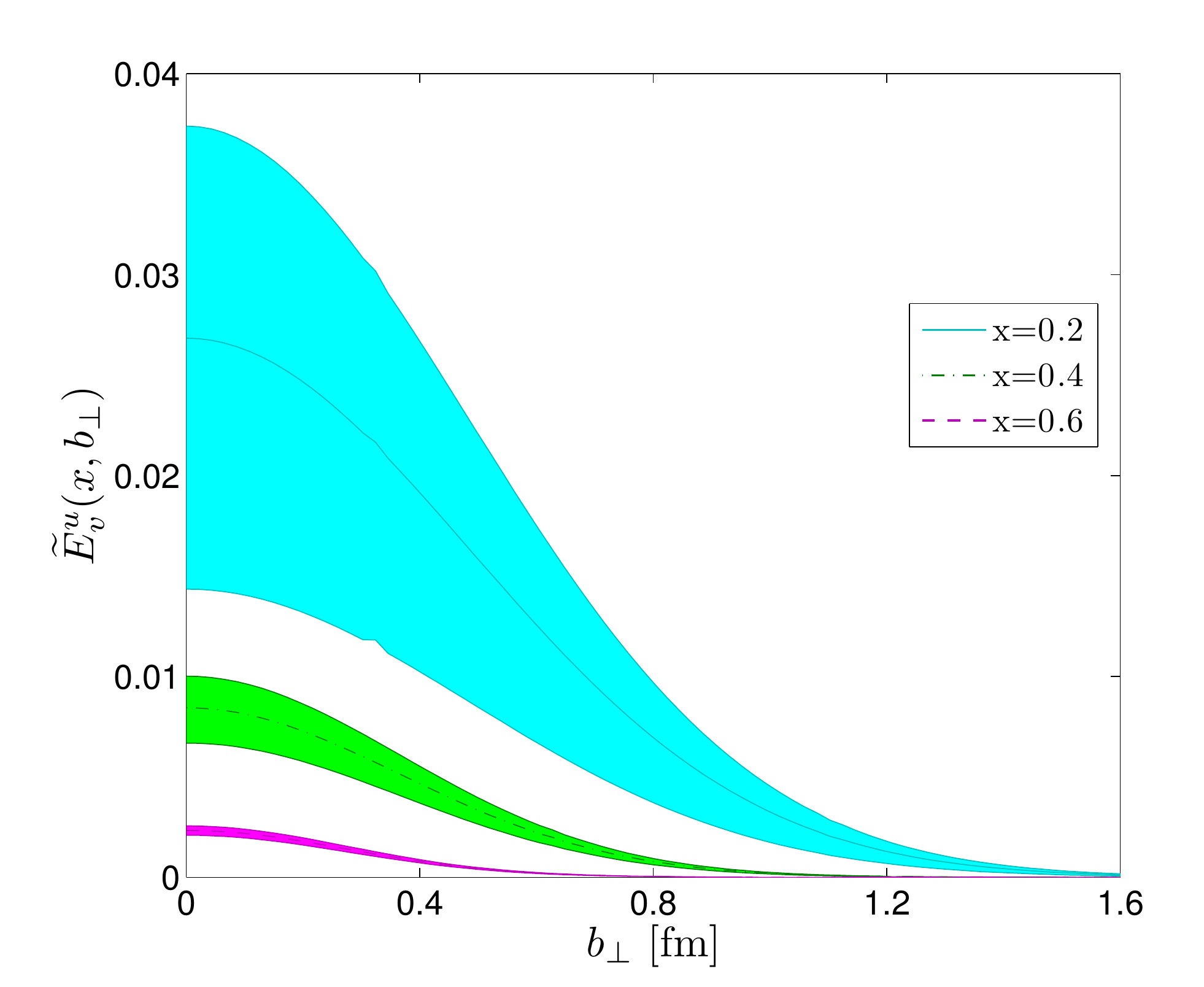}
\hspace{0.1cm}%
\small{(h)}\includegraphics[width=7cm,clip]{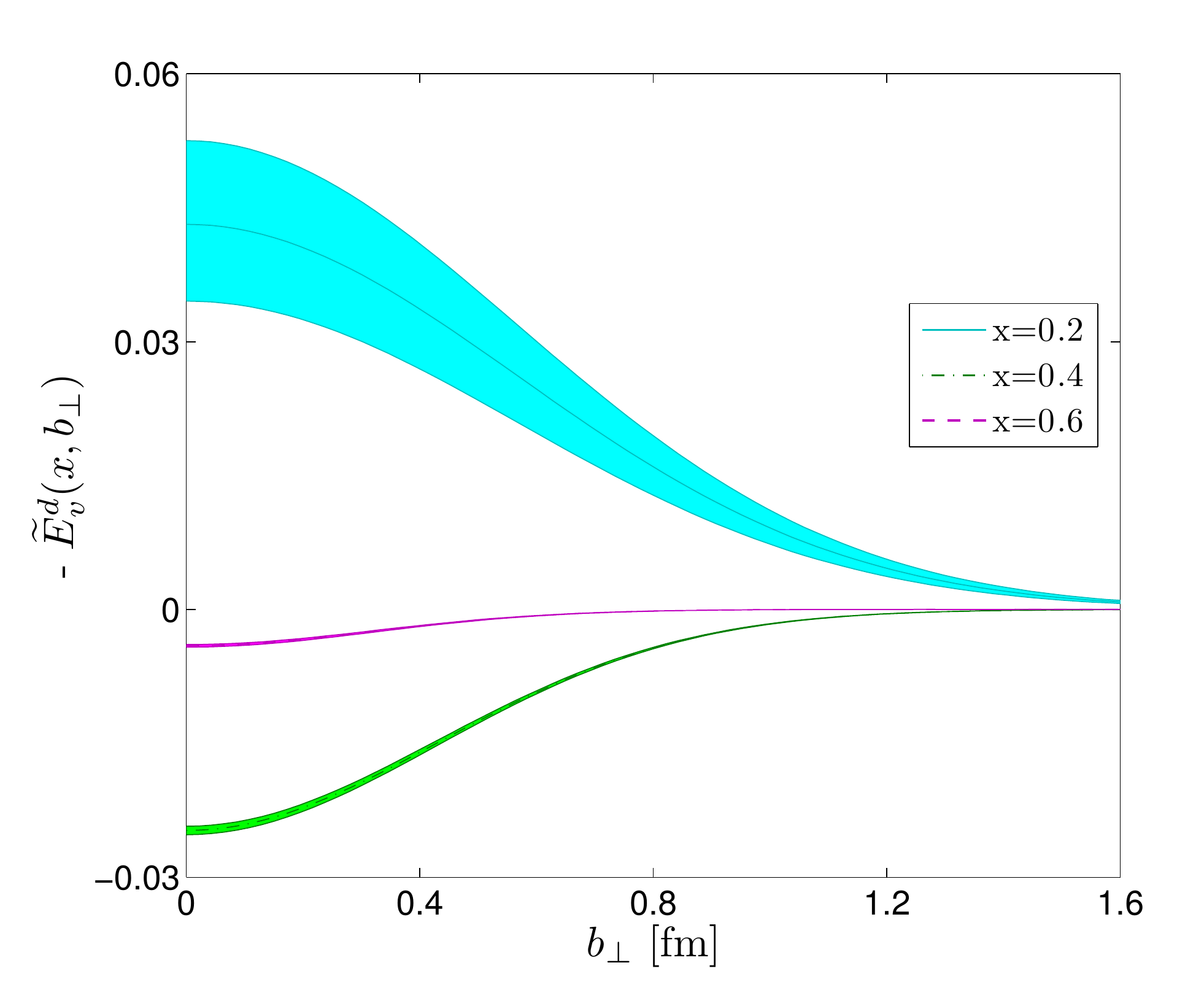}
\end{minipage}
\caption{\label{impact_b}(Color online) Plots of chiral-even GPDs in impact parameter space as functions 
of $b=|b_{\perp}|$ for different values of $x=0.2,~0.4,~0.6$ at $\mu^2=10$ GeV$^2$. Left panel 
for $u$ quark and right panel for $d$ quark. The error bands correpsond to $2\sigma$ error in the model parameters.}
\end{figure}

\begin{figure}[H]
\begin{minipage}[c]{0.98\textwidth}
\small{(a)}
\includegraphics[width=7cm,clip]{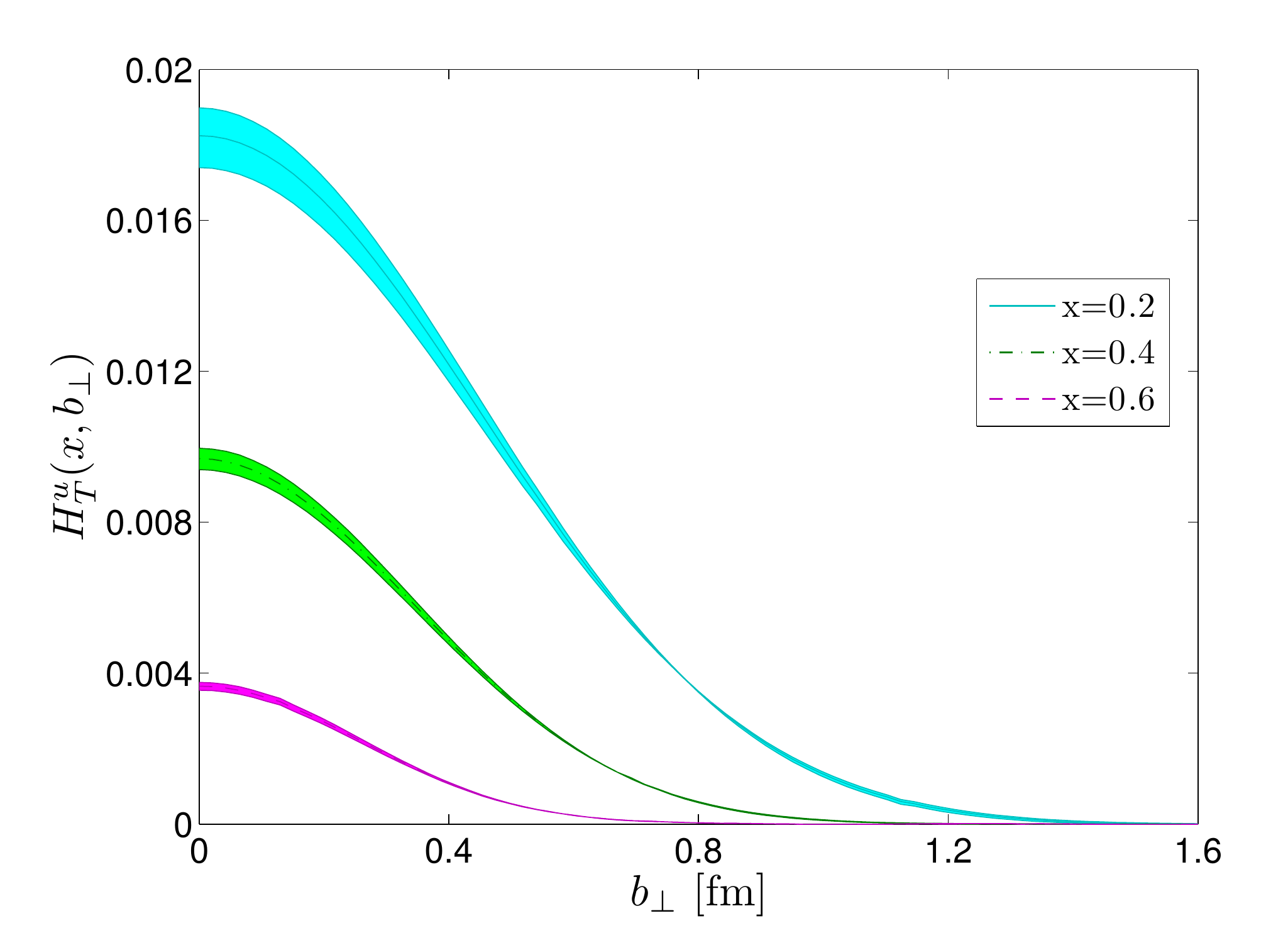}
\hspace{0.1cm}%
\small{(b)}\includegraphics[width=7cm,clip]{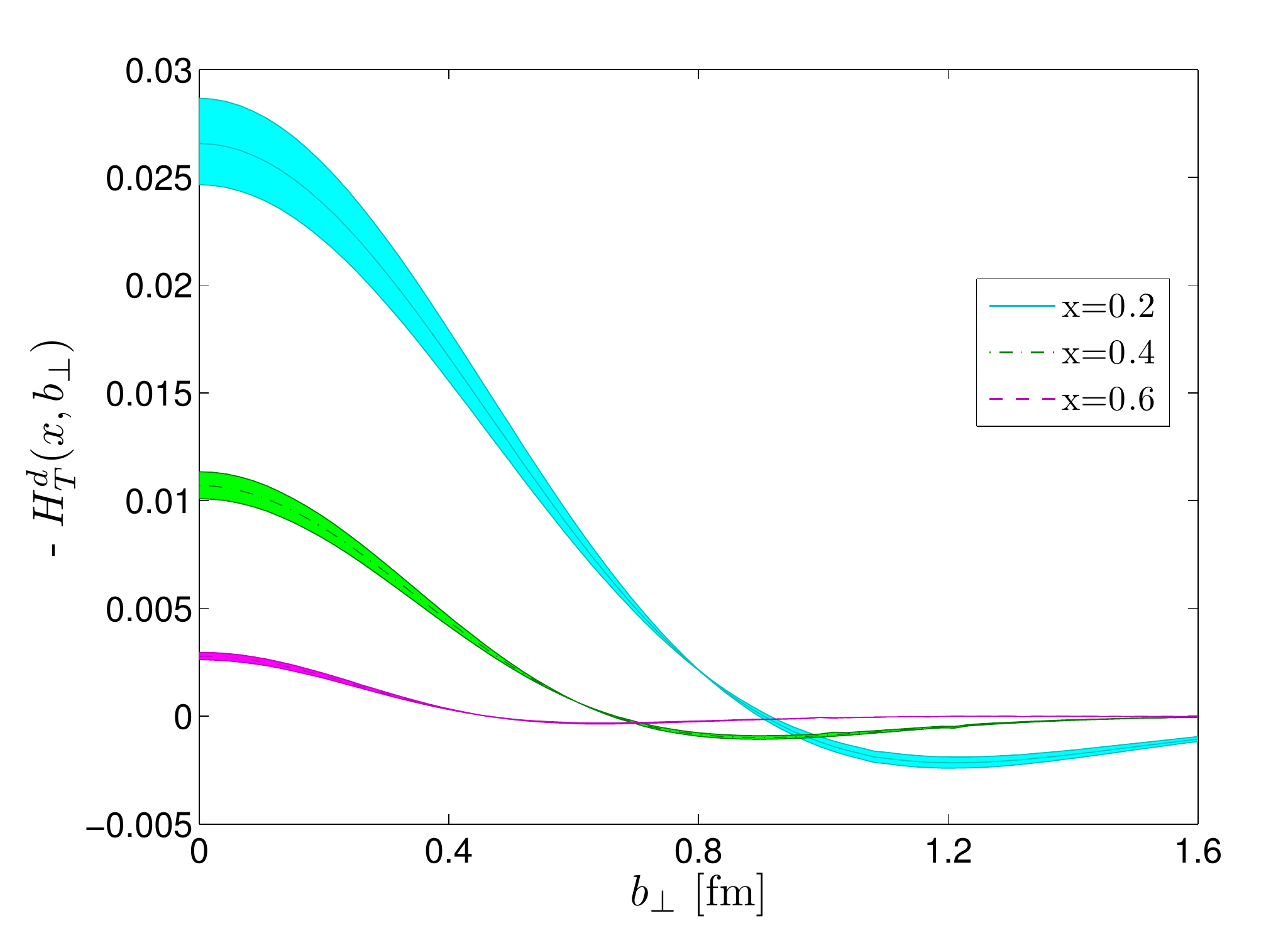}
\end{minipage}
\begin{minipage}[c]{0.98\textwidth}
\small{(c)}\includegraphics[width=7cm,clip]{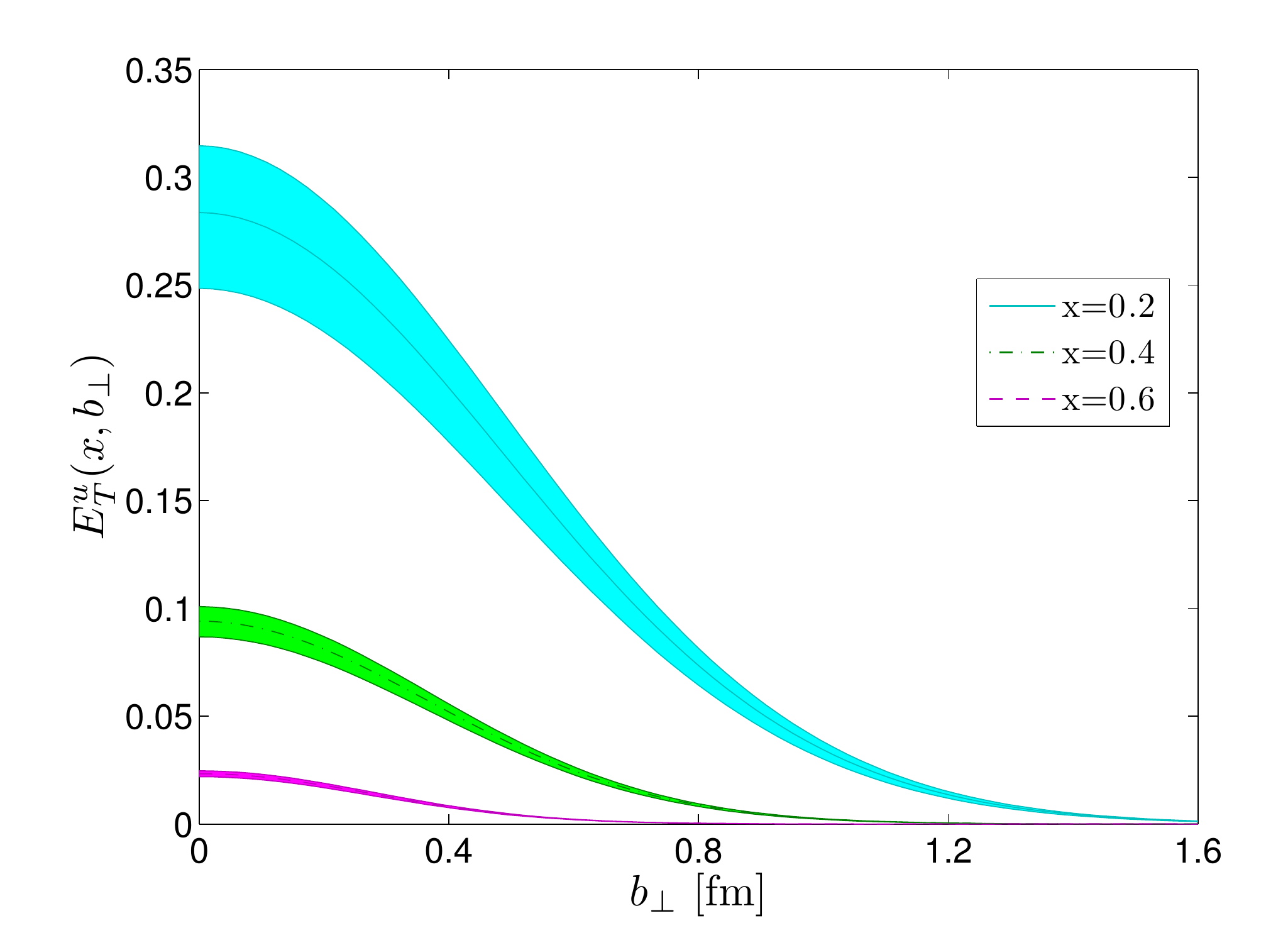}
\hspace{0.1cm}%
\small{(d)}\includegraphics[width=7cm,clip]{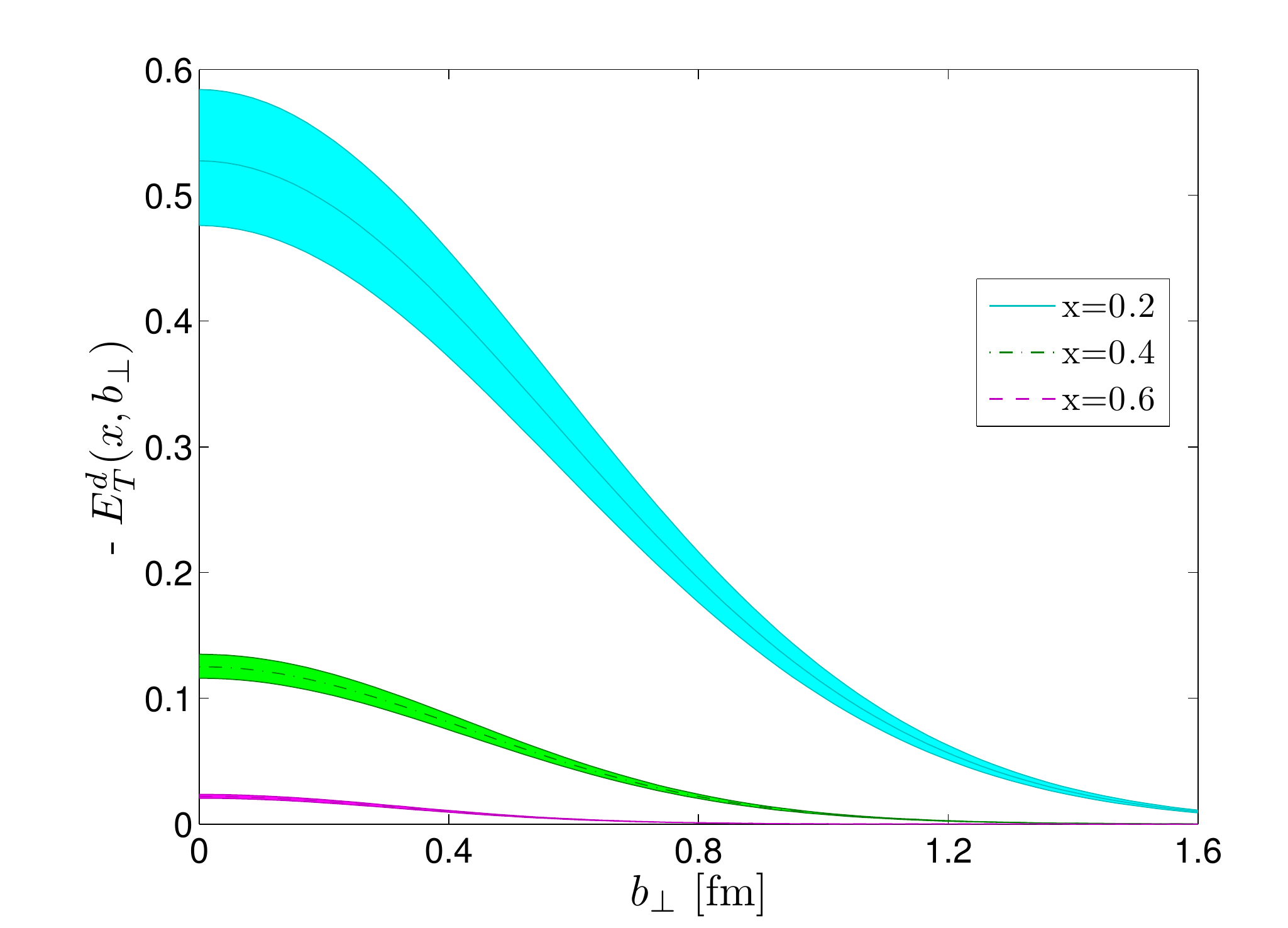}
\end{minipage}
\begin{minipage}[c]{0.98\textwidth}
\small{(e)}
\includegraphics[width=7cm,clip]{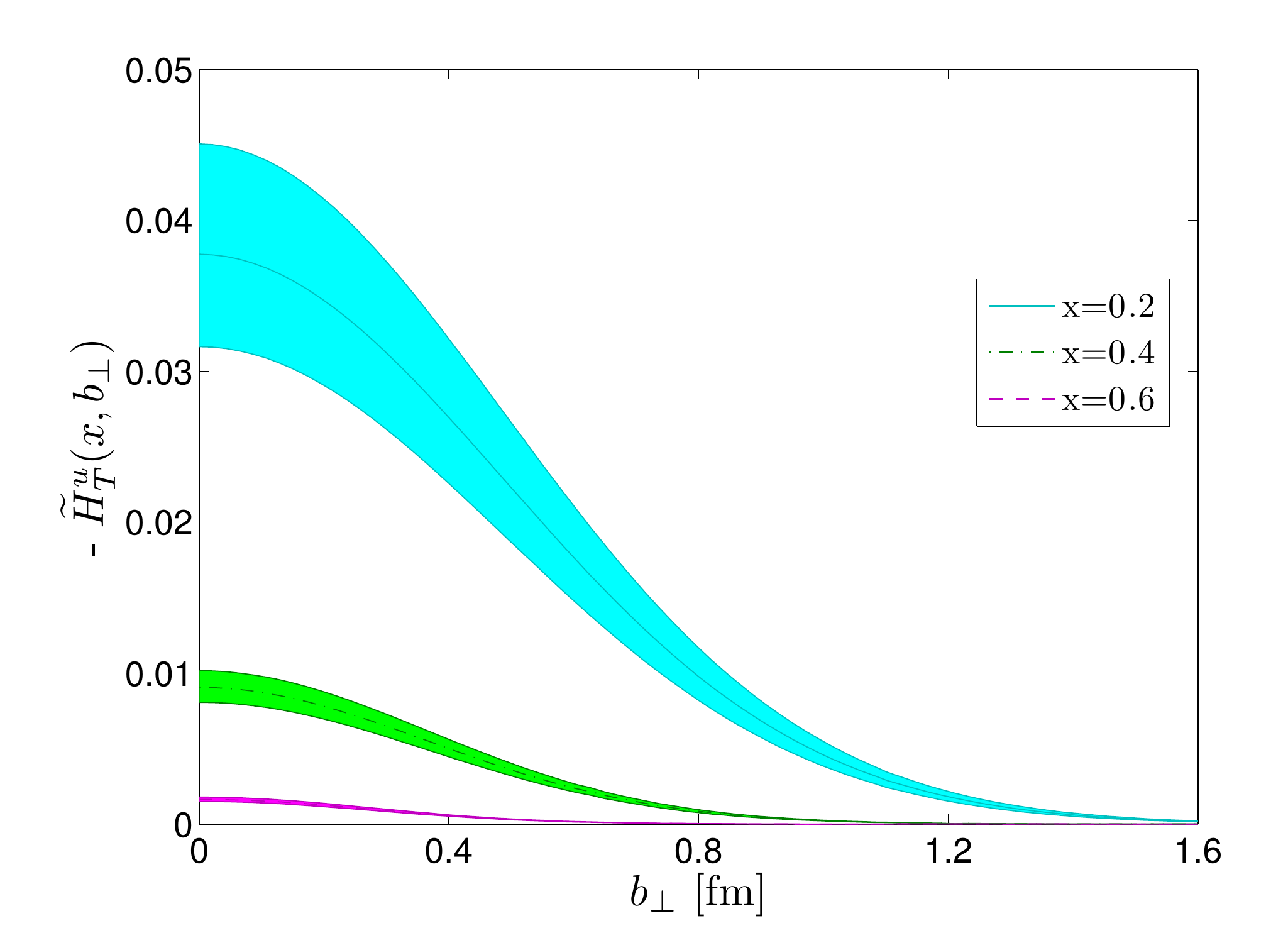}
\hspace{0.1cm}%
\small{(f)}\includegraphics[width=7cm,clip]{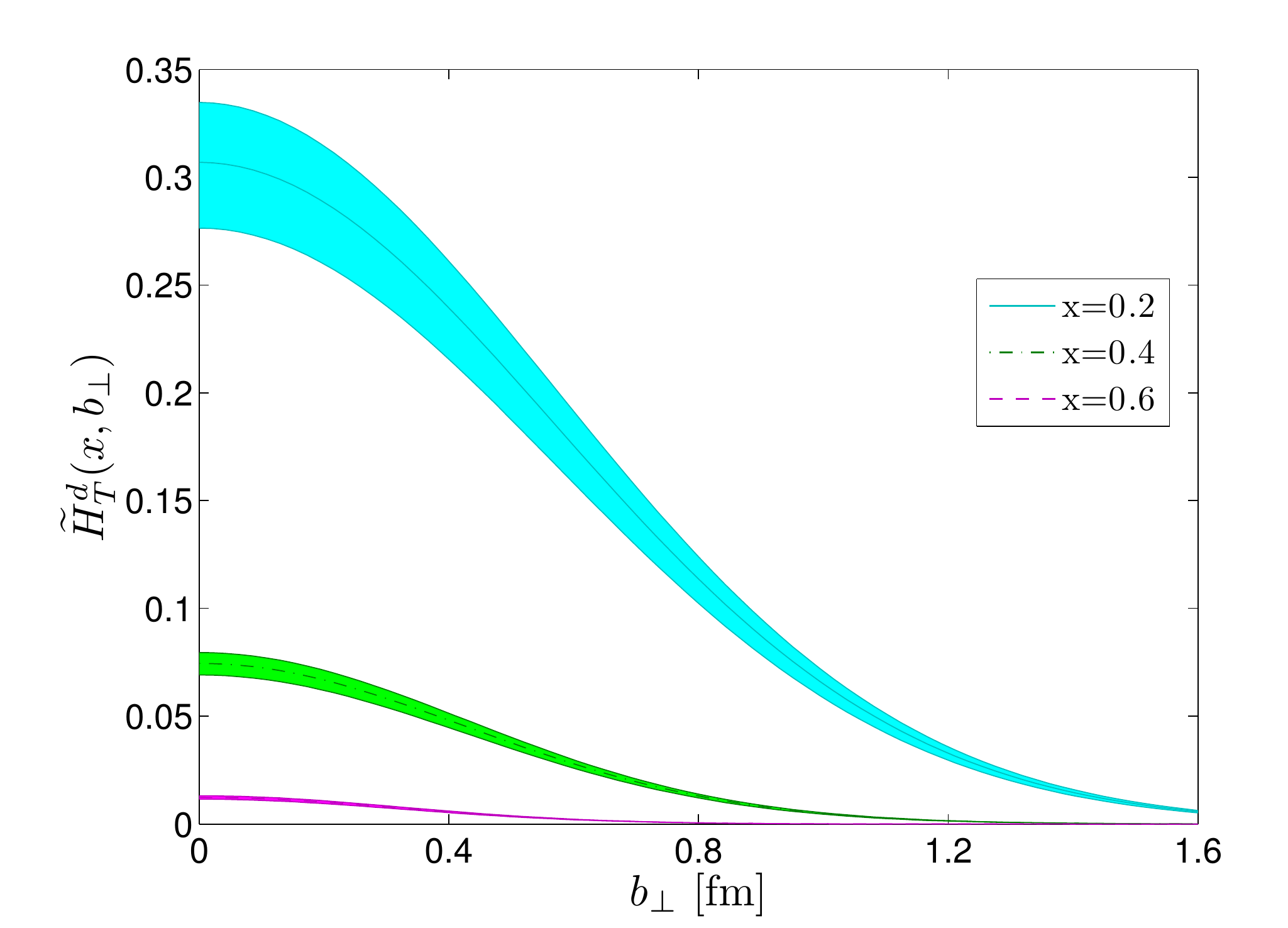}
\end{minipage}
\caption{\label{odd_impact_b}(Color online) Plots of chiral-odd GPDs in impact parameter space as functions of $b=|b_{\perp}|$ 
for different values of $x=0.2,~0.4,~0.6$ at $\mu^2=10$ GeV$^2$. Left panel for $u$ quark and right panel 
for $d$ quark.The error bands correpsond to $2\sigma$ error in the model parameters. }
\end{figure}

The impact parameter dependent   GPDs at zero skewness  
are defined as the Fourier transform of the GPDs with respect to the transverse momentum transferred in the process:
\be
f(x,{b}^2) = \int\frac{d^2{\tvec\Delta}_\perp}{(2\pi)^2} \,e^{-i{\svec b}_\perp\cdot{\svec\Delta}_\perp} \,f(x,\xi=0,t=-{\tvec\Delta}_\perp^2),
\label{eq:fourier}
\ee
where $b=|\bfb|$ is the transverse impact parameter. In Fig.\ref{impact_b} and Fig.\ref{odd_impact_b}, we show chiral
even and chiral odd GPDs respectively in transverse impact parameter space for zero skewness at
experimentally accessible scale $\mu^2=10$ GeV$^2$. $\widetilde{E}_T^q $ being an 
odd function of $\xi$ is zero for $\xi=0$. The parameters in the model are fitted to the form factor data with least $\chi^2 $ error\cite{MC}. 
The error bands in the plots correspond to $2\sigma$ errors in the least $\chi^2$ fitting.
\subsection{Evolution of GPDs}
\begin{figure}[H]
\begin{minipage}[c]{0.98\textwidth}
\small{(a)}
\includegraphics[width=7.5cm,clip]{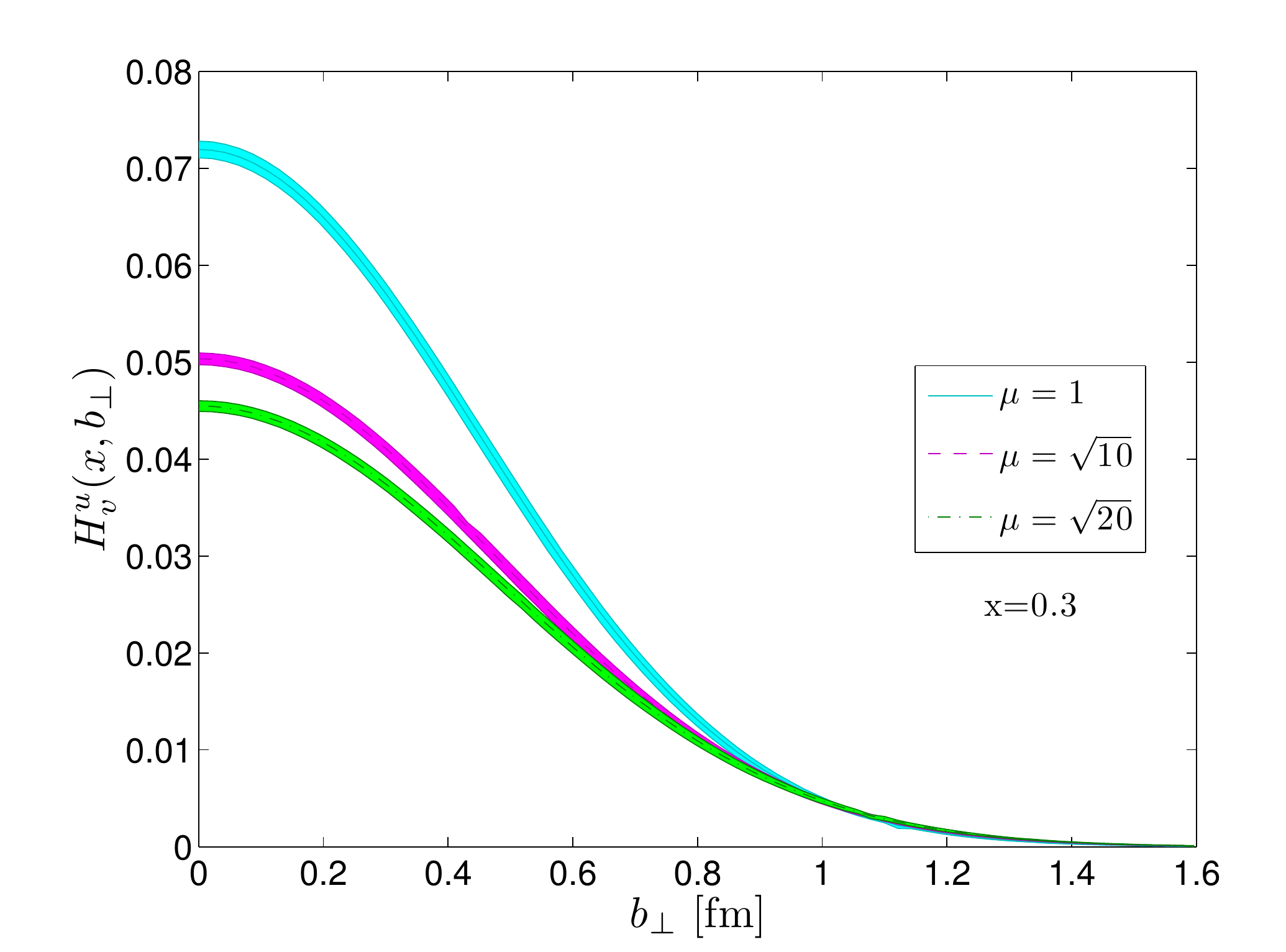}
\hspace{0.1cm}%
\small{(b)}\includegraphics[width=7.5cm,clip]{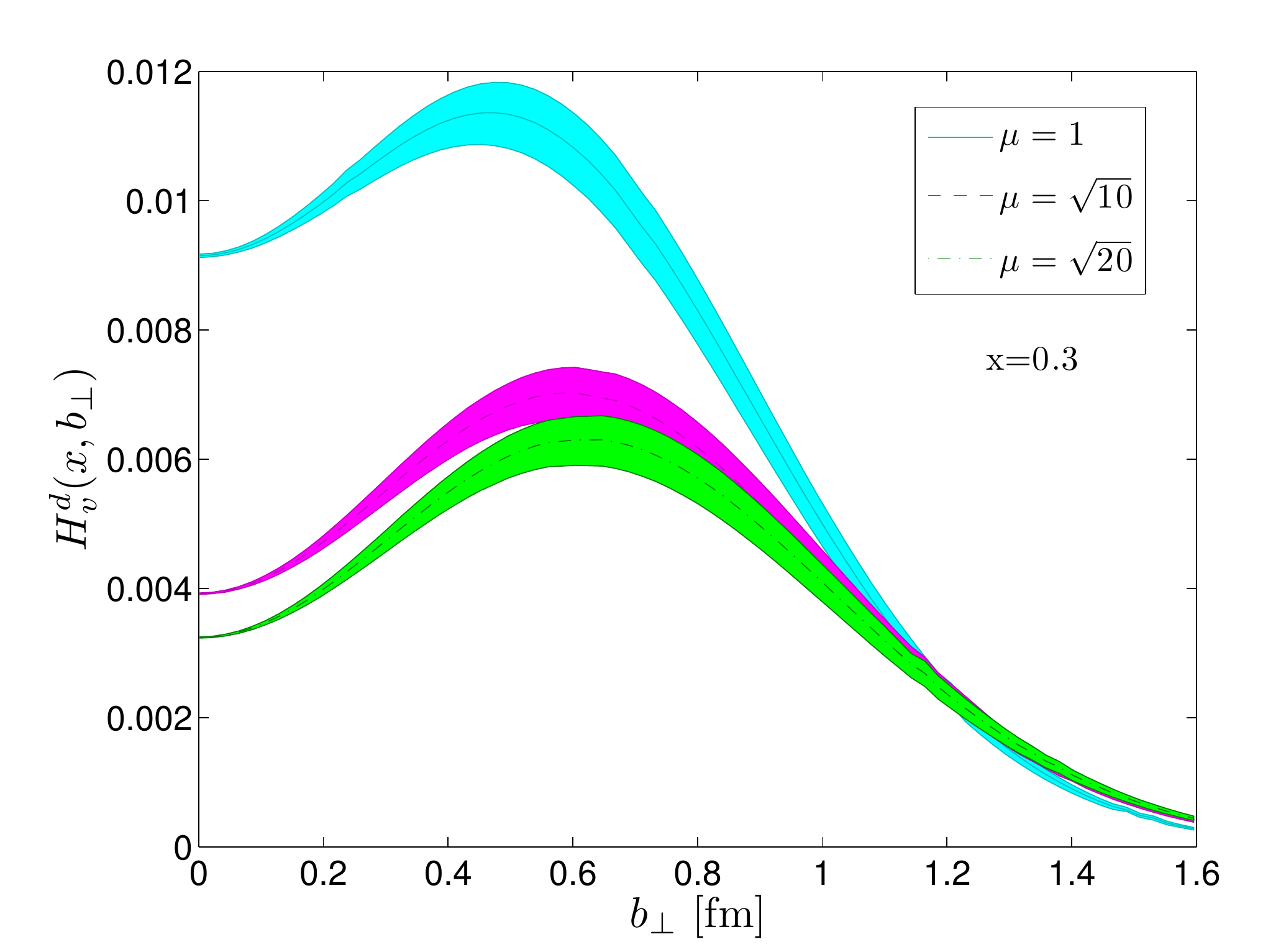}
\end{minipage}
\begin{minipage}[c]{0.98\textwidth}
\small{(c)}\includegraphics[width=7.5cm,clip]{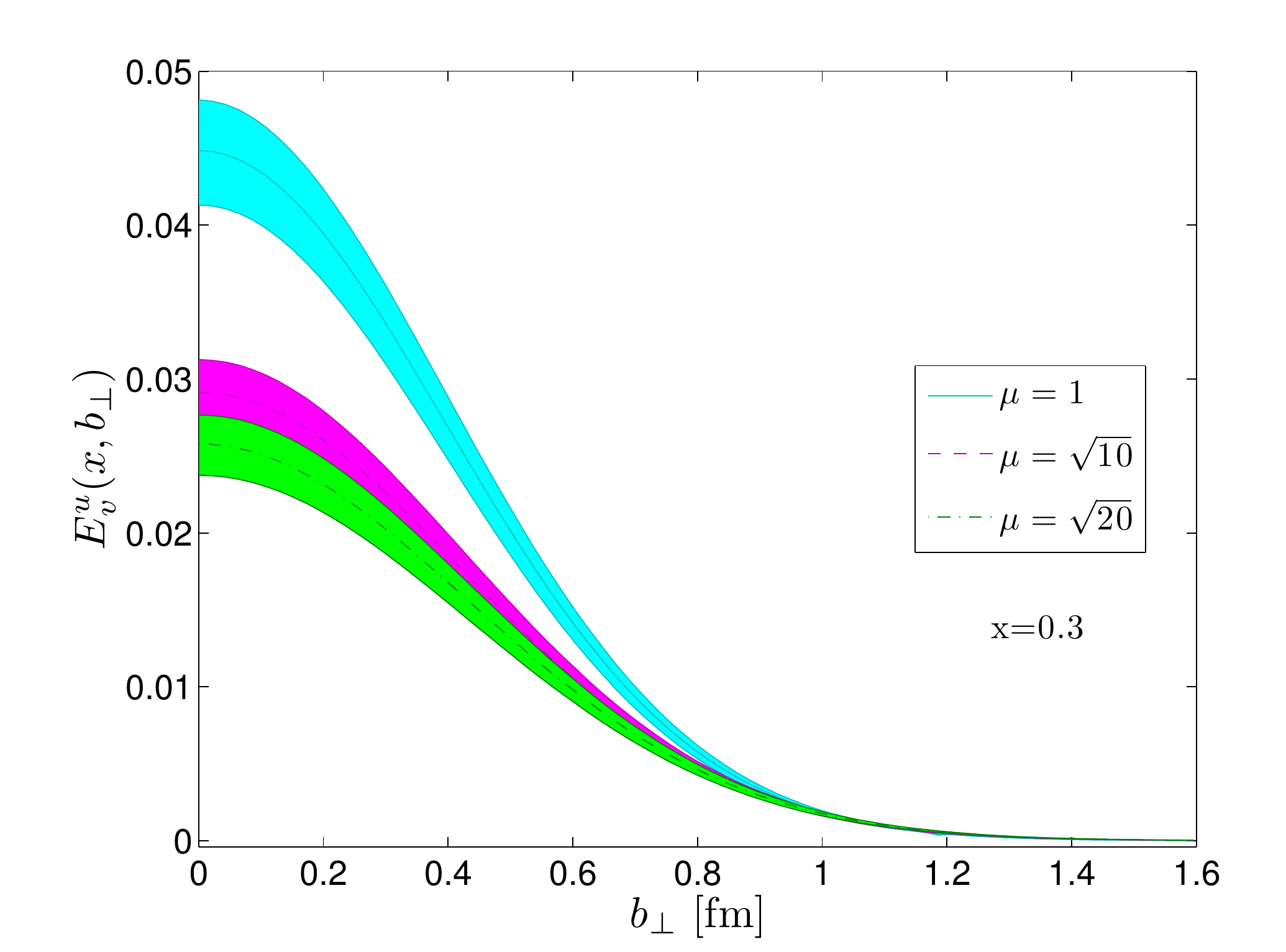}
\hspace{0.1cm}%
\small{(d)}\includegraphics[width=7.5cm,clip]{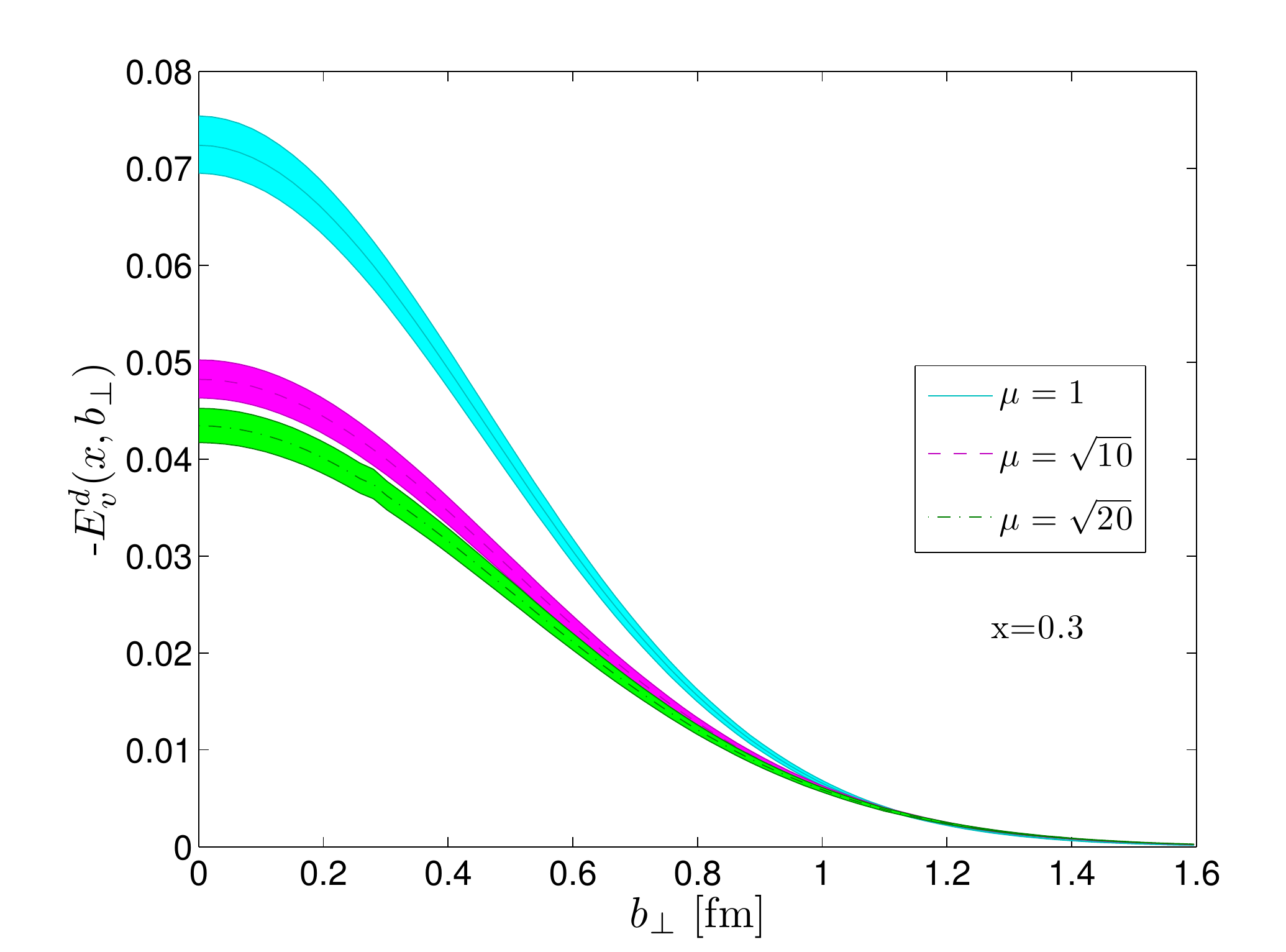}
\end{minipage}
\begin{minipage}[c]{0.98\textwidth}
\small{(e)}
\includegraphics[width=7.5cm,clip]{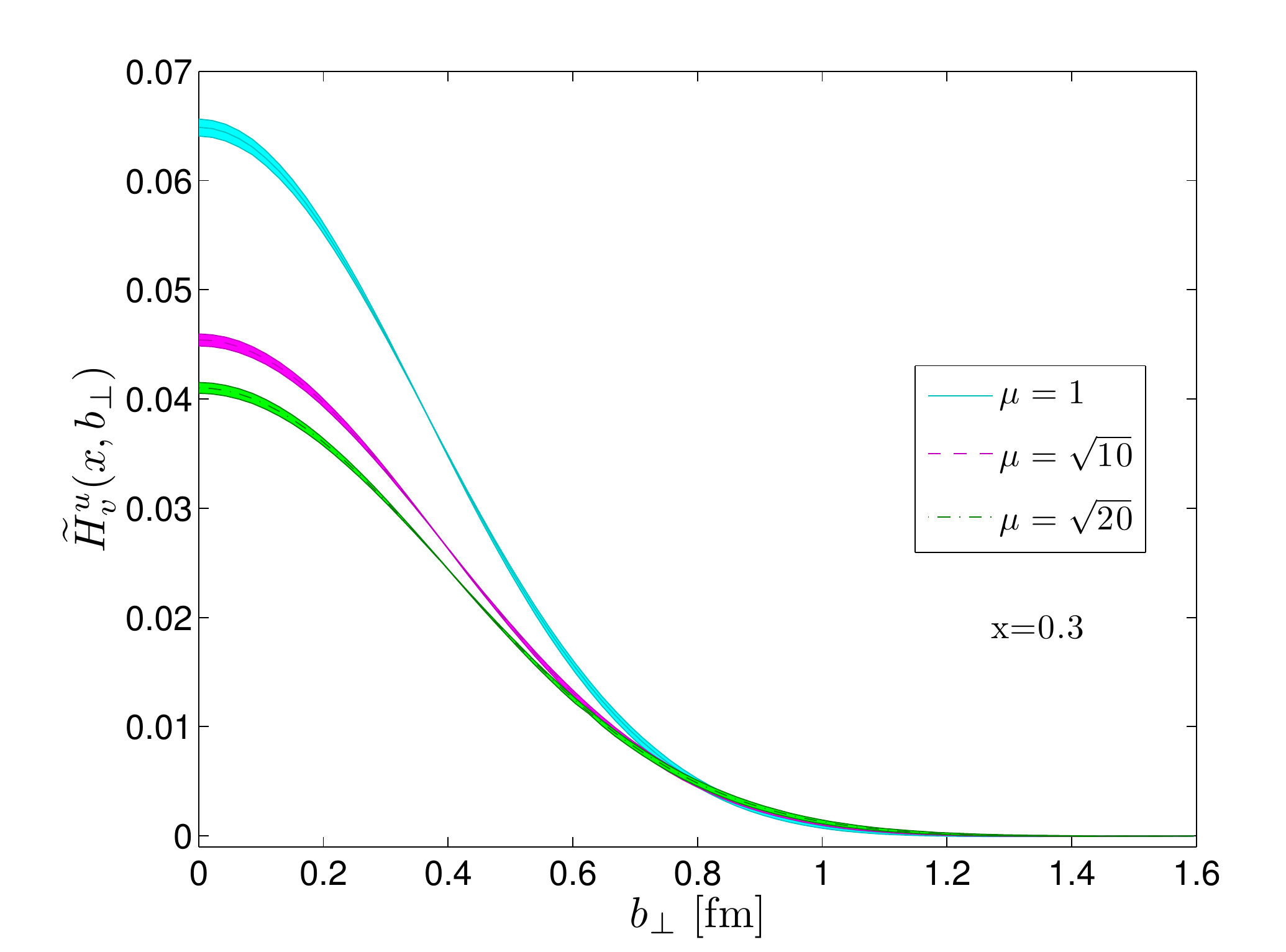}
\hspace{0.1cm}%
\small{(f)}\includegraphics[width=7.5cm,clip]{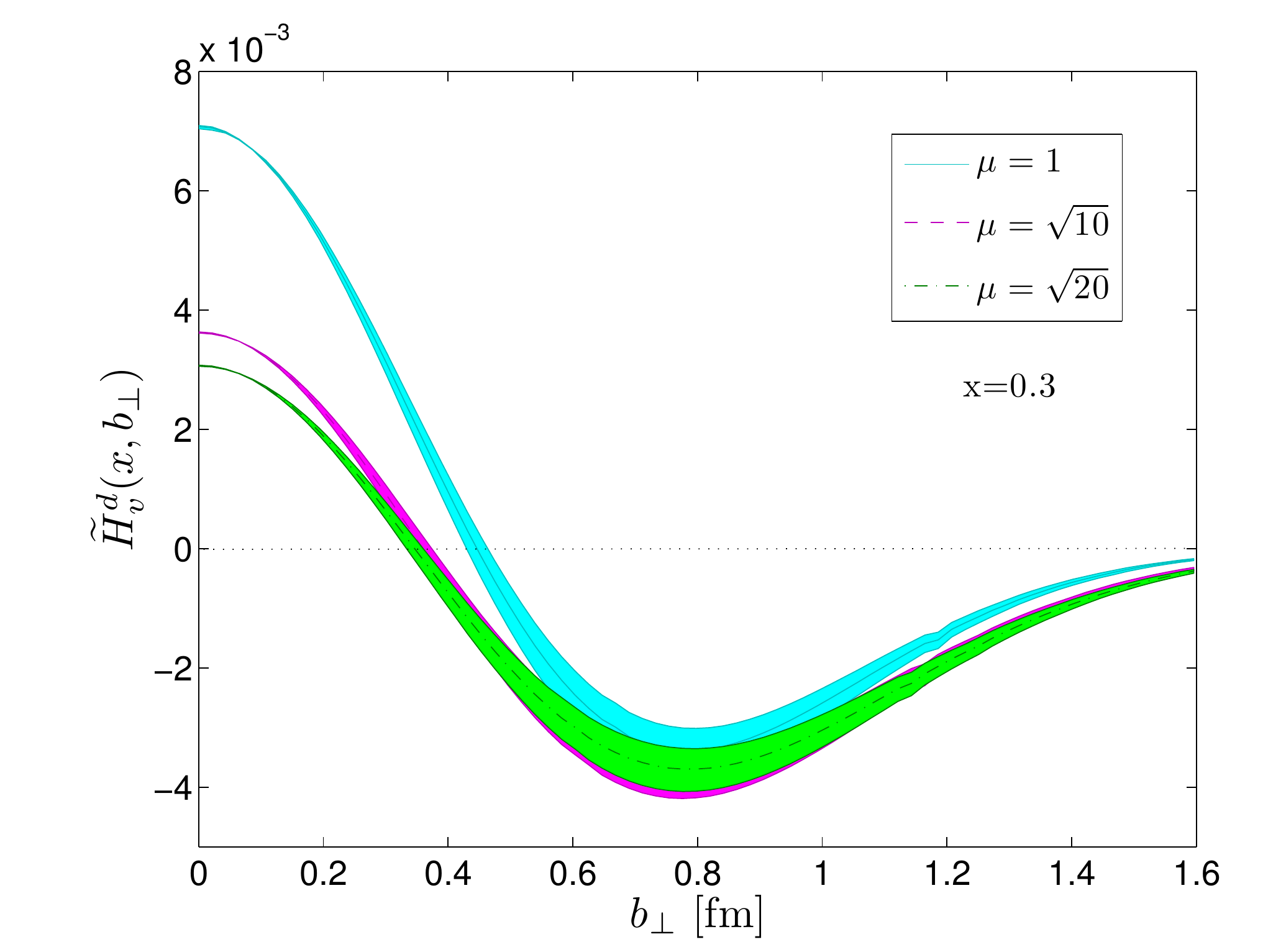}
\end{minipage}
\begin{minipage}[c]{0.98\textwidth}
\small{(g)}\includegraphics[width=7.5cm,clip]{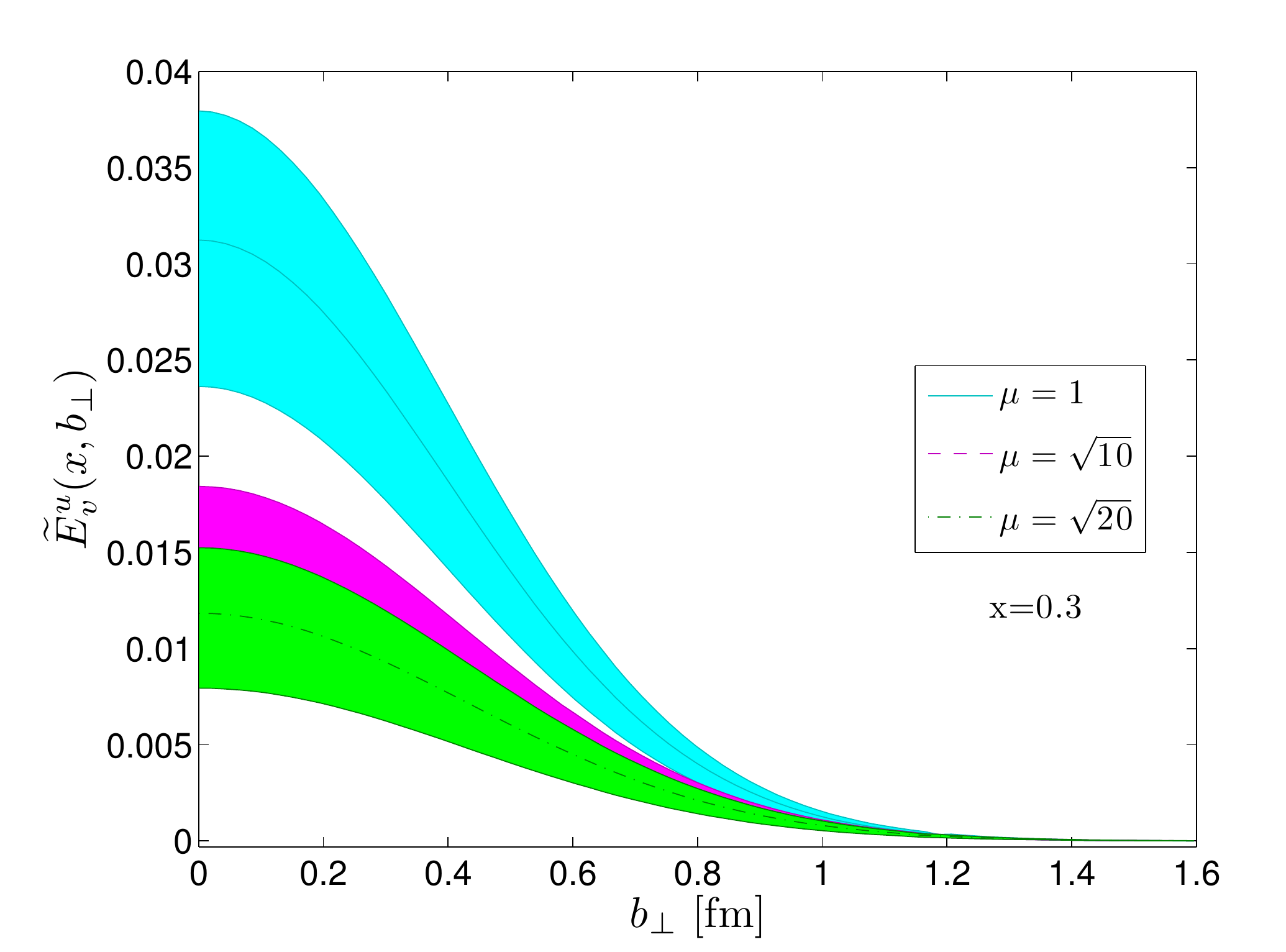}
\hspace{0.1cm}%
\small{(h)}\includegraphics[width=7.5cm,clip]{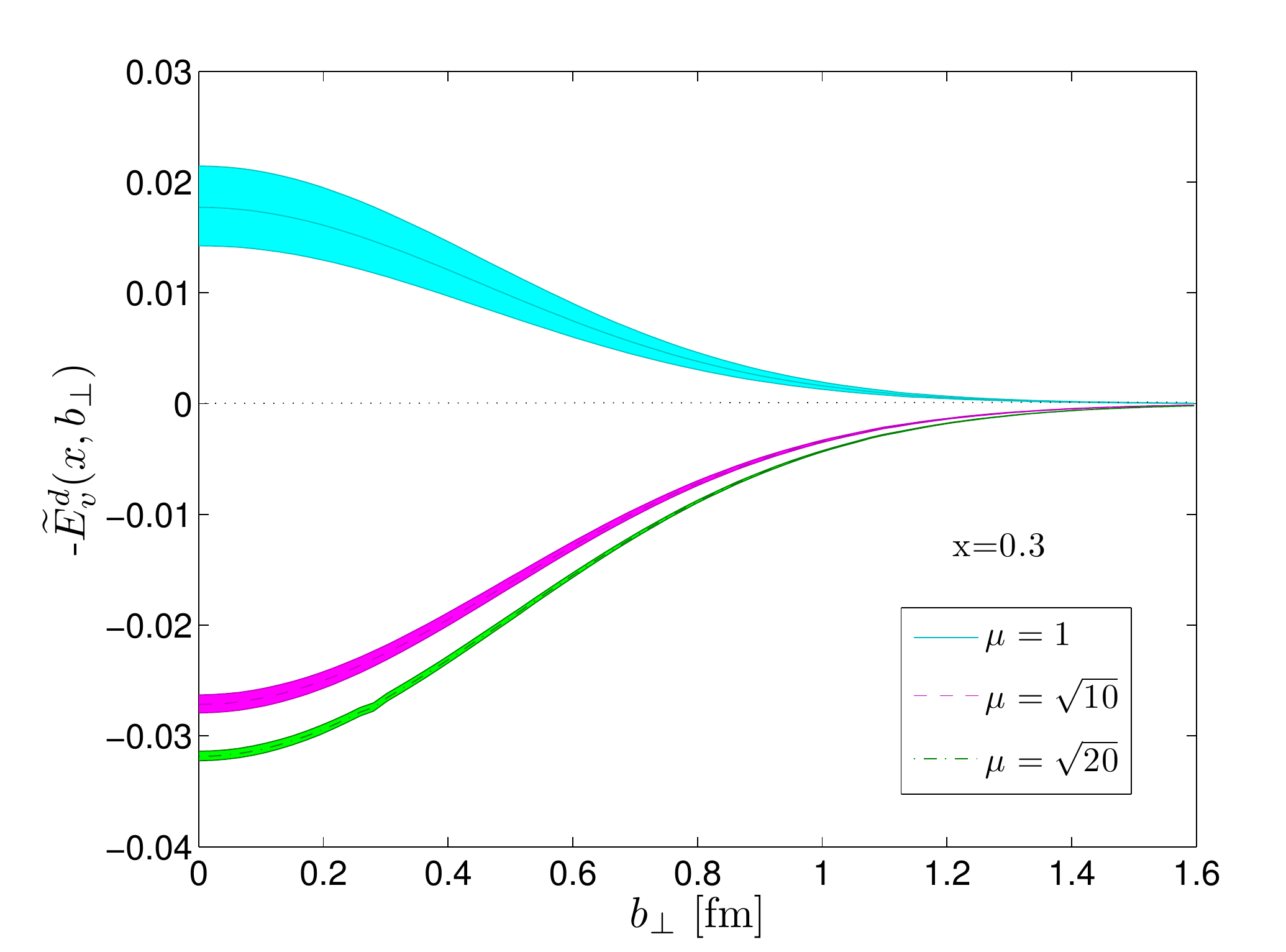}
\end{minipage}
\caption{\label{impact_b_evolve}(Color online) Plots of evolved chiral-even GPDs in impact parameter space as functions 
of $b=|b_{\perp}|$ for different scales $\mu=1,~\sqrt{10},~\sqrt{20}~GeV$ and fixed value of $x=0.3$.}
\end{figure}

\begin{figure}[H]
\begin{minipage}[c]{0.98\textwidth}
\small{(a)}
\includegraphics[width=7.5cm,clip]{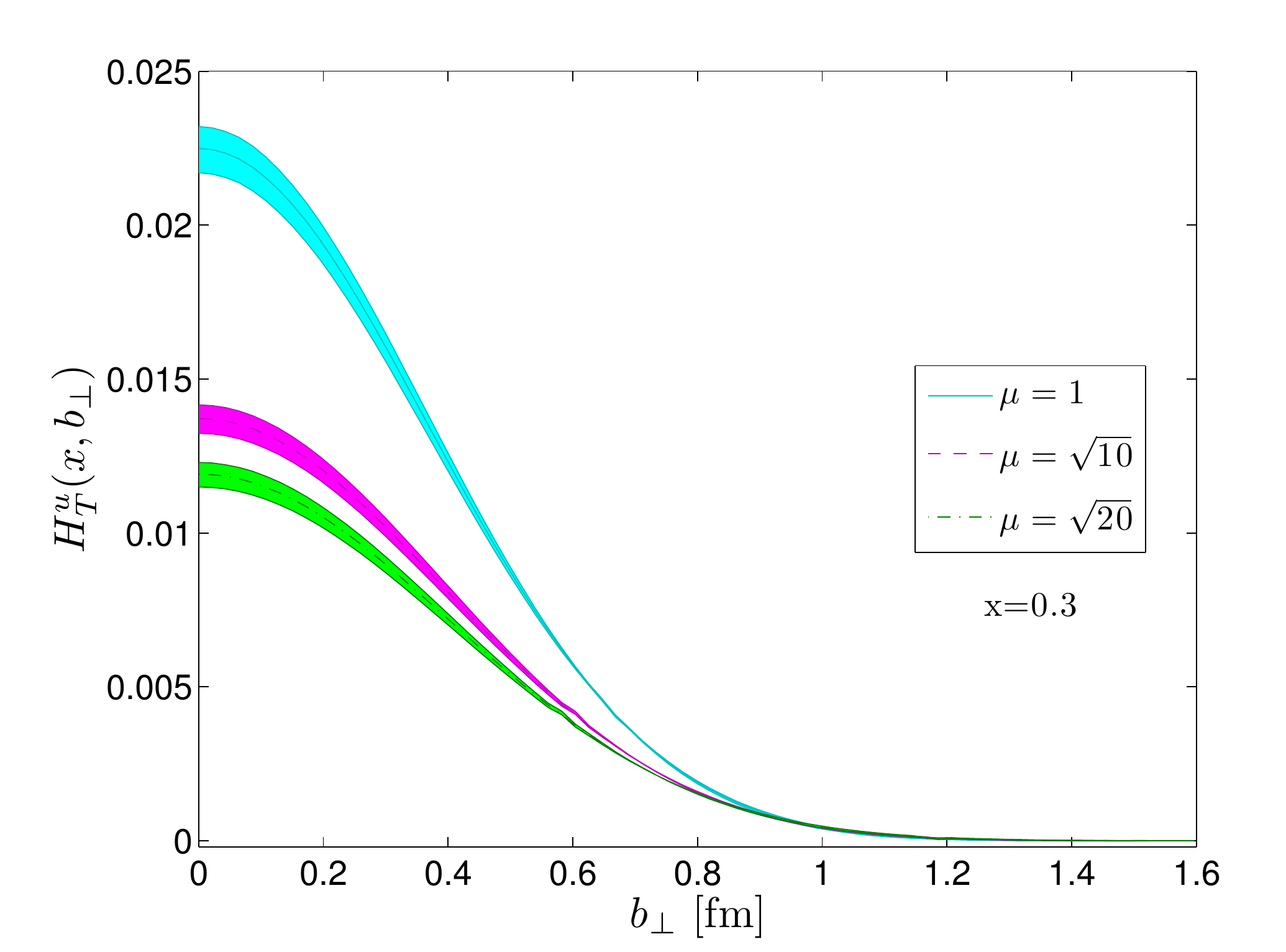}
\hspace{0.1cm}%
\small{(b)}\includegraphics[width=7.5cm,clip]{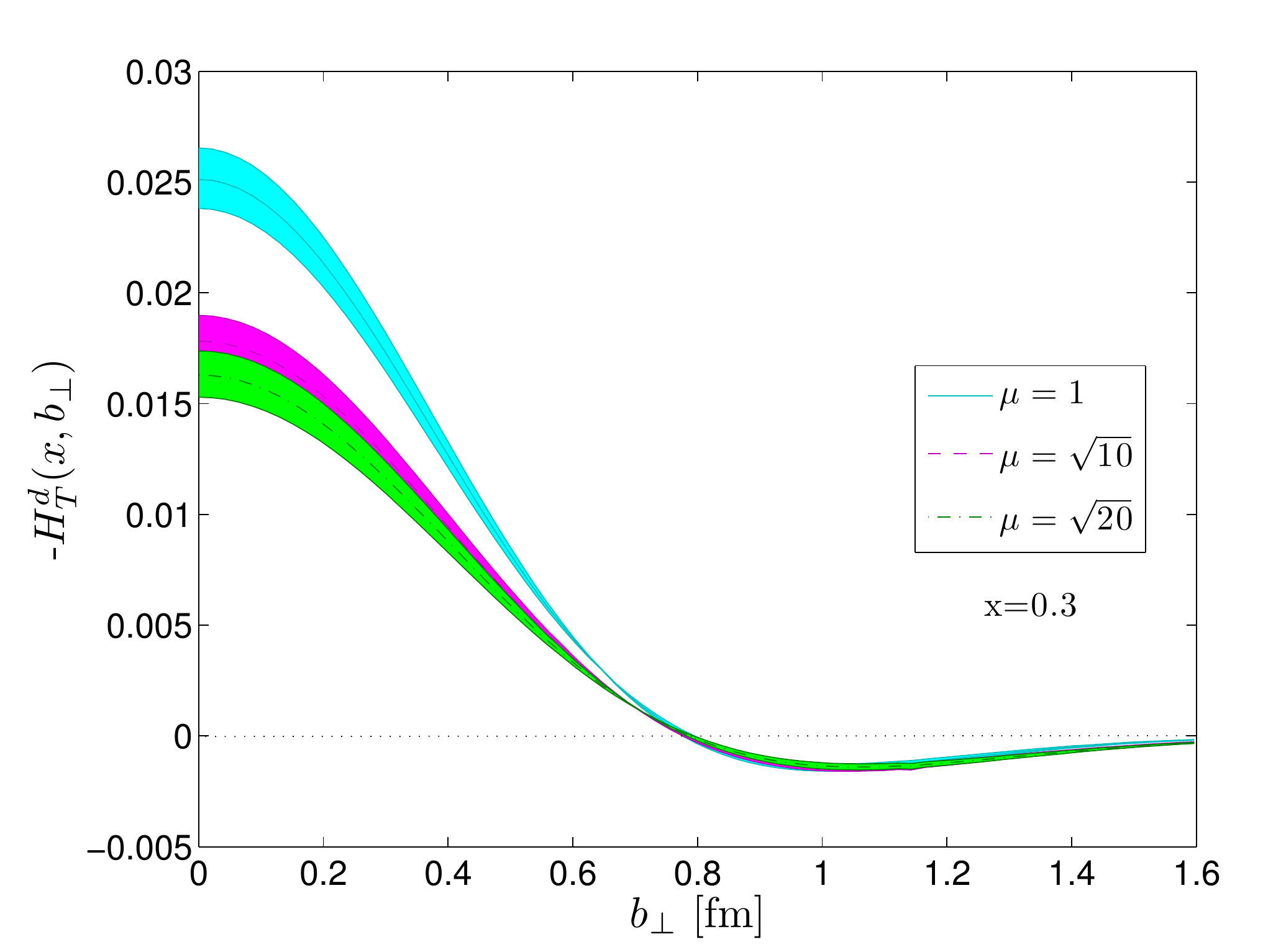}
\end{minipage}
\begin{minipage}[c]{0.98\textwidth}
\small{(c)}\includegraphics[width=7.5cm,clip]{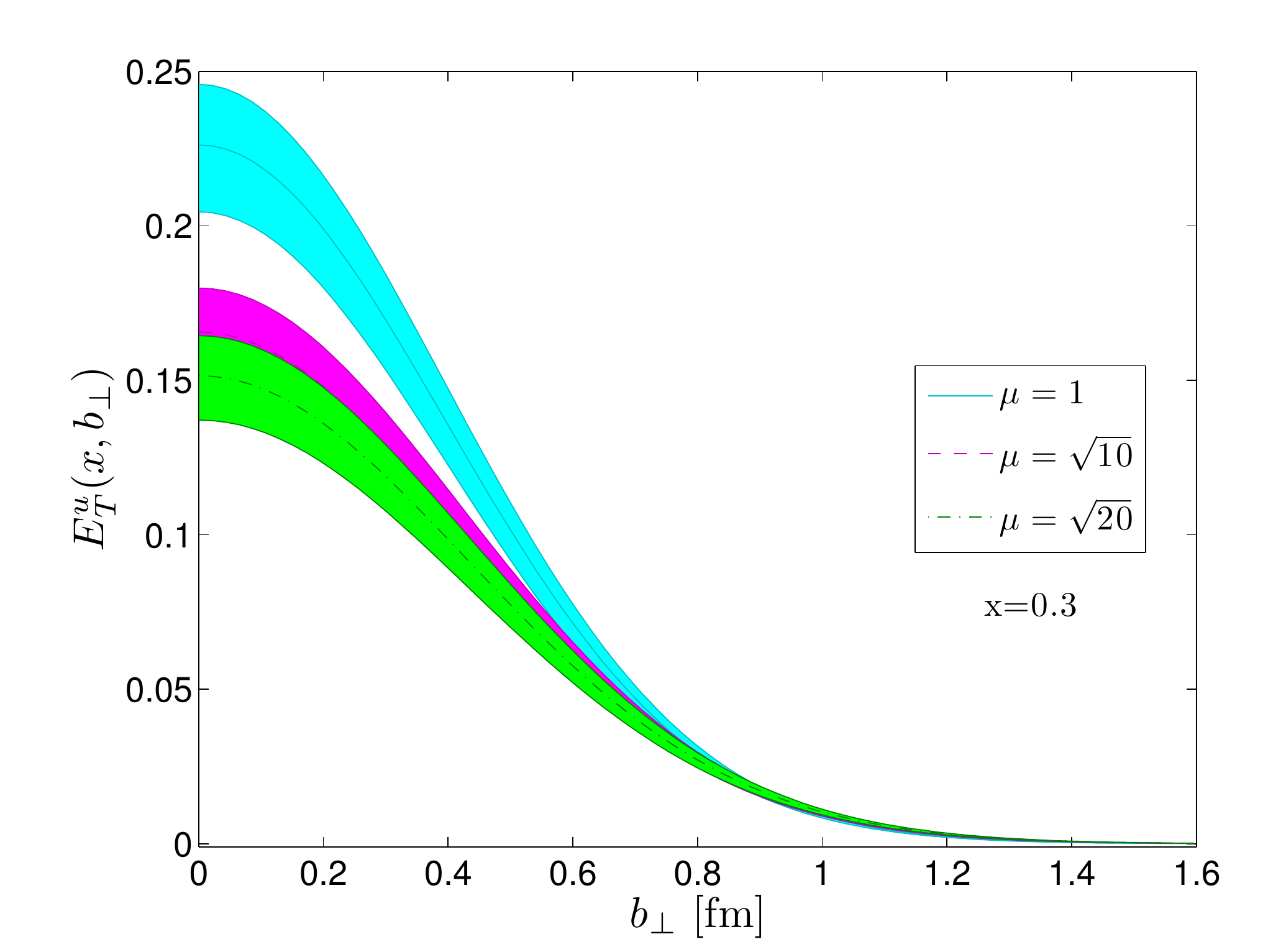}
\hspace{0.1cm}%
\small{(d)}\includegraphics[width=7.5cm,clip]{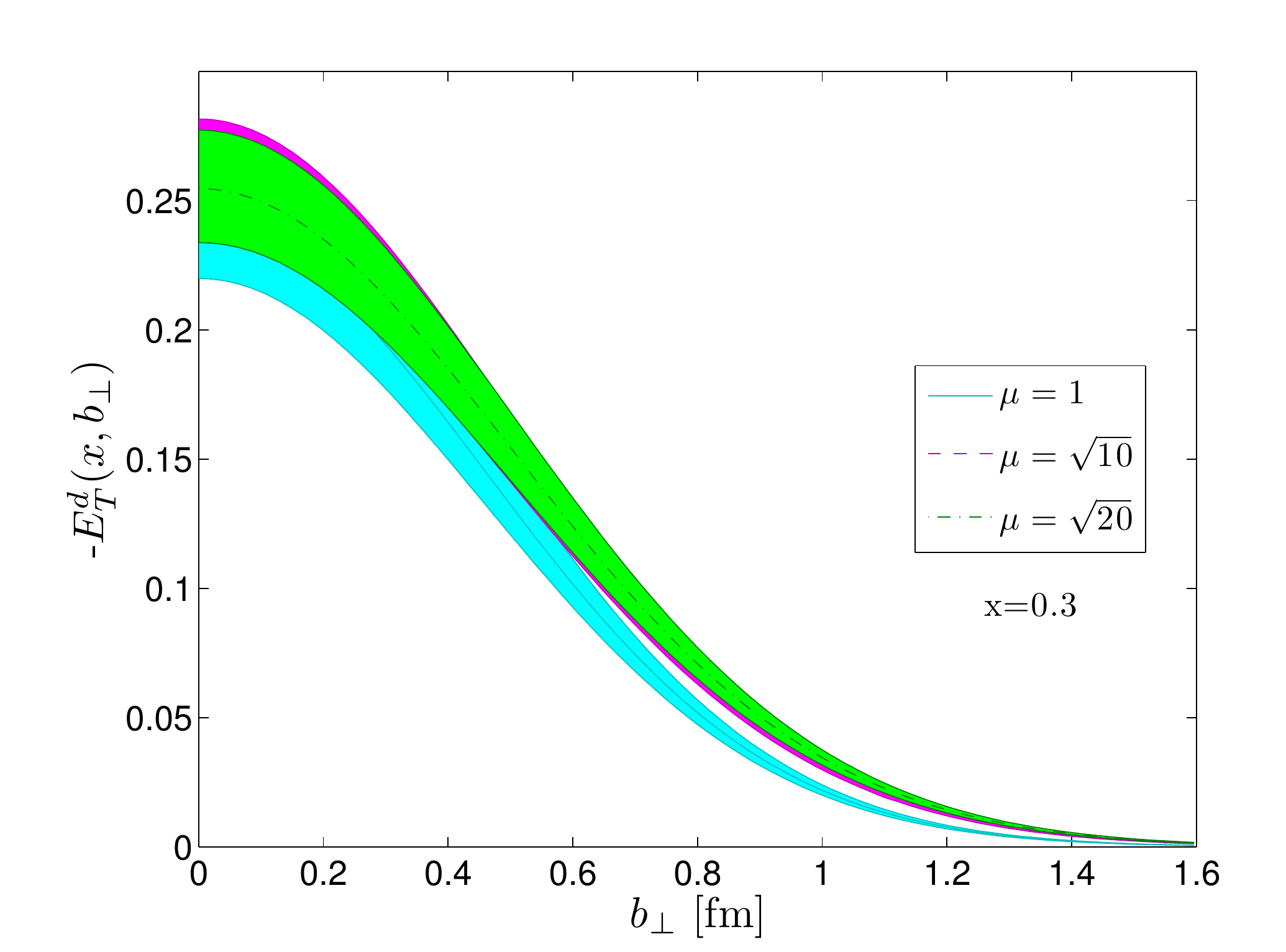}
\end{minipage}
\begin{minipage}[c]{0.98\textwidth}
\small{(e)}
\includegraphics[width=7.5cm,clip]{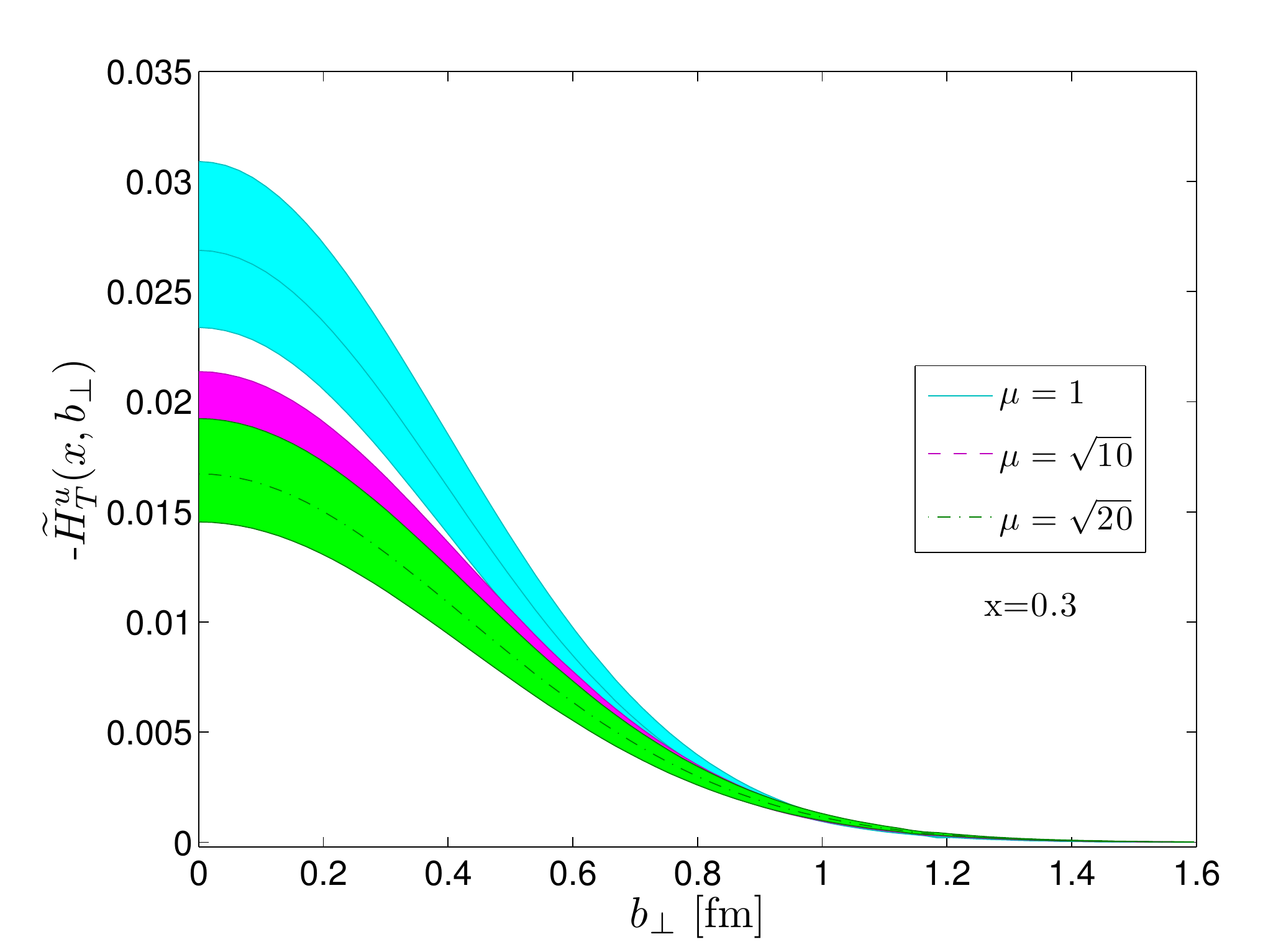}
\hspace{0.1cm}%
\small{(f)}\includegraphics[width=7.5cm,clip]{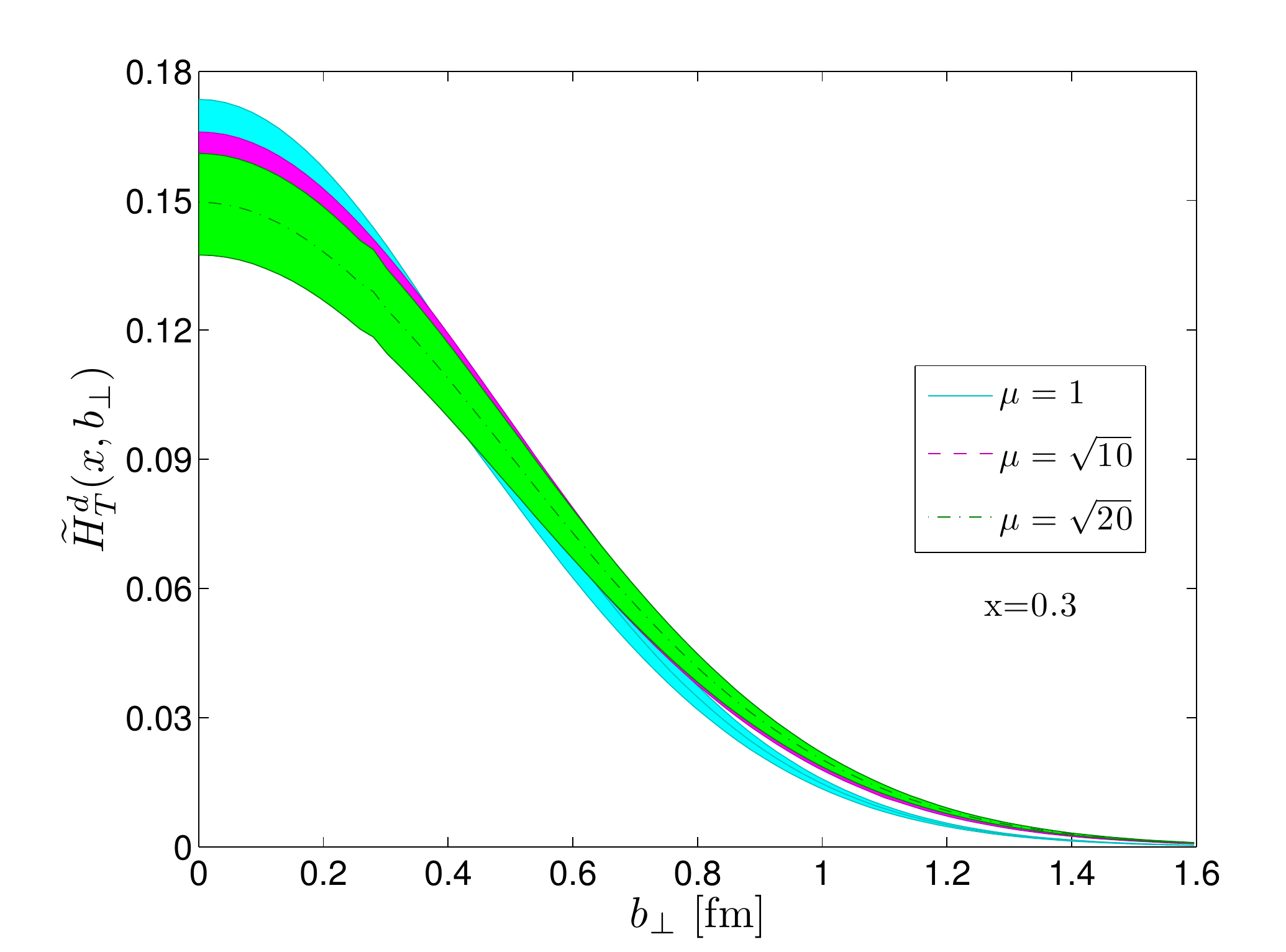}
\end{minipage}
\caption{\label{odd_impact_b_evolve}(Color online) Plots of evolved chiral-odd GPDs in impact parameter space as functions 
of $b=|b_{\perp}|$ for different scales $\mu=1,~\sqrt{10},~\sqrt{20}~GeV$ and fixed value of $x=0.3$.}
\end{figure}

The scale evolution of GPDs is governed by DGLAP equation \cite{Freund:2001bf,Diehl:2004cx}. In the quark-diquark model  used in this work, the DGLAP evolution of the unpolarized PDF is
generated by evolving the parameters in the model\cite{MC}. Since the GPDs also follow the DGLAP equation, we use the same parameter evolution as PDFs. 
In this model, the DGLAP evolution information is encoded into the parameters $a_i, b_i,$ and $\delta_i$ by fitting the unpolarised PDFs 
for the scale range $0 \leq \mu^2\leq 150~GeV^2 $ and the evolution is found to be consistent with the DGLAP evolution  upto 
scale $\mu^2=10^4~GeV^2$.
The parameters vary with the scale as
\be 
a_i^\nu(\mu)&=&a_i^\nu(\mu_0) + A^\nu_{i}(\mu), \label{a_im}\\
b_i^\nu(\mu)&=&b_i^\nu(\mu_0) - B^\nu_{i}(\mu)\frac{4C_F}{\beta_0}\ln\bigg(\frac{\alpha_s(\mu^2)}{\alpha_s(\mu_0^2)}\bigg),\label{b_im}\\
\delta^\nu(\mu)&=& \exp\bigg[\delta^\nu_1\bigg(\ln(\mu^2/\mu_0^2)\bigg)^{\delta^\nu_2}\bigg],\label{DL}
\ee
Where $A^\nu_{i}(\mu)$ and $B^\nu_{i}(\mu)$  can be combinedly written as 
\be 
P^\nu_{i}(\mu)&=&\alpha^\nu_{P,i} ~\mu^{2\beta^\nu_{P,i}}\bigg[\ln\bigg(\frac{\mu^2}{\mu_0^2}\bigg)\bigg]^{\gamma^\nu_{P,i}}\bigg|_{i=1,2} ,\label{Pi_evolu}
\ee
 The detail of the evolution fit and the values of the parameters are given in Ref. \cite{MC}. In Fig.\ref{impact_b_evolve} and Fig.\ref{odd_impact_b_evolve}, we show the scale evolution of chiral-even and chiral-odd GPDs respectively at different scales $\mu^2=1,~10$ and $20~GeV^2$.

\section{Proton spin densities}
For $\xi=0$,  GPDs in the impact parameter space can be interpreted as densities of quarks with longitudinal momentum fraction $x$ and transverse location $\bfb$ with respect to the nucleon center of momentum. Depending on the polarization of both the active quark and the parent nucleon, one can define three-dimensional densities $\rho(x,{\bfb}, \lambda,\Lambda)$ and  $\rho(x,{\bfb},{\tvec s},{\tvec S})$ representing the probability to find a quark with longitudinal momentum fraction $x$ and transverse position $\bfb$ either with light-cone helicity $\lambda$ ($=\pm 1$) in the nucleon with longitudinal polarization $\Lambda$ ($=\pm 1$) or with transverse spin $\tvec s$ in the nucleon with transverse spin $\tvec S$. The densities are given as
\be
\rho(x,{\bfb}, \lambda,\Lambda) =  \frac{1}{2} \left[ H(x,{b}^2)
  + b^j\varepsilon^{ji} S^i  \frac{1}{M}\,
       E'(x,{b}^2)
  + \lambda \Lambda \widetilde{H}(x,{b}^2) \,\right] ,
 \label{eq:long}
 \ee
\be
\rho(x,{\bfb},{\tvec s},{\tvec S})
&= & \frac{1}{2}\Big[ H(x,{b}^2)  + s^iS^i\left( H_T(x,{b}^2)  -\frac{1}{4M^2} \Delta_b \widetilde H_T(x,{b}^2) \right) .
\nonumber\\
&+& \frac{b^j\varepsilon^{ji}}{M}\left(
S^iE'(x,{b}^2)  + s^i\left[ E'_T(x,{b}^2)  + 2 \widetilde H'_T(x,{b}^2) \right]\right)
\nonumber\\
&+& s^i(2b^ib^j - b^2\delta_{ij}) S^j\frac{1}{M^2} \widetilde H''_T(x,{b}^2) \Big] ,
\label{eq:tr}
 \ee
where $\varepsilon^{ij}$ is the two-dimensional antisymmetric tensor with $\varepsilon^{12} = -\varepsilon^{21} = 1$ and $\varepsilon^{11} = \varepsilon^{22} = 0$.  We use the shorthand notations
$
f' = \frac{\partial}{\partial b^2}\, f ,
~
f''= \Big( \frac{\partial}{\partial b^2} \Big)^2 f,
~
\Delta_b f
= \frac{\partial}{\partial b^i}\, \frac{\partial}{\partial b^i}\, f
= 4\, \frac{\partial}{\partial b^2}
    \Big( b^2 \frac{\partial}{\partial b^2} \Big) f .$
    
For zero skewness, $H(x,b^2)$  in impact parameter space gives the description of the density of unpolarized quarks in the unpolarized proton. $E(x,b^2)$ is responsible for a deformation of the density when the proton is transversely polarized. $\tilde H(x,b^2)$ provides the information of the difference in the density of quarks with helicity equal or opposite to the proton helicity. One may also interpret the chiral-odd GPDs( at $\xi=0$) as a density in transverse impact parameter space like chiral-even GPDs depending on the polarization of both the active quark and the nucleon. The particular combination  $({E}_T+2\widetilde{{H}}_T)$  plays a role  similar to $E(x,b^2)$ and is responsible for a deformation in the transversely polarized quark density in an unpolarized target~\cite{Burk3,Diehl05,Pasquini2,Dahiya07}. On the other hand a combination of ${H}_T(x,b^2)$ and $\widetilde{{H}}_T(x,b^2)$ provides a distortion in the density when the active quark and the nucleon are transversely polarized~\cite{Diehl05,Pasquini2}. One can notice that with increasing $x$, the width of all the distributions in transverse impact parameter space  decreases, which implies that the distributions are more localized and the quark is more concentrated near the center of momentum(i.e., at small $b$) for higher values of $x$.  
\begin{figure}[htp]
\begin{minipage}[c]{0.98\textwidth}
\includegraphics[width=7.2cm,height=5.8cm,clip]{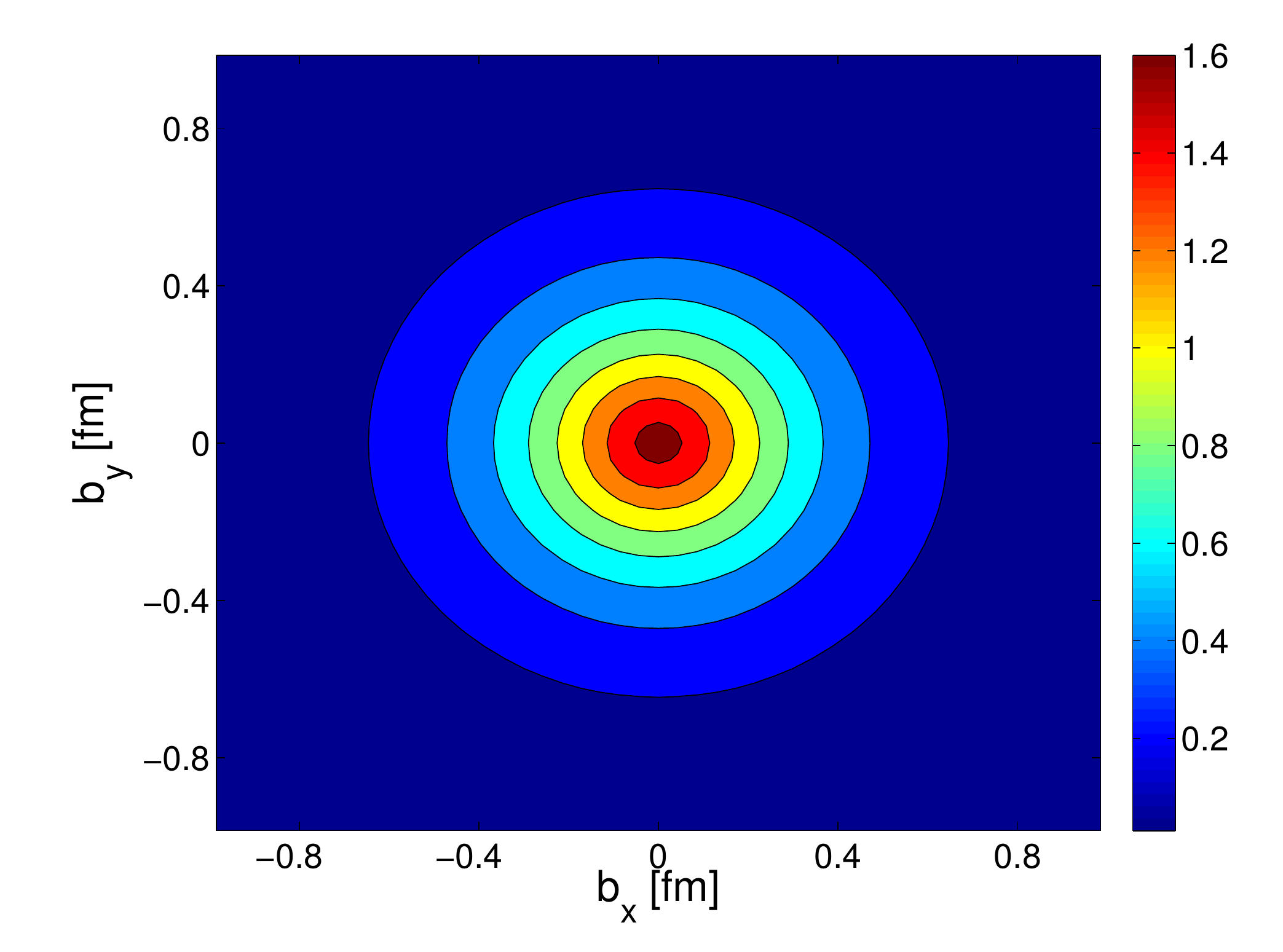}
\includegraphics[width=7.2cm,height=5.8cm,clip]{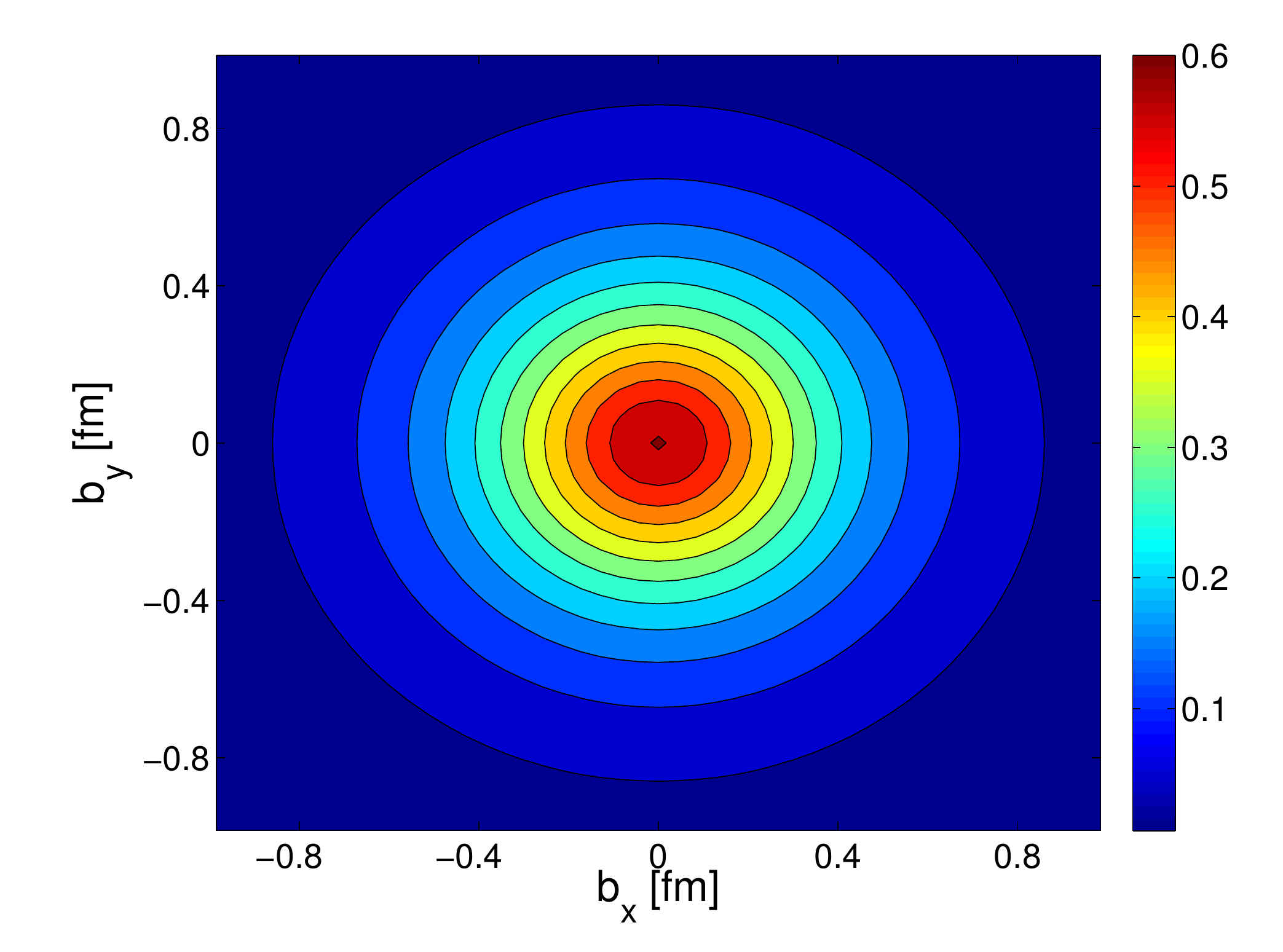}
\end{minipage}
\begin{minipage}[c]{0.98\textwidth}
\includegraphics[width=7.2cm,height=5.8cm,clip]{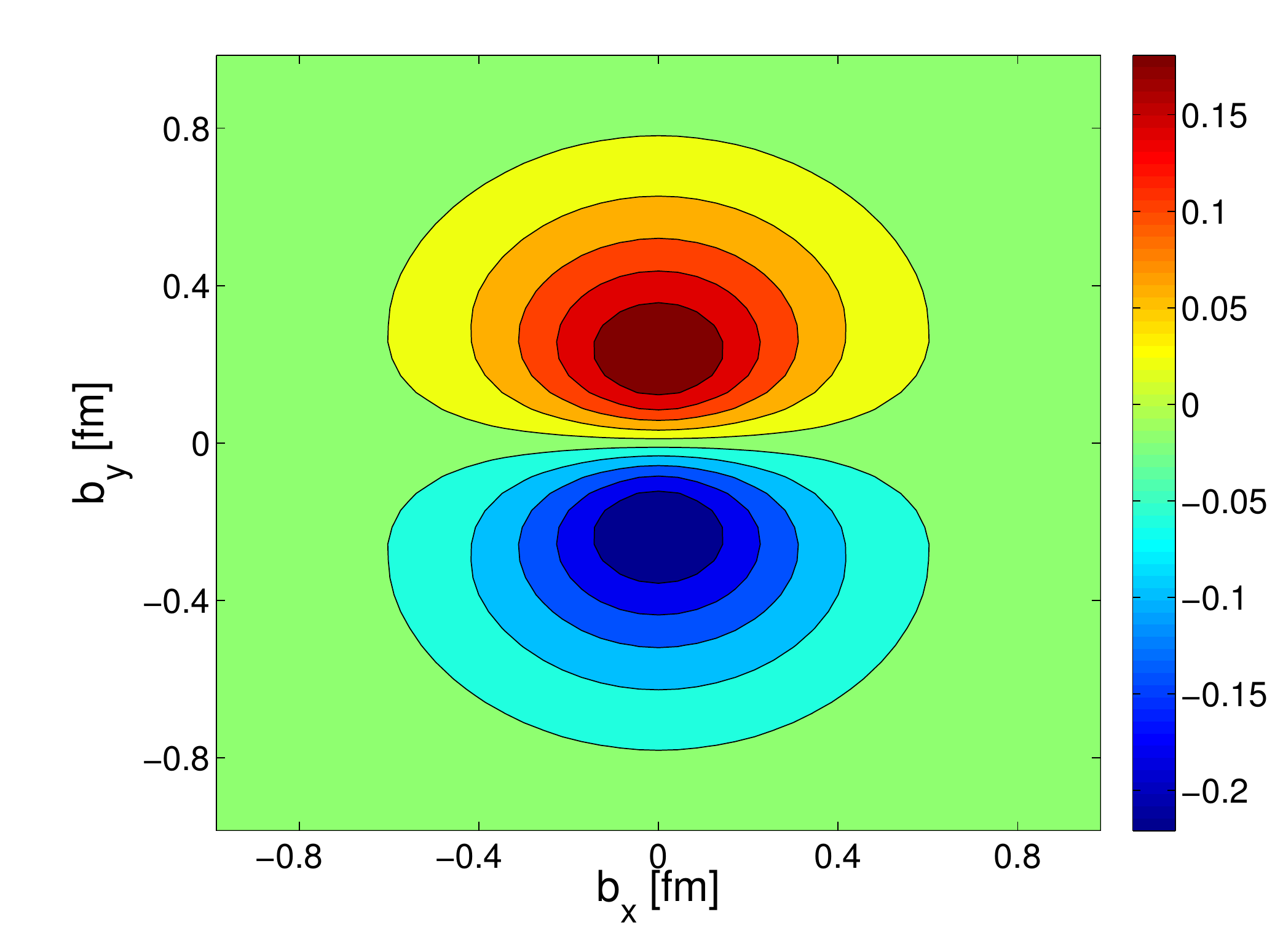}
\includegraphics[width=7.2cm,height=5.8cm,clip]{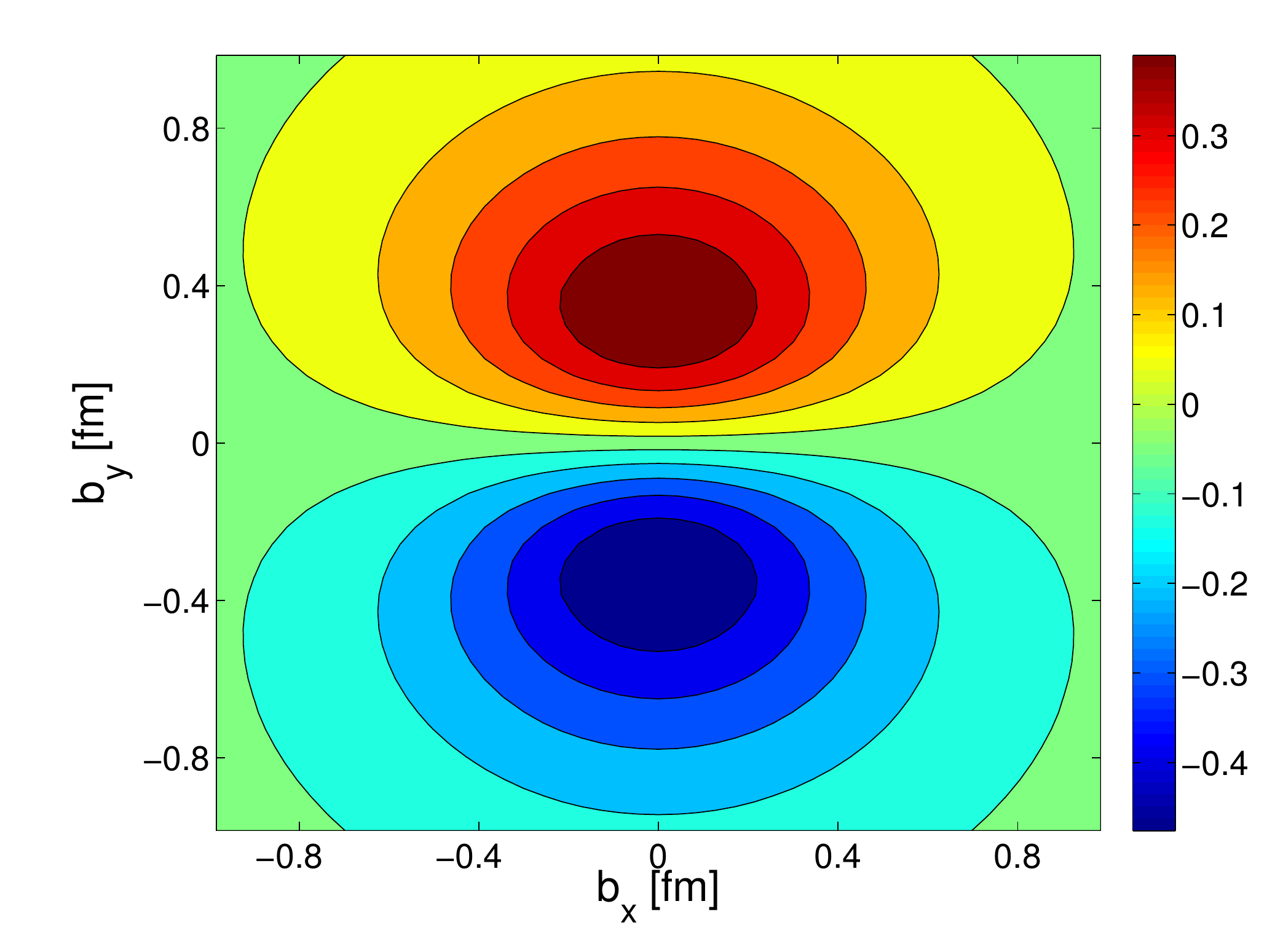}
\end{minipage}
\begin{minipage}[c]{0.98\textwidth}
\includegraphics[width=7.2cm,height=5.8cm,clip]{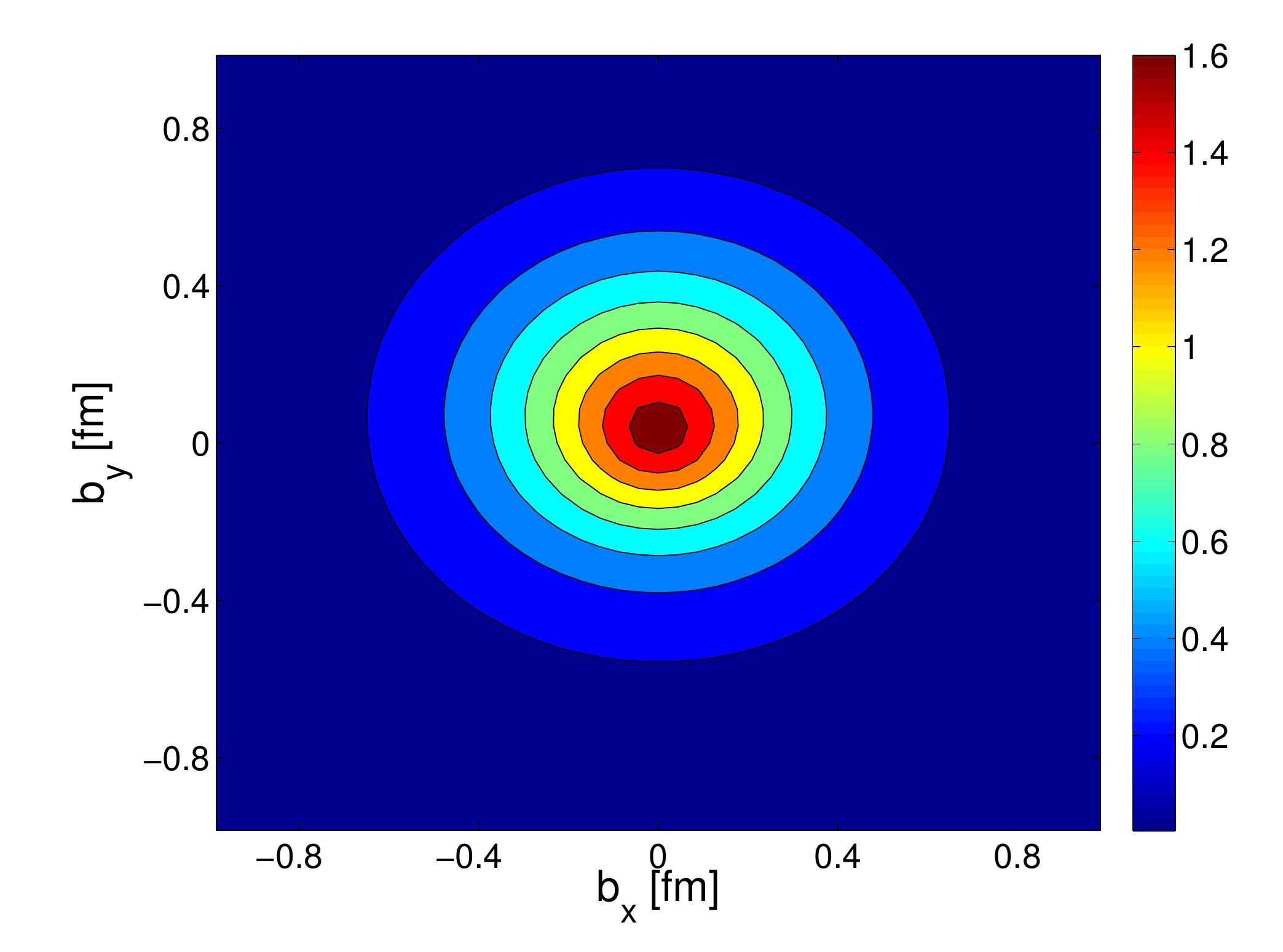}
\includegraphics[width=7.2cm,height=5.8cm,clip]{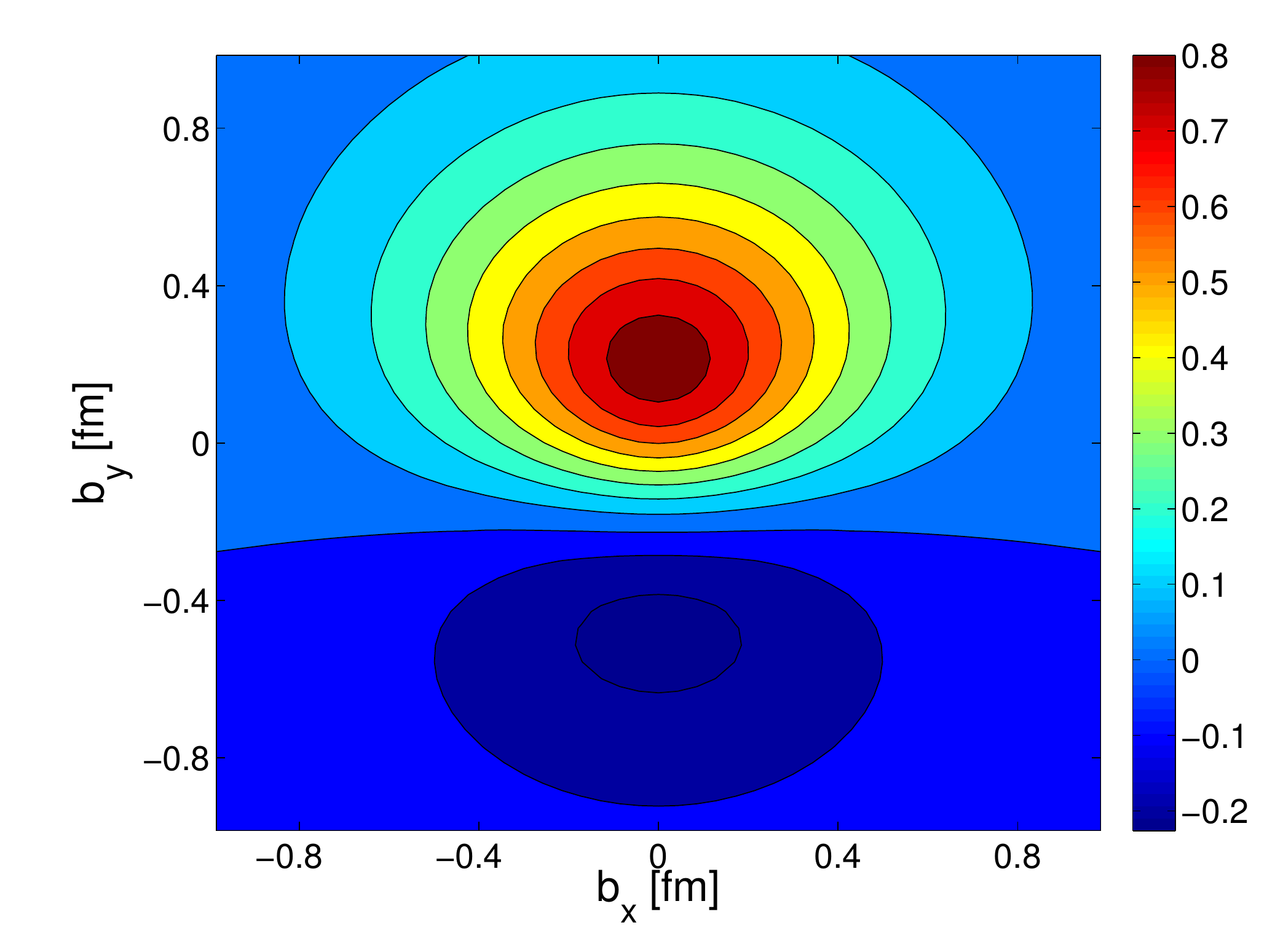}
\end{minipage}
\caption{\label{Fig1}(Color online) In an unpolarized proton : the monopole contribution $\frac{1}{2}H$ (top) for unpolarized quarks, the dipole contribution $-\frac{1}{2} s_x b_y (E'_T+2\widetilde H'_T)/M$ (middle) for transversely polarized quarks along $\hat x$-direction, and the sum of both (lower). The left (right) panel gives the results for $u$ ($d$) quark.} 
\end{figure}
\begin{figure}[htp]
\begin{minipage}[c]{0.98\textwidth}
\includegraphics[width=7.2cm,height=5.8cm,clip]{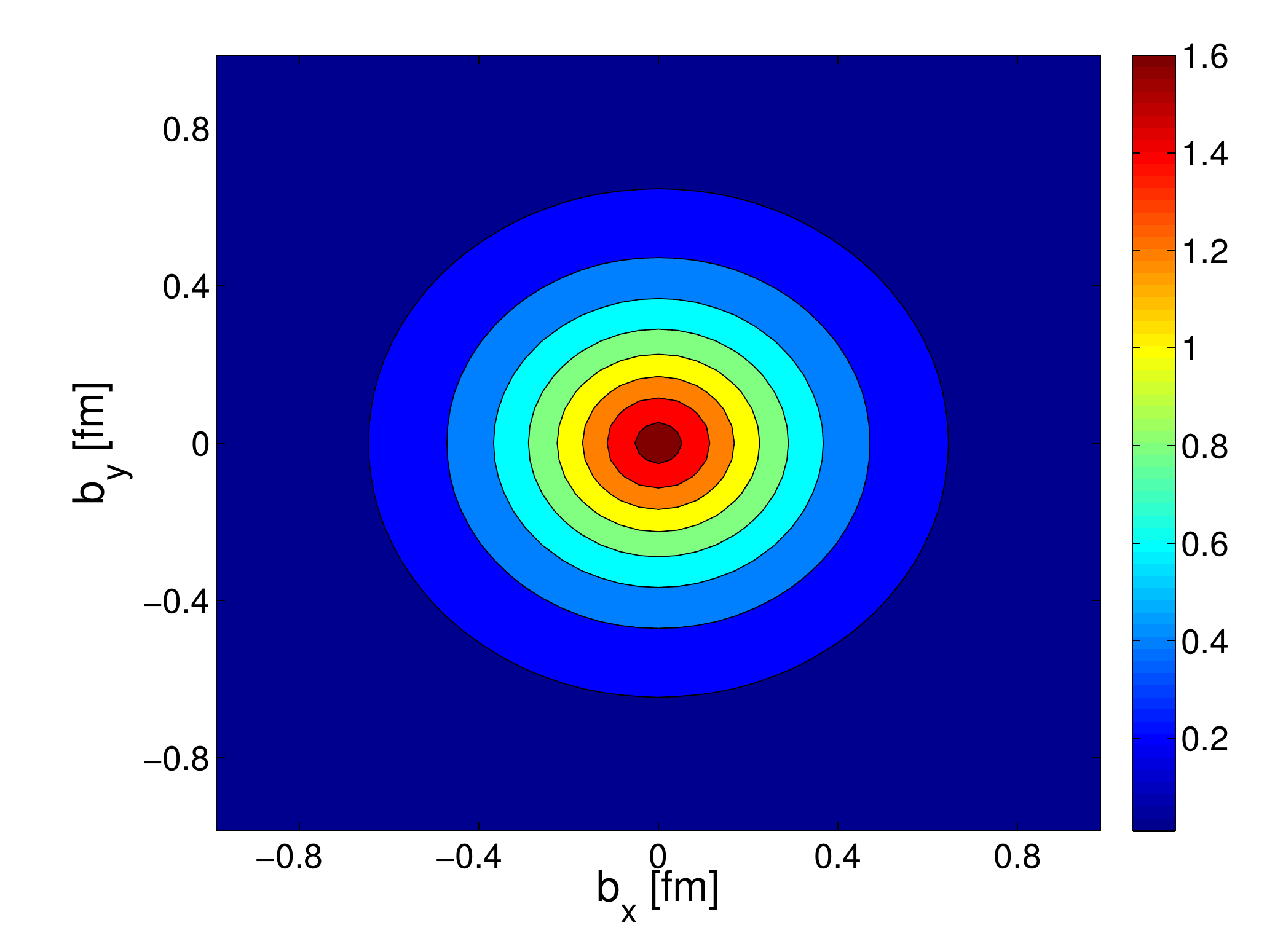}
\includegraphics[width=7.2cm,height=5.8cm,clip]{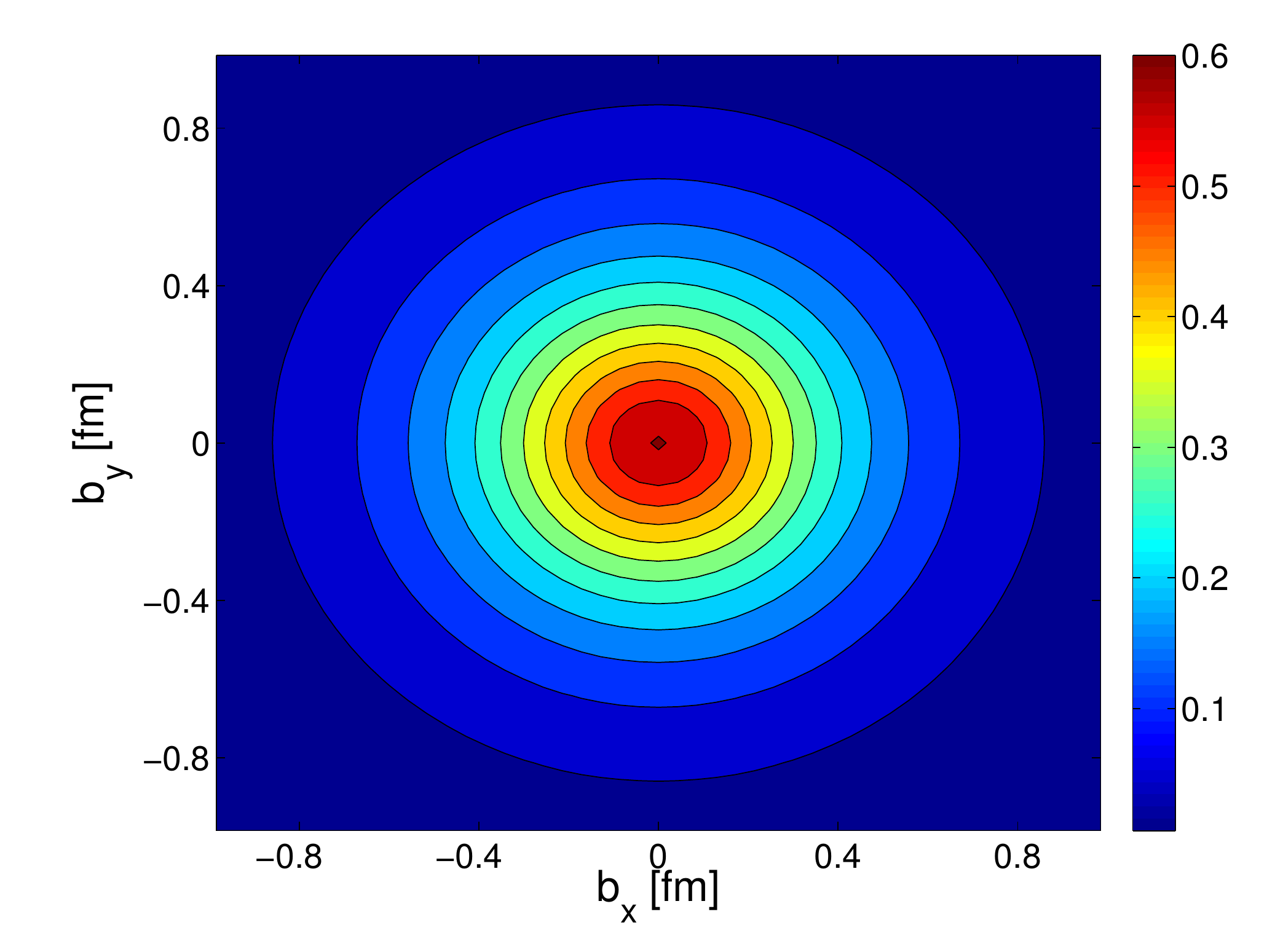}
\end{minipage}
\begin{minipage}[c]{0.98\textwidth}
\includegraphics[width=7.2cm,height=5.8cm,clip]{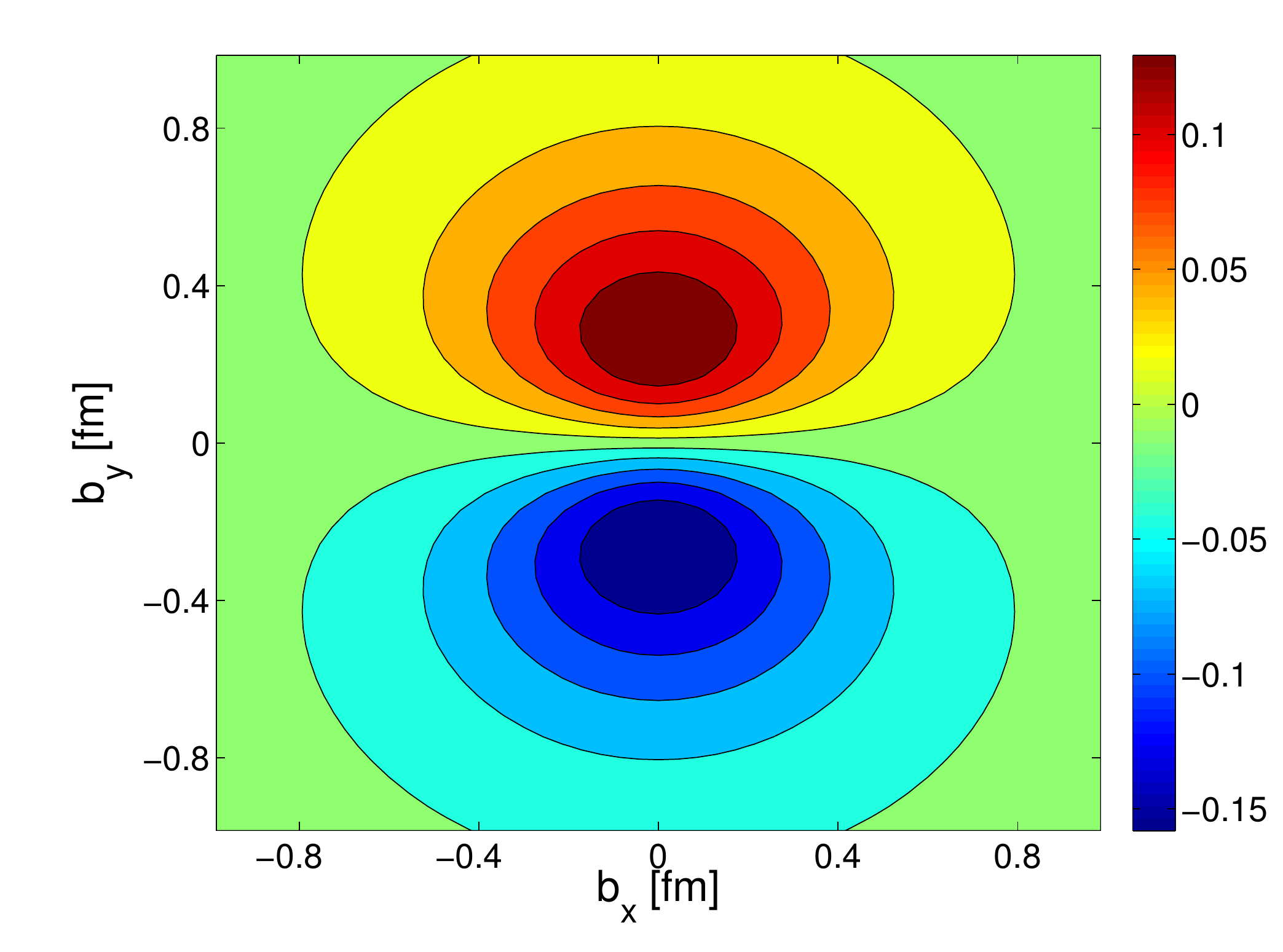}
\includegraphics[width=7.2cm,height=5.8cm,clip]{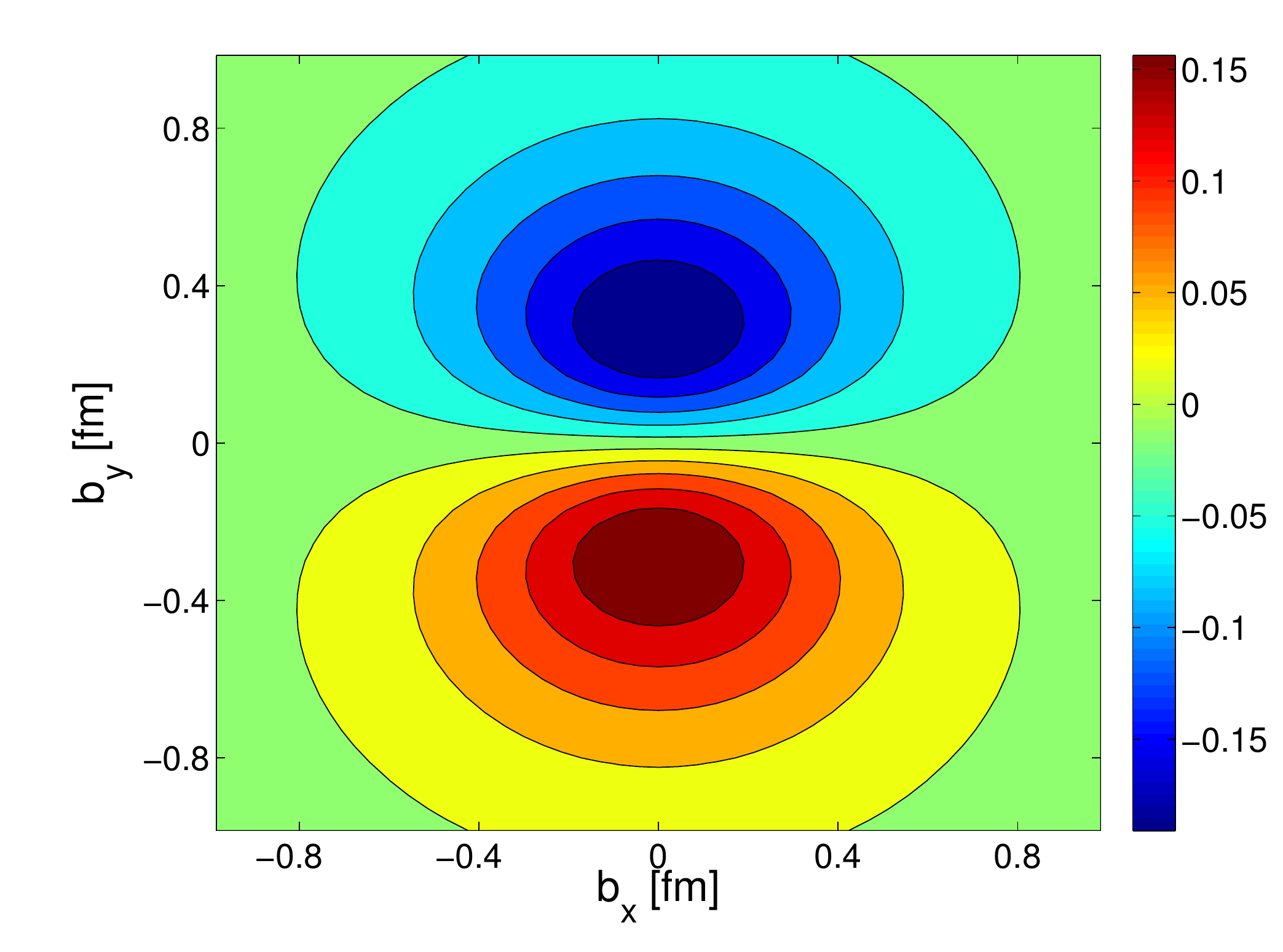}
\end{minipage}
\begin{minipage}[c]{0.98\textwidth}
\includegraphics[width=7.2cm,height=5.8cm,clip]{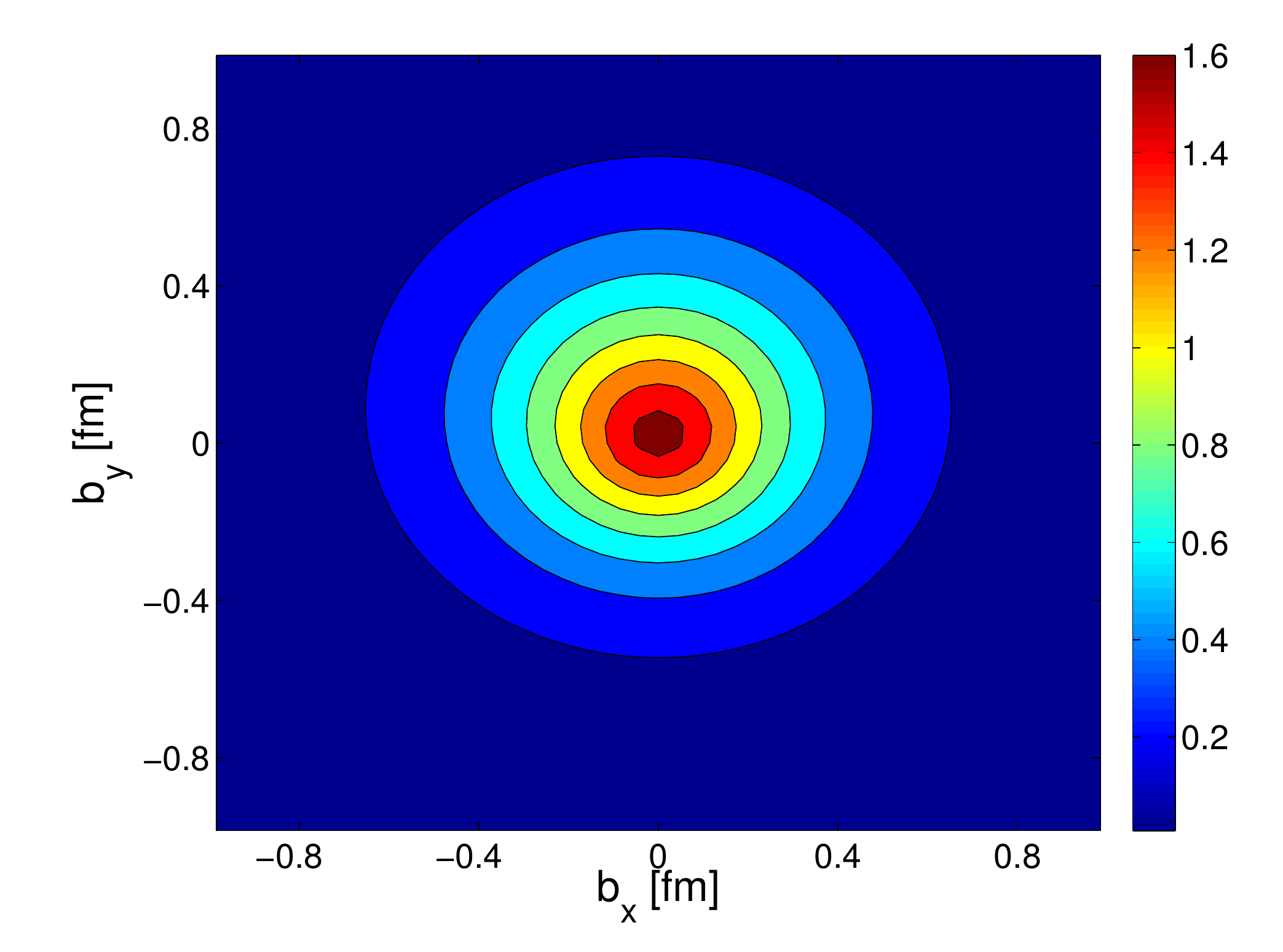}
\includegraphics[width=7.2cm,height=5.8cm,clip]{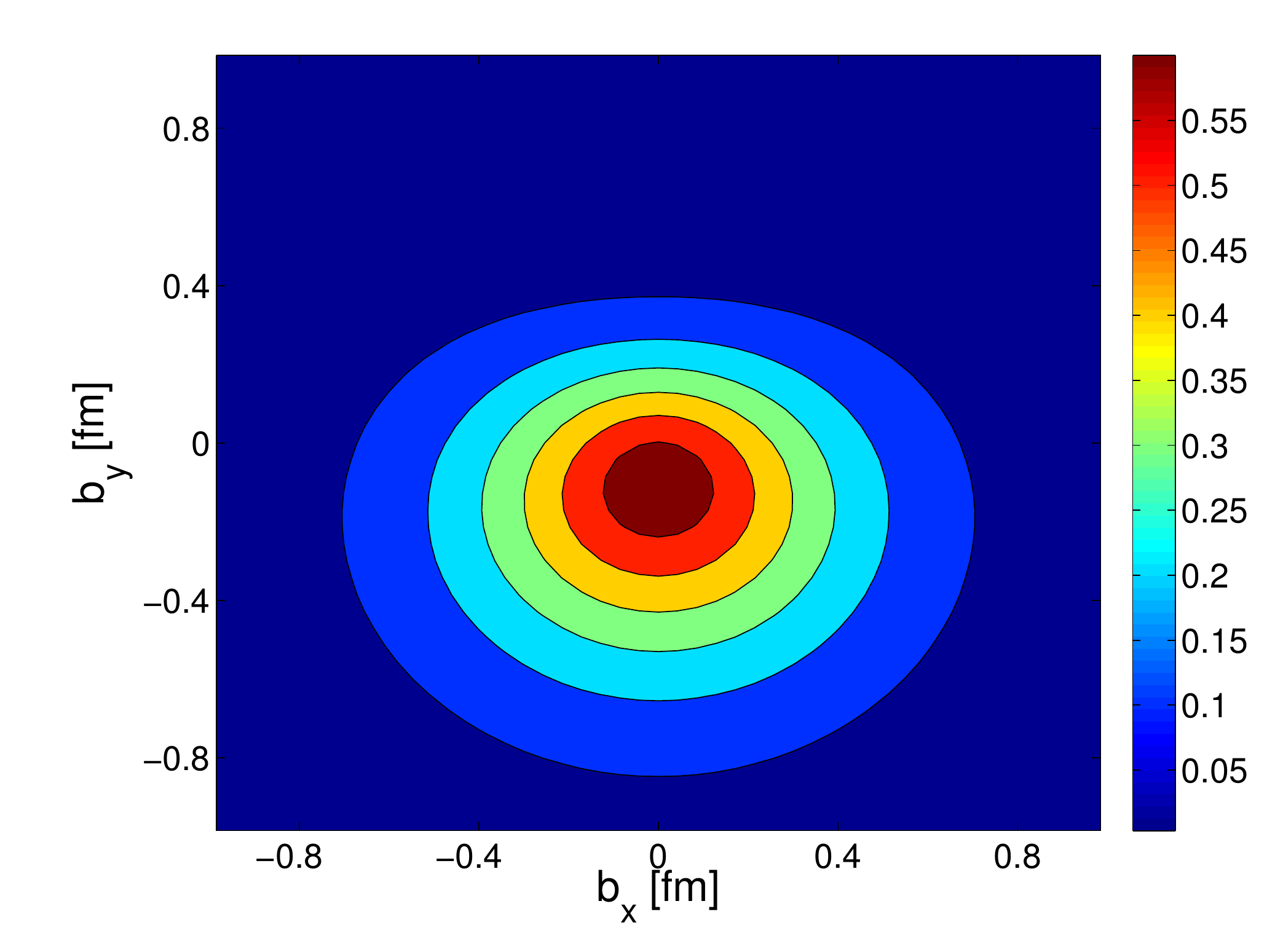}
\end{minipage}
\caption{\label{Fig2}(Color online) For unpolarized quarks : the monopole contribution $\frac{1}{2}H$ (top) for an unpolarized proton, the dipole contribution $-\frac{1}{2} S_x b_y E'/M$ (middle) for a transversely polarized proton along $\hat x$ , and the sum of both (lower). The left (right) panel gives the results for $u$ ($d$) quarks.} 
\end{figure}
\begin{figure}[htp]
\begin{minipage}[c]{0.98\textwidth}
\includegraphics[width=7.2cm,height=5.8cm,clip]{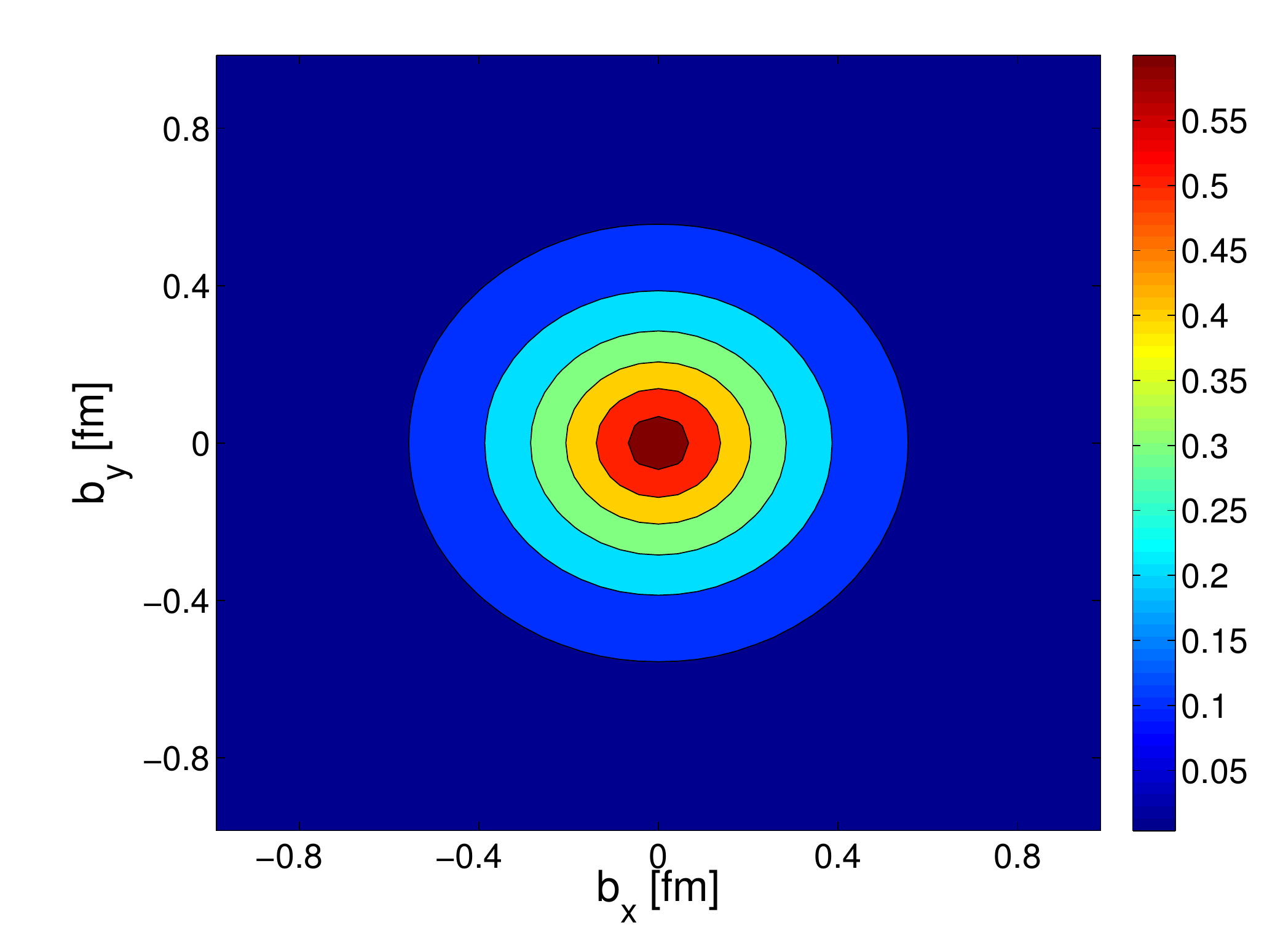}
\includegraphics[width=7.2cm,height=5.8cm,clip]{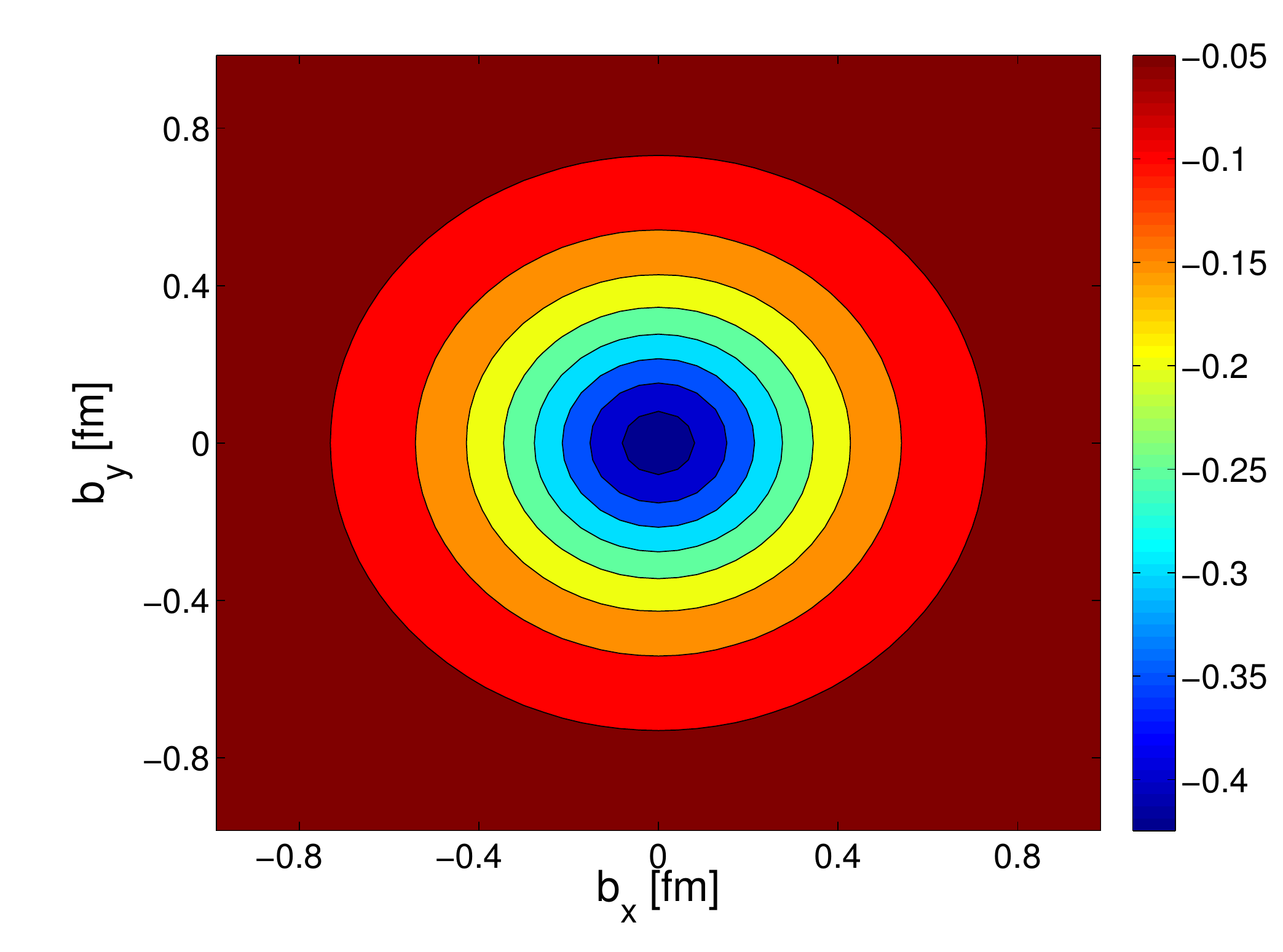}
\end{minipage}
\begin{minipage}[c]{0.98\textwidth}
\includegraphics[width=7.2cm,height=5.8cm,clip]{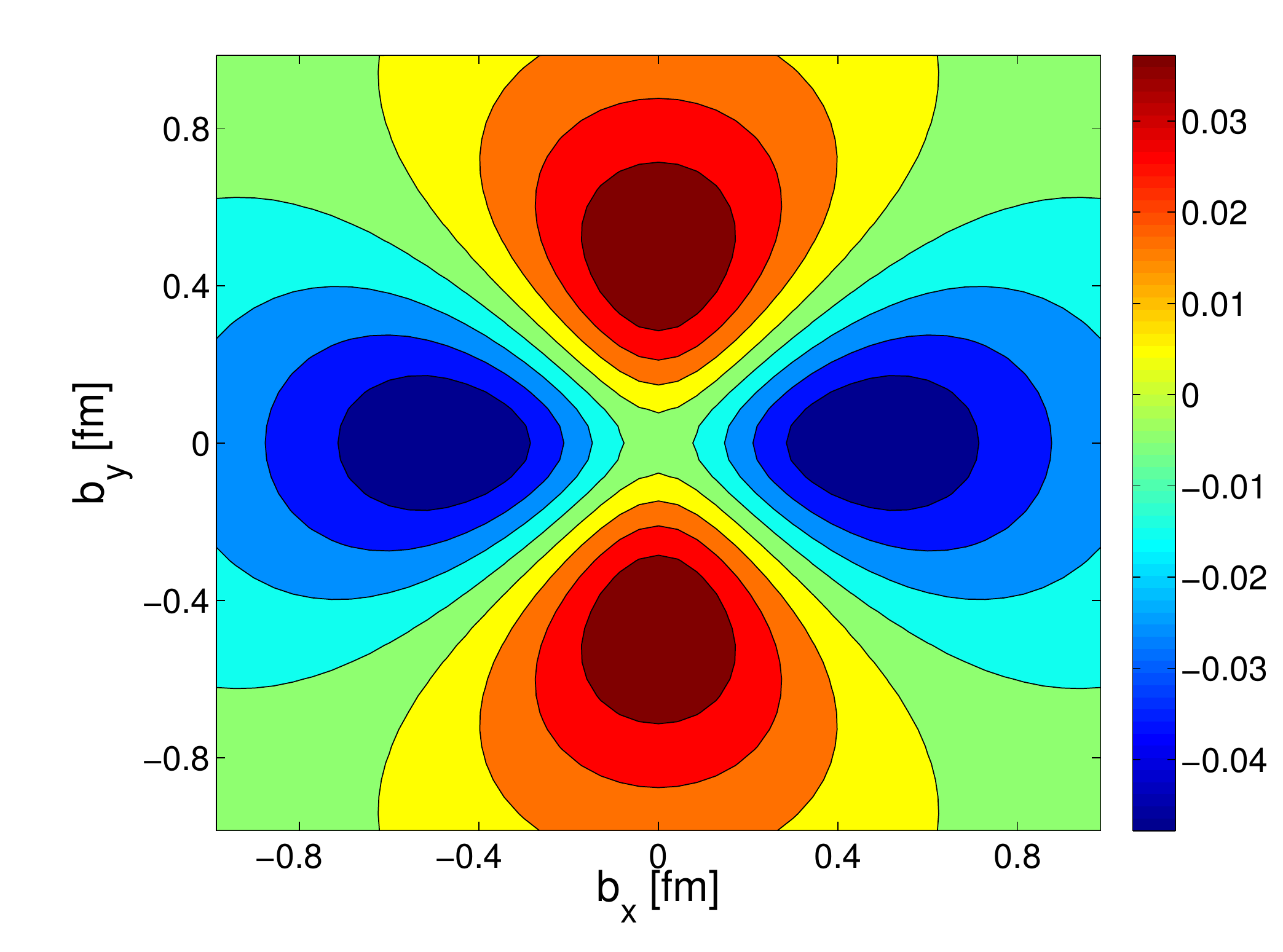}
\includegraphics[width=7.2cm,height=5.8cm,clip]{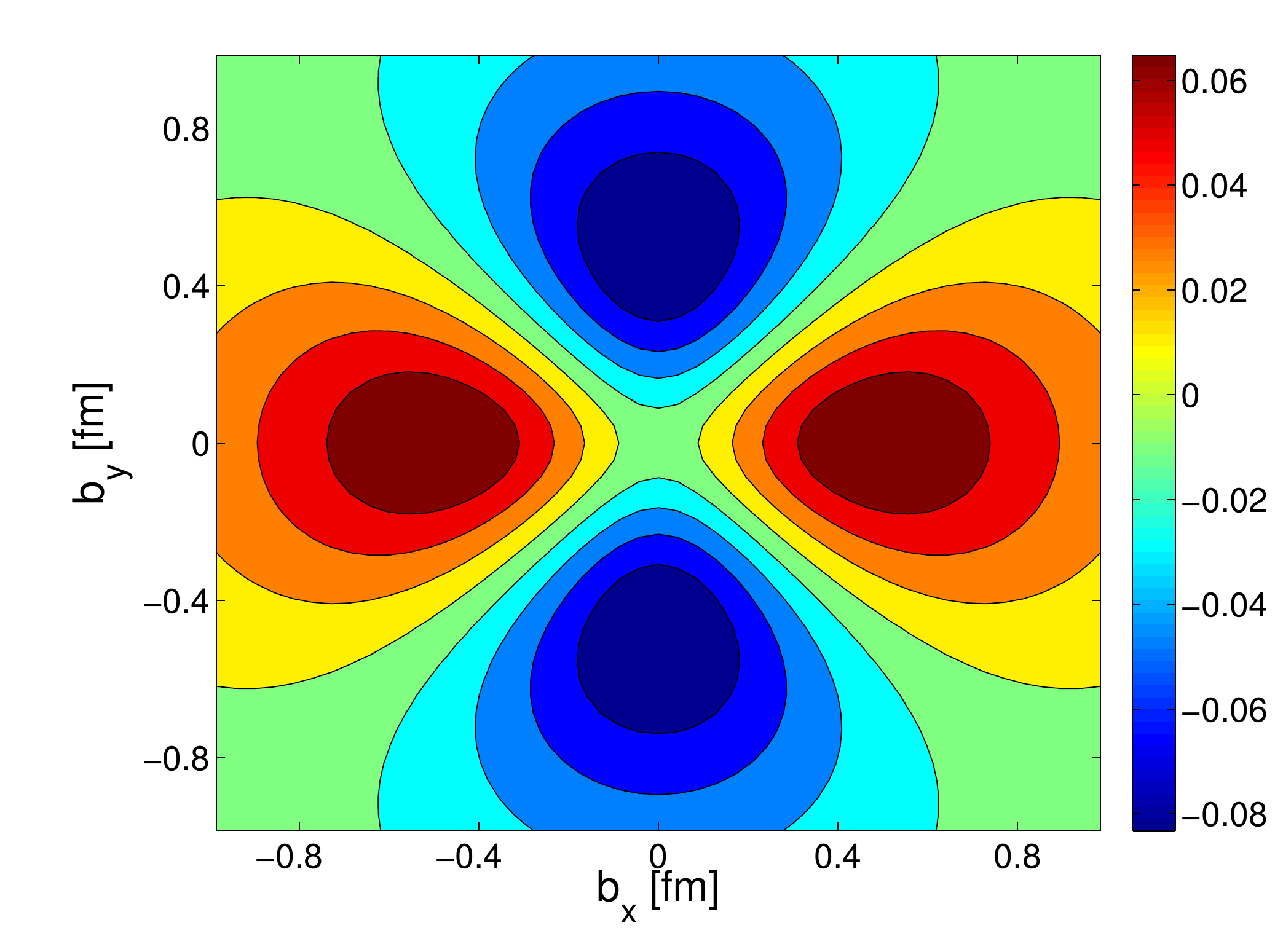}
\end{minipage}
\caption{\label{Fig3}(Color online) The monopole contribution $\frac{1}{2} s_xS_x(H_T-\Delta_b\widetilde{H}_T/4M^2)$ (top) and the quadrupole contribution $\frac{1}{2} s_xS_x(b_x^2-b_y^2)\widetilde H''_T/M^2$ (lower) for both quarks and the nucleon transversely polarized along $\hat x$. The left (right) panel gives the results for $u$ ($d$) quarks.} 
\end{figure}
\begin{figure}[htp]
\begin{minipage}[c]{0.98\textwidth}
\includegraphics[width=7.2cm,height=5.8cm,clip]{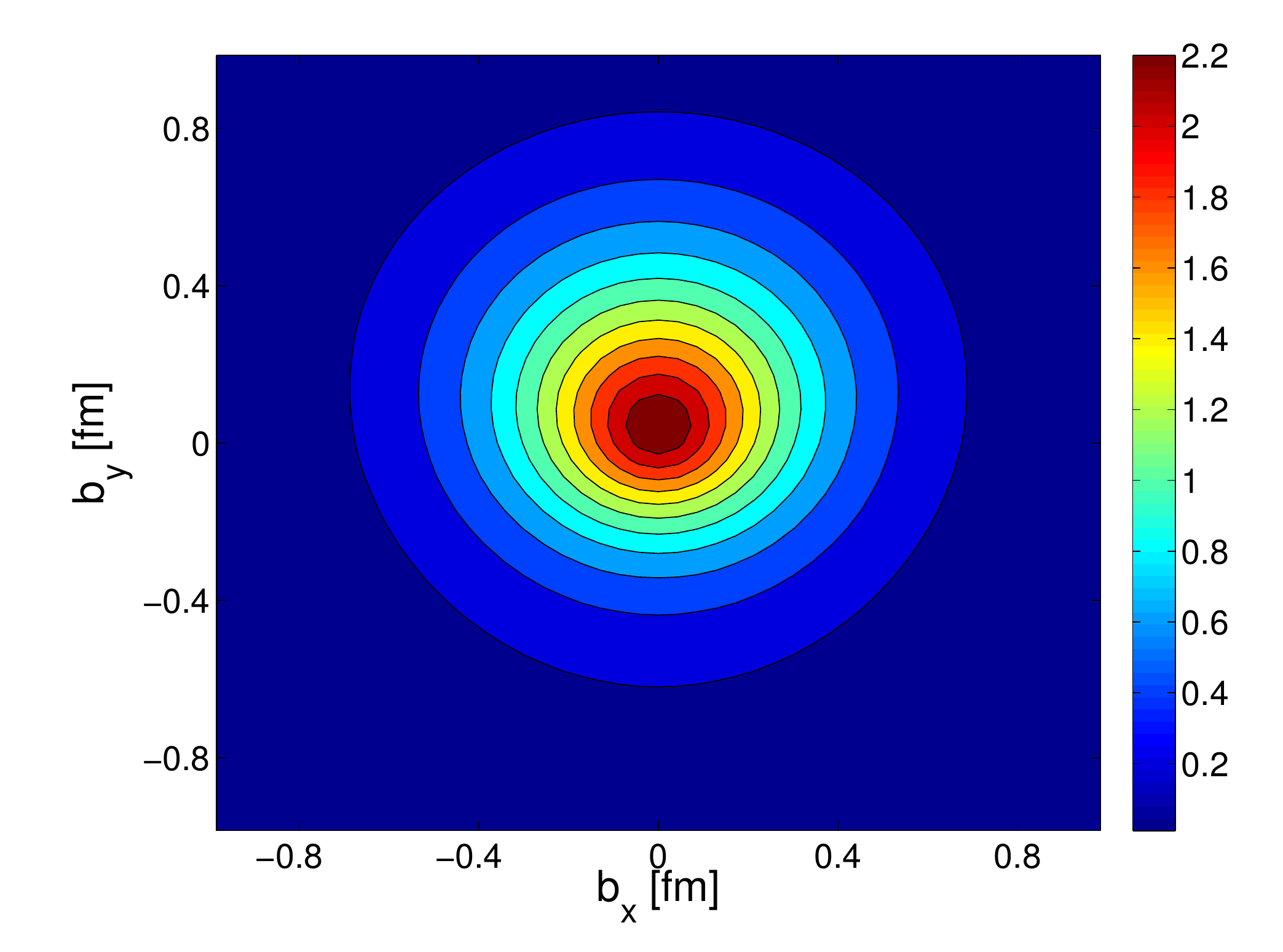}
\includegraphics[width=7.2cm,height=5.8cm,clip]{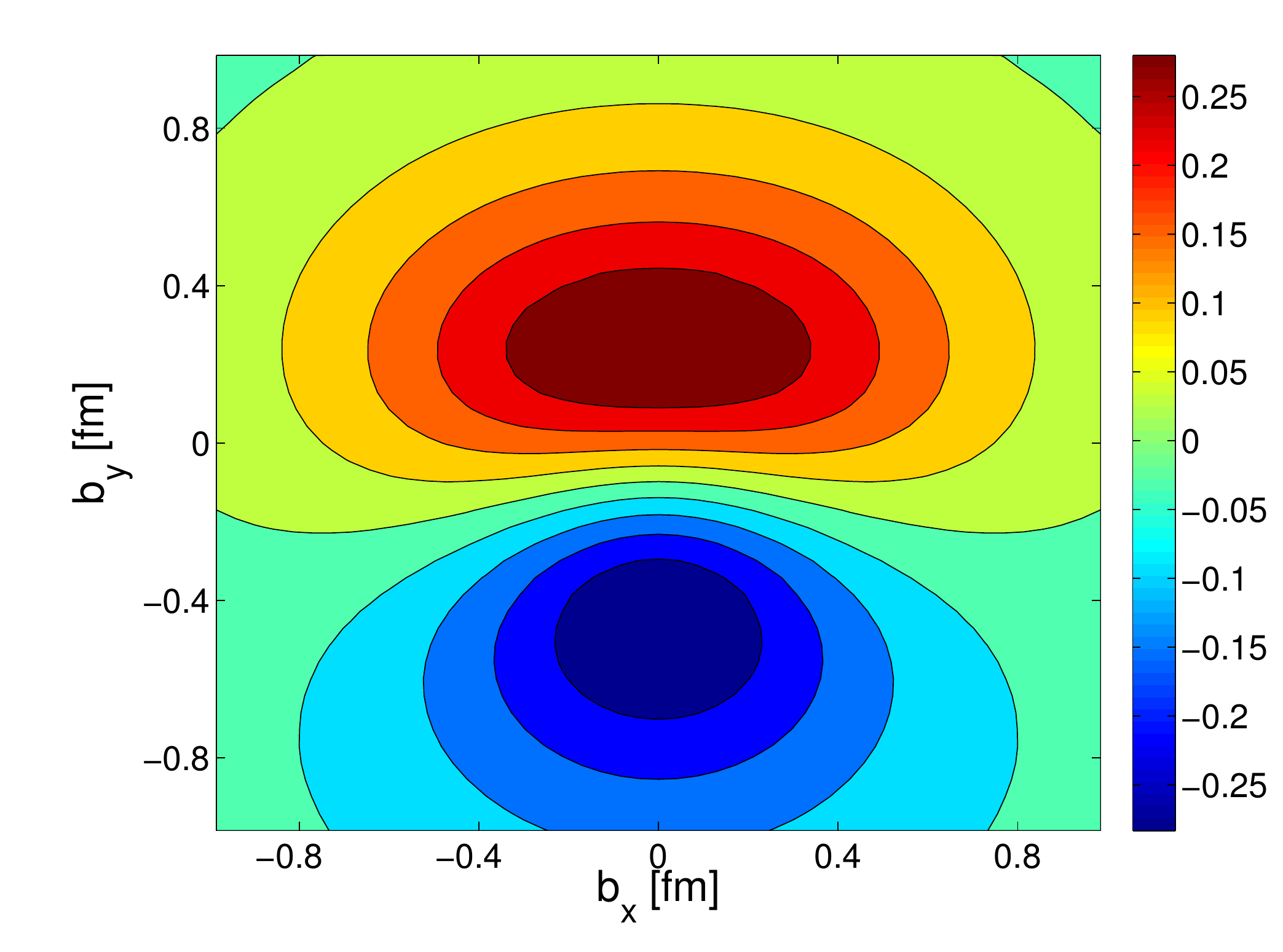}
\end{minipage}
\caption{\label{Fig4}(Color online) The total spin distribution as a sum of monopole, dipole and quadrupole terms,  for both quarks and proton polarized along $\hat x$; left (right) panel for $u$ ($d$) quarks.} 
\end{figure}
\begin{figure}[htp]
\begin{minipage}[c]{0.98\textwidth}
\includegraphics[width=7.2cm,height=5.8cm,clip]{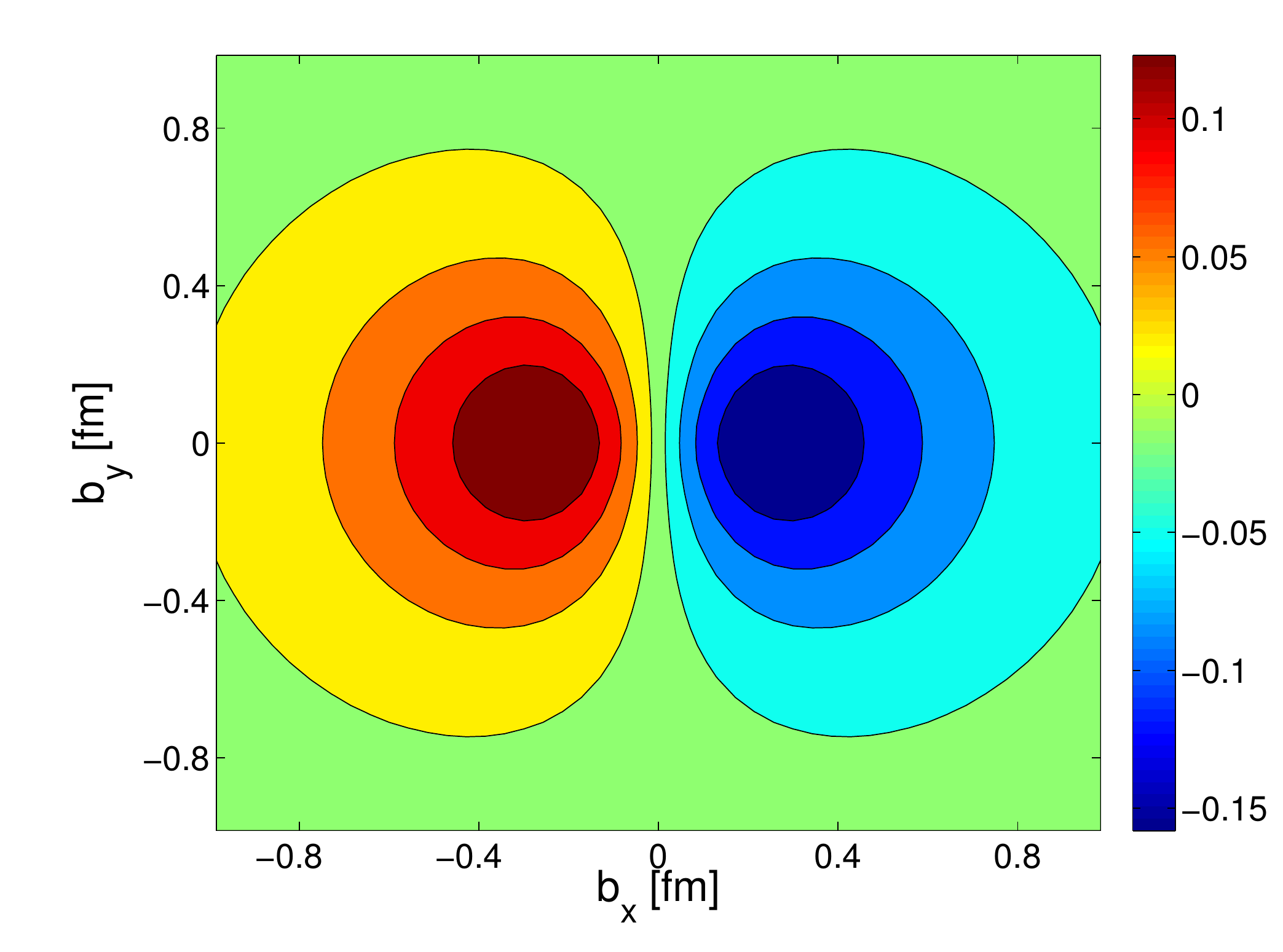}
\includegraphics[width=7.2cm,height=5.8cm,clip]{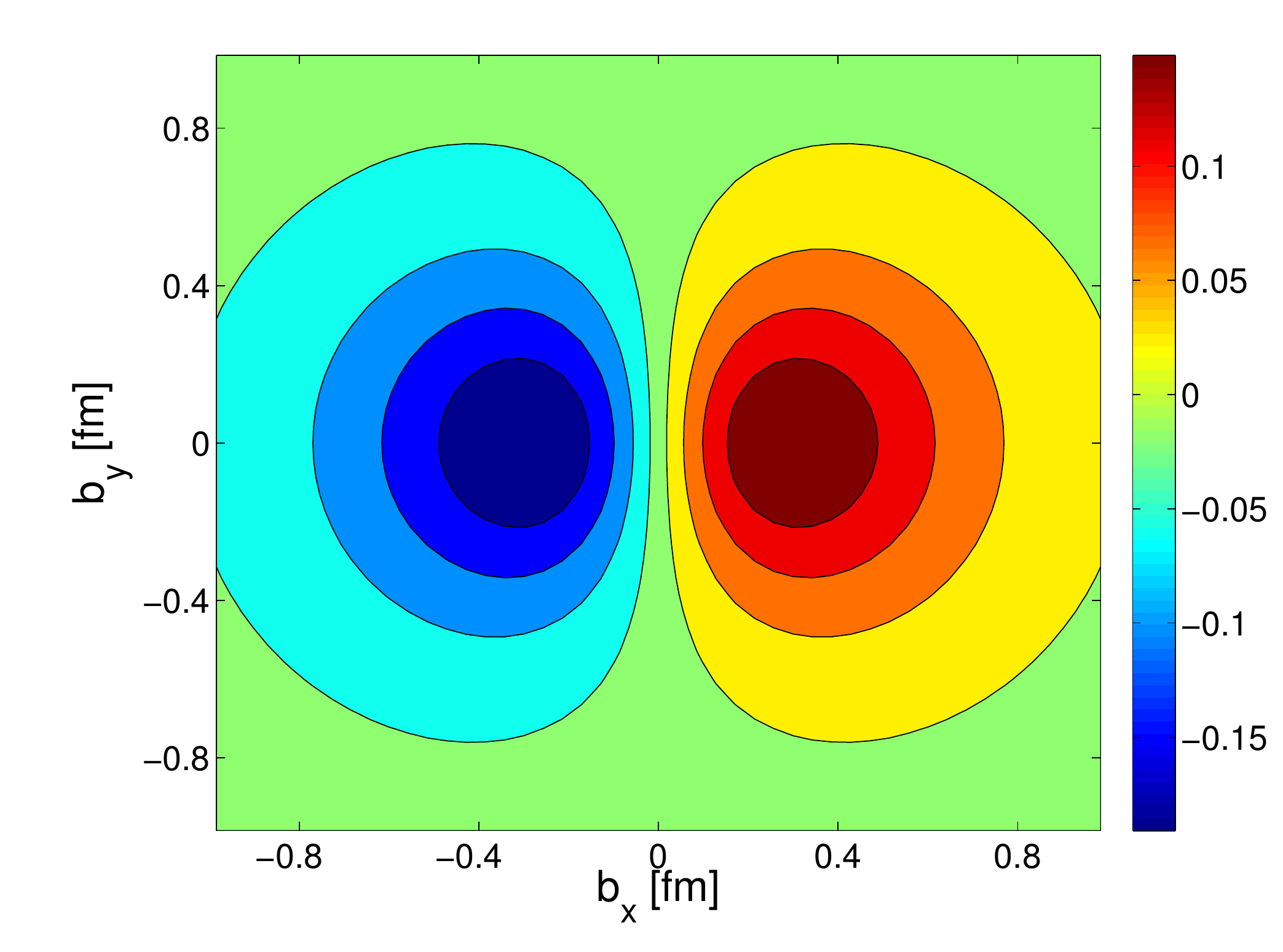}
\end{minipage}
\begin{minipage}[c]{0.98\textwidth}
\includegraphics[width=7.2cm,height=5.8cm,clip]{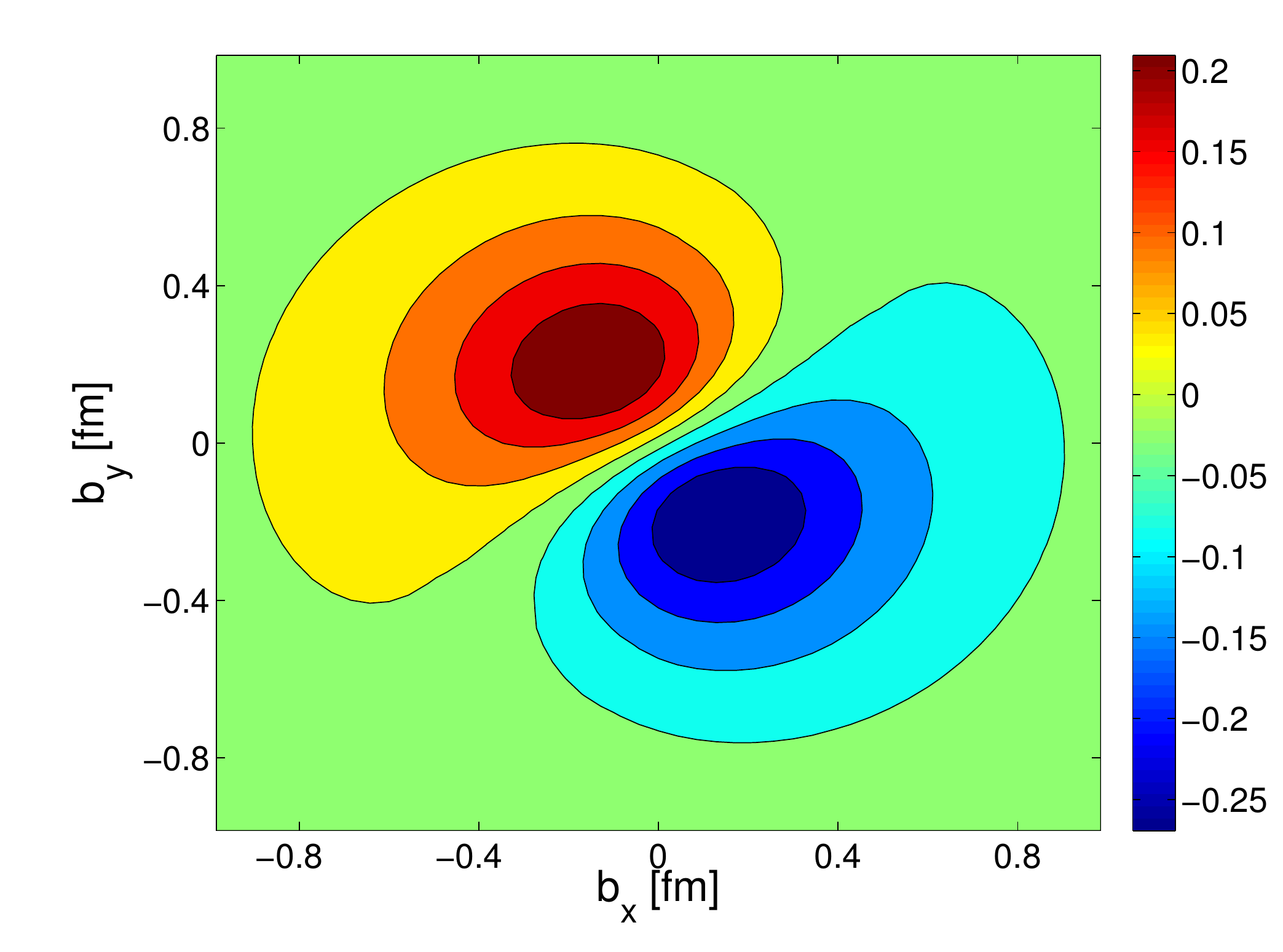}
\includegraphics[width=7.2cm,height=5.8cm,clip]{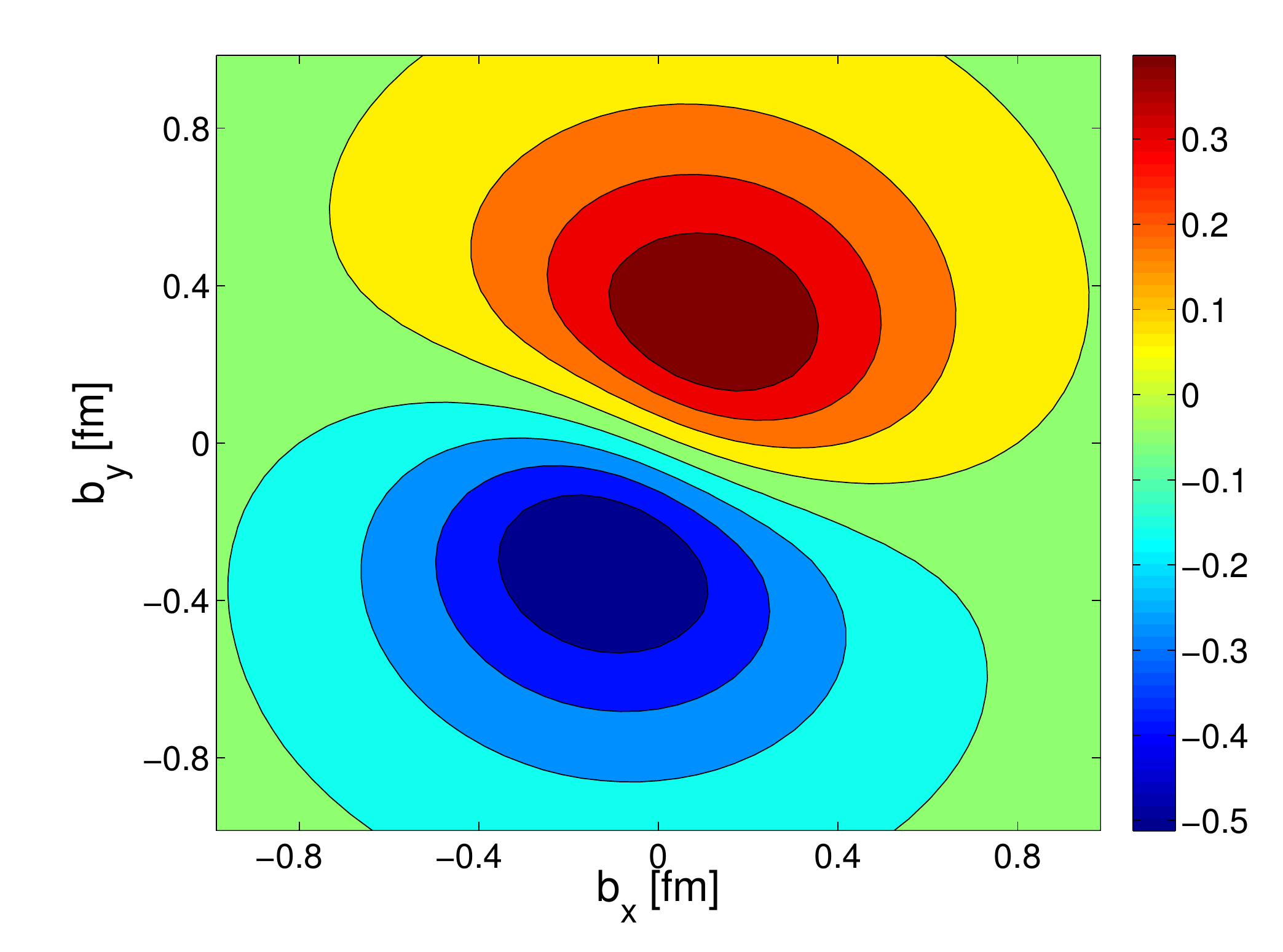}
\end{minipage}
\begin{minipage}[c]{0.98\textwidth}
\includegraphics[width=7.2cm,height=5.8cm,clip]{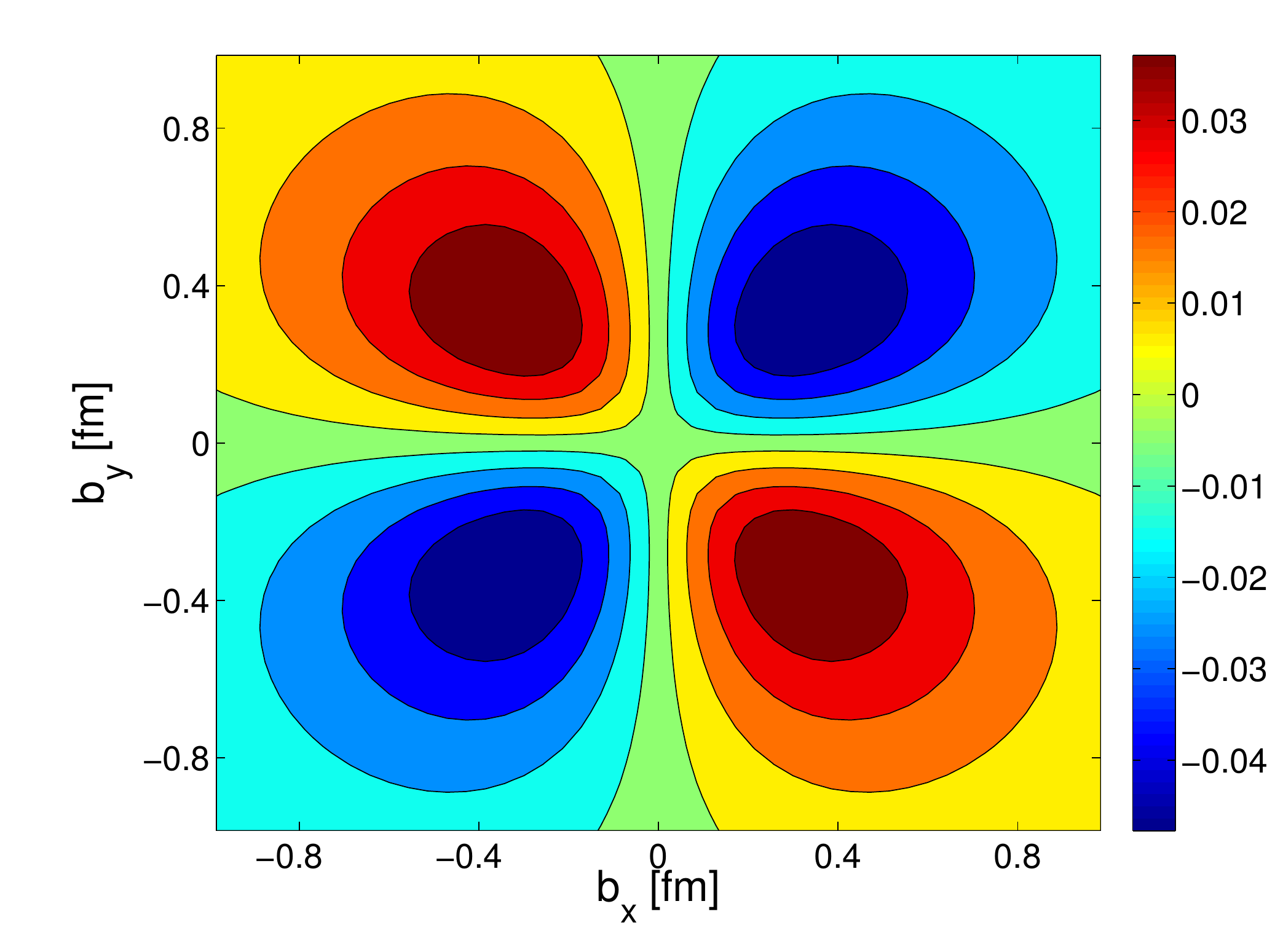}
\includegraphics[width=7.2cm,height=5.8cm,clip]{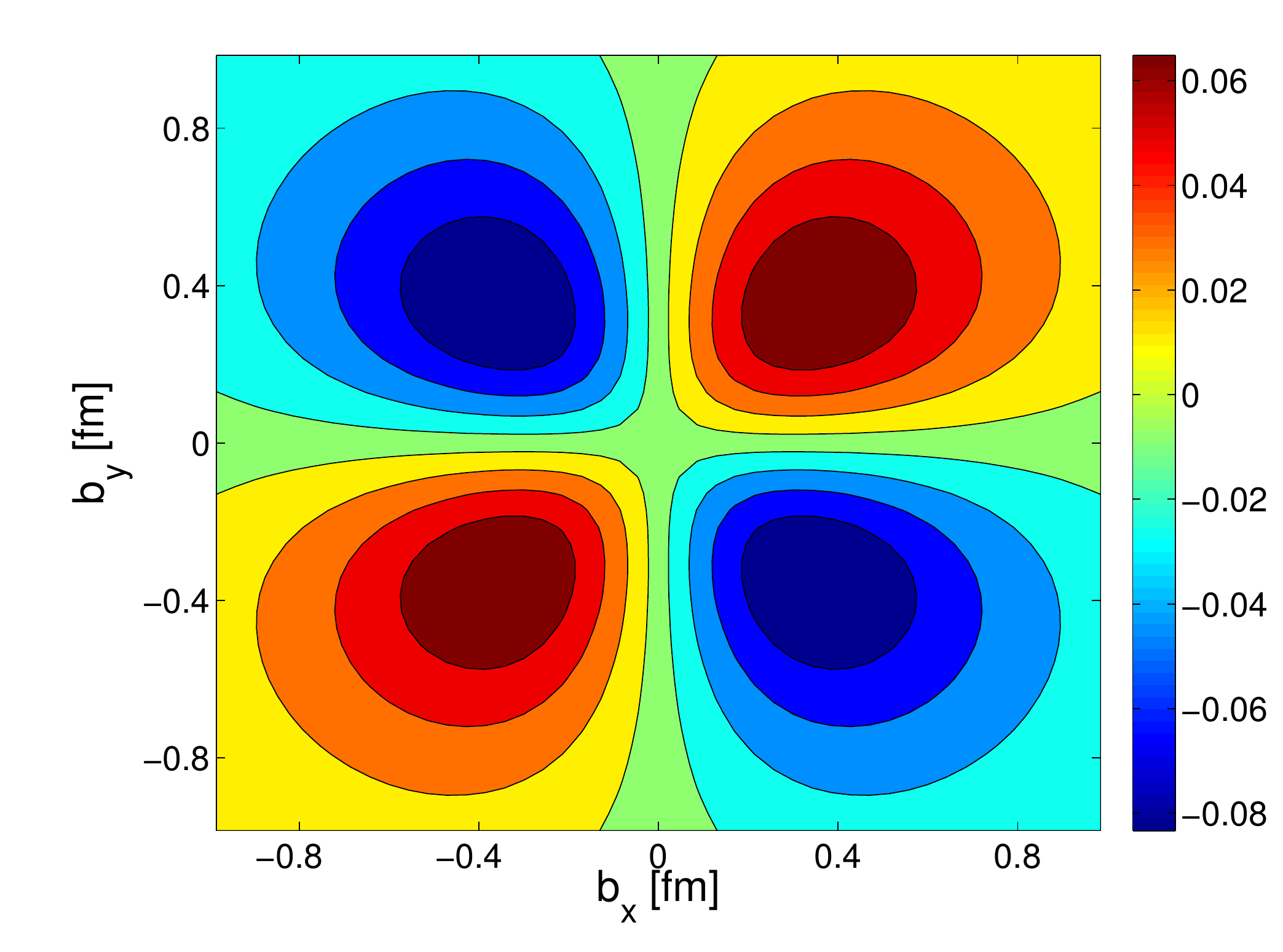}
\end{minipage}
\caption{\label{Fig5}(Color online) The dipole contribution $\frac{1}{2} S_yb_xE'/M$ (top), the total dipole contribution $\frac{1}{2}[S_yb_xE'-s_xb_y(E'_T+2\tilde H'_T)]/M$  (middle) and the quadrupole contribution $s_xS_yb_{x}b_{y}\tilde H''_T/M^2$ (lower) for quarks polarized along $\hat x$ in a nucleon transversely polarized along $\hat y$ direction. Left (right) panel for $u$ ($d$) quarks.} 
\end{figure}

\begin{figure}[htp]
\begin{minipage}[c]{0.98\textwidth}
\includegraphics[width=7.2cm,height=5.8cm,clip]{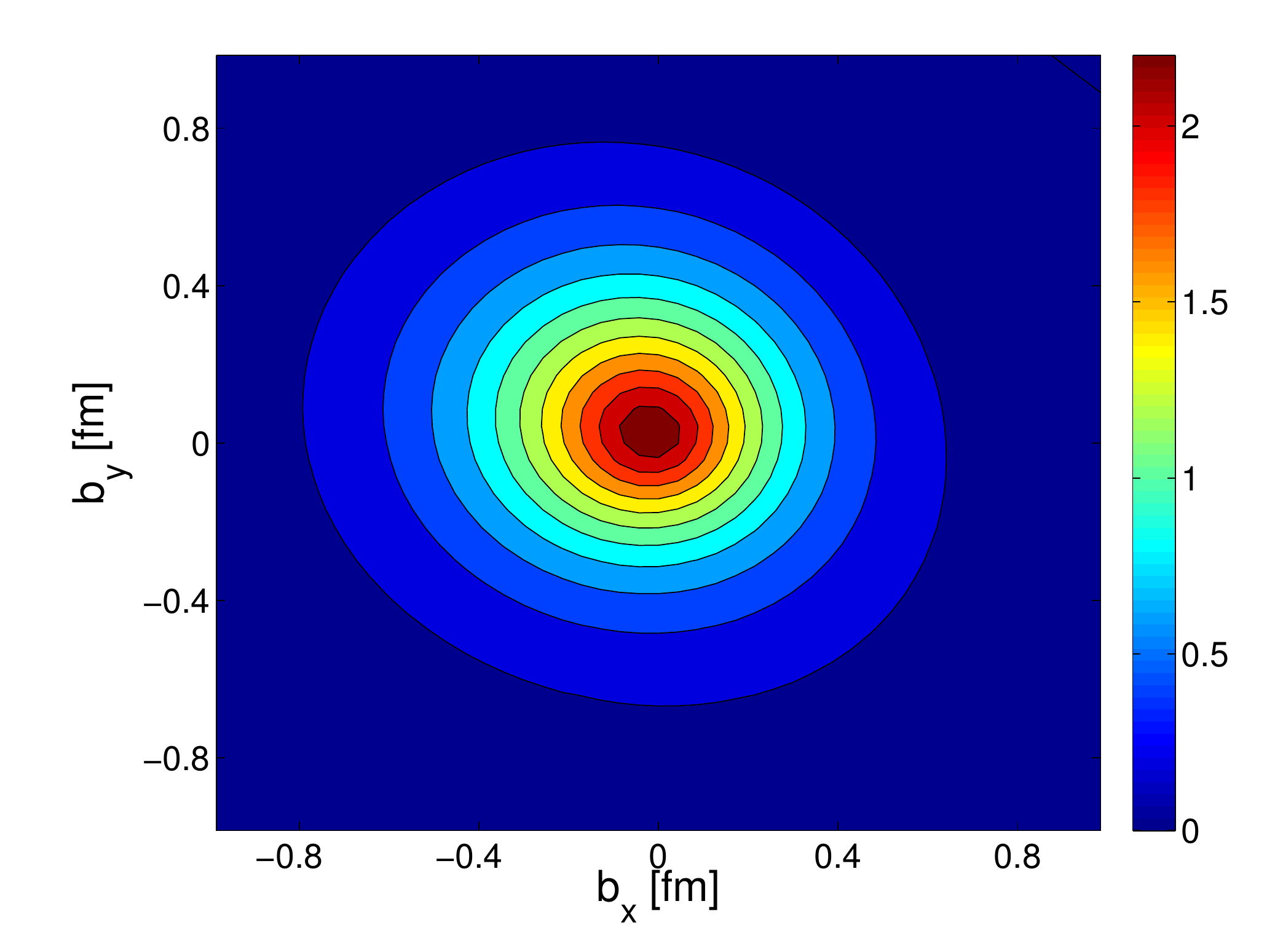}
\includegraphics[width=7.2cm,height=5.8cm,clip]{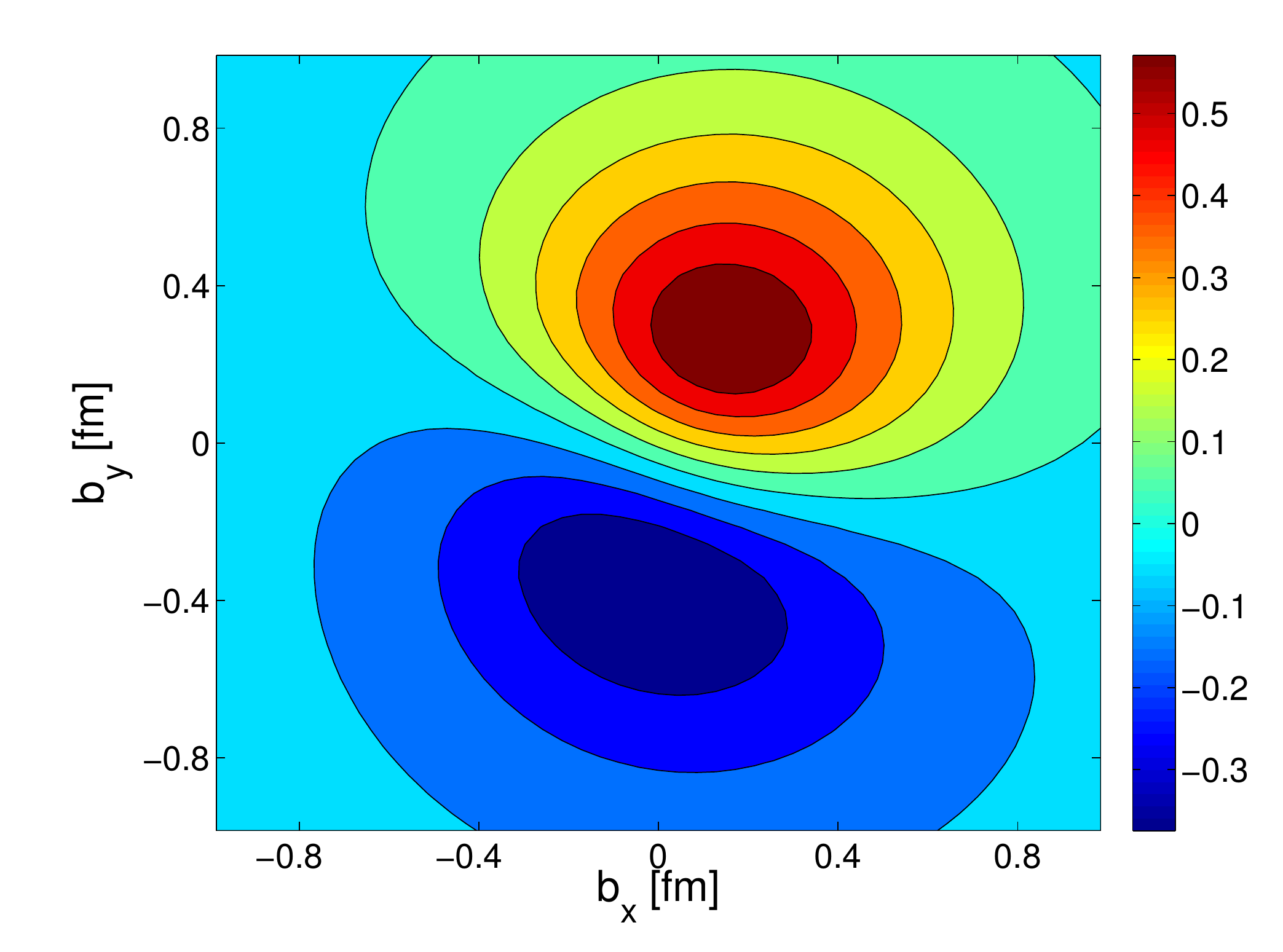}
\end{minipage}
\caption{\label{Fig6}(Color online) The total spin distribution as a sum of monopole, dipole and quadrupole terms, for quarks polarized along $\hat x$ in a nucleon transversely polarized along $\hat y$ direction; left (right) panel for $u$ ($d$) quarks.} 
\end{figure}
\begin{figure}[htp]
\begin{minipage}[c]{0.98\textwidth}
\includegraphics[width=7.2cm,height=5.8cm,clip]{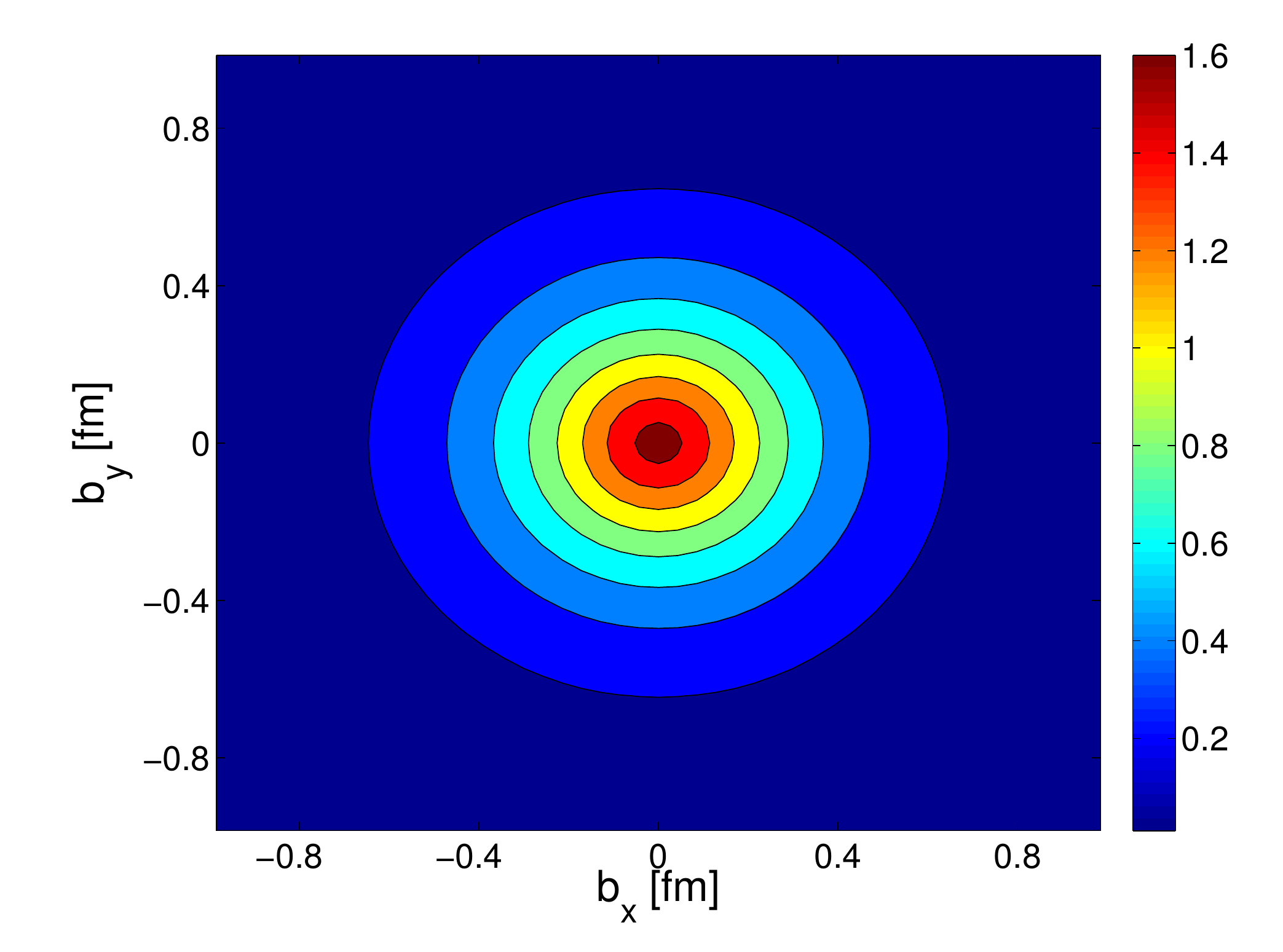}
\includegraphics[width=7.2cm,height=5.8cm,clip]{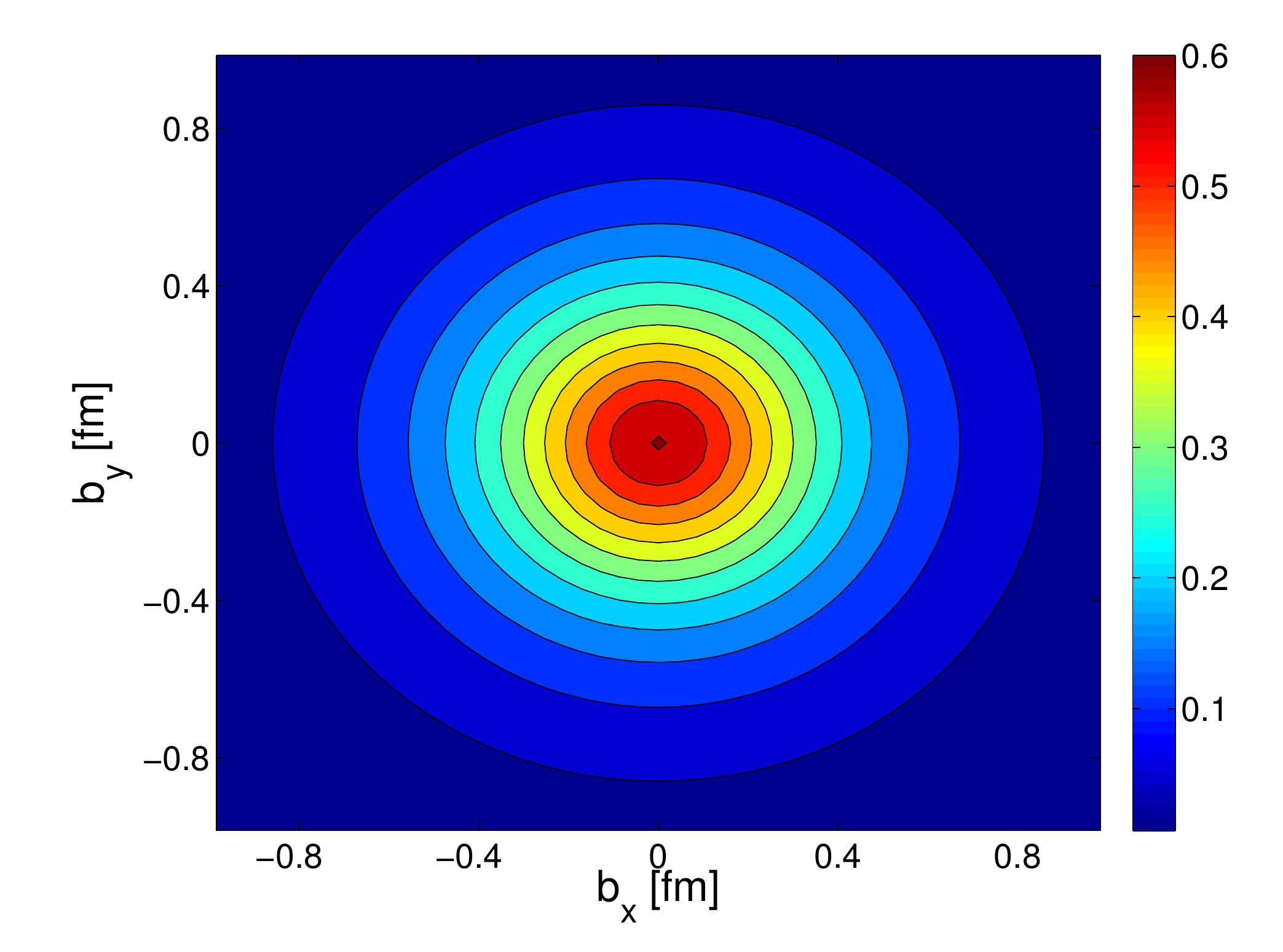}
\end{minipage}
\begin{minipage}[c]{0.98\textwidth}
\includegraphics[width=7.2cm,height=5.8cm,clip]{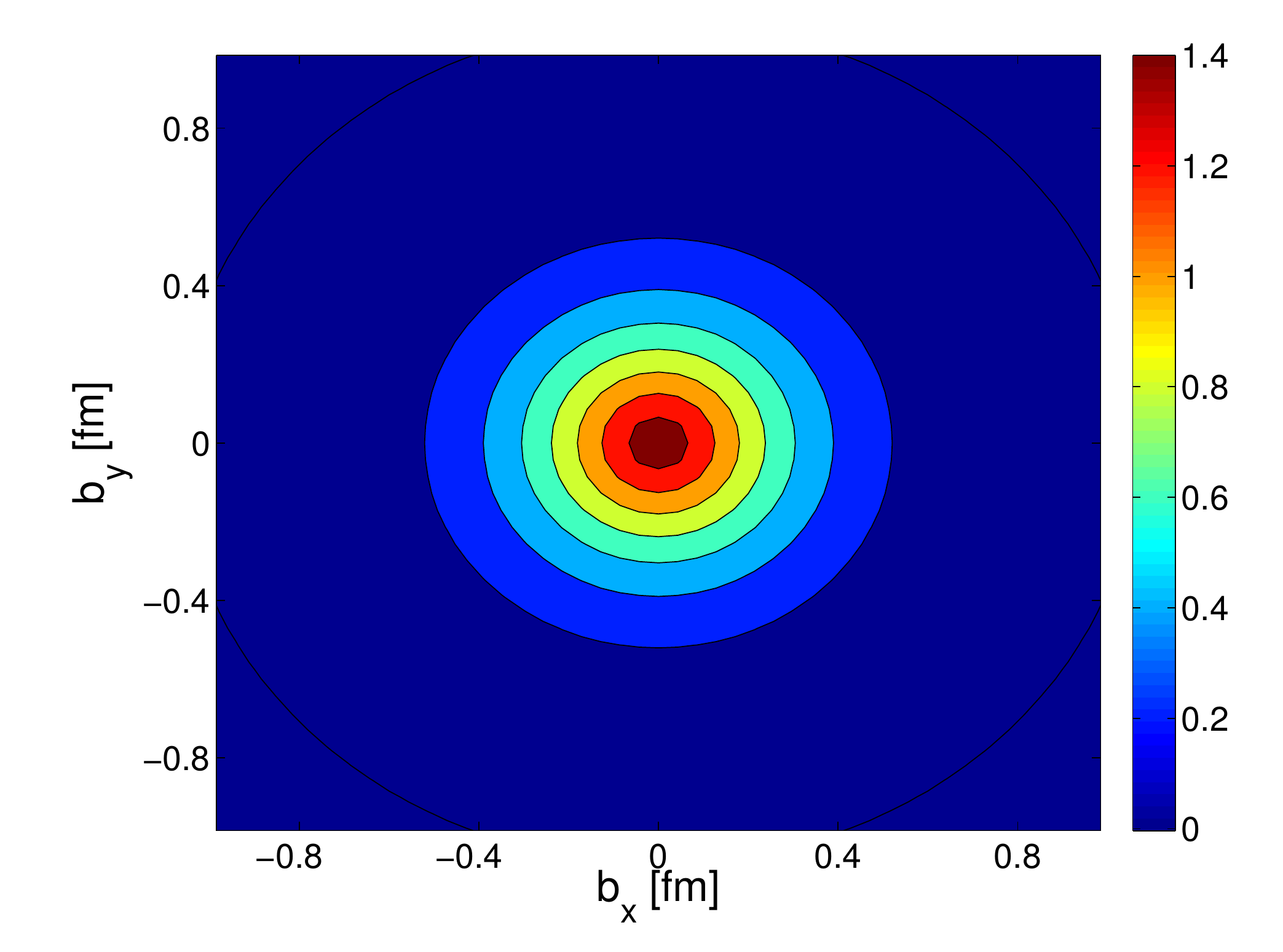}
\includegraphics[width=7.2cm,height=5.8cm,clip]{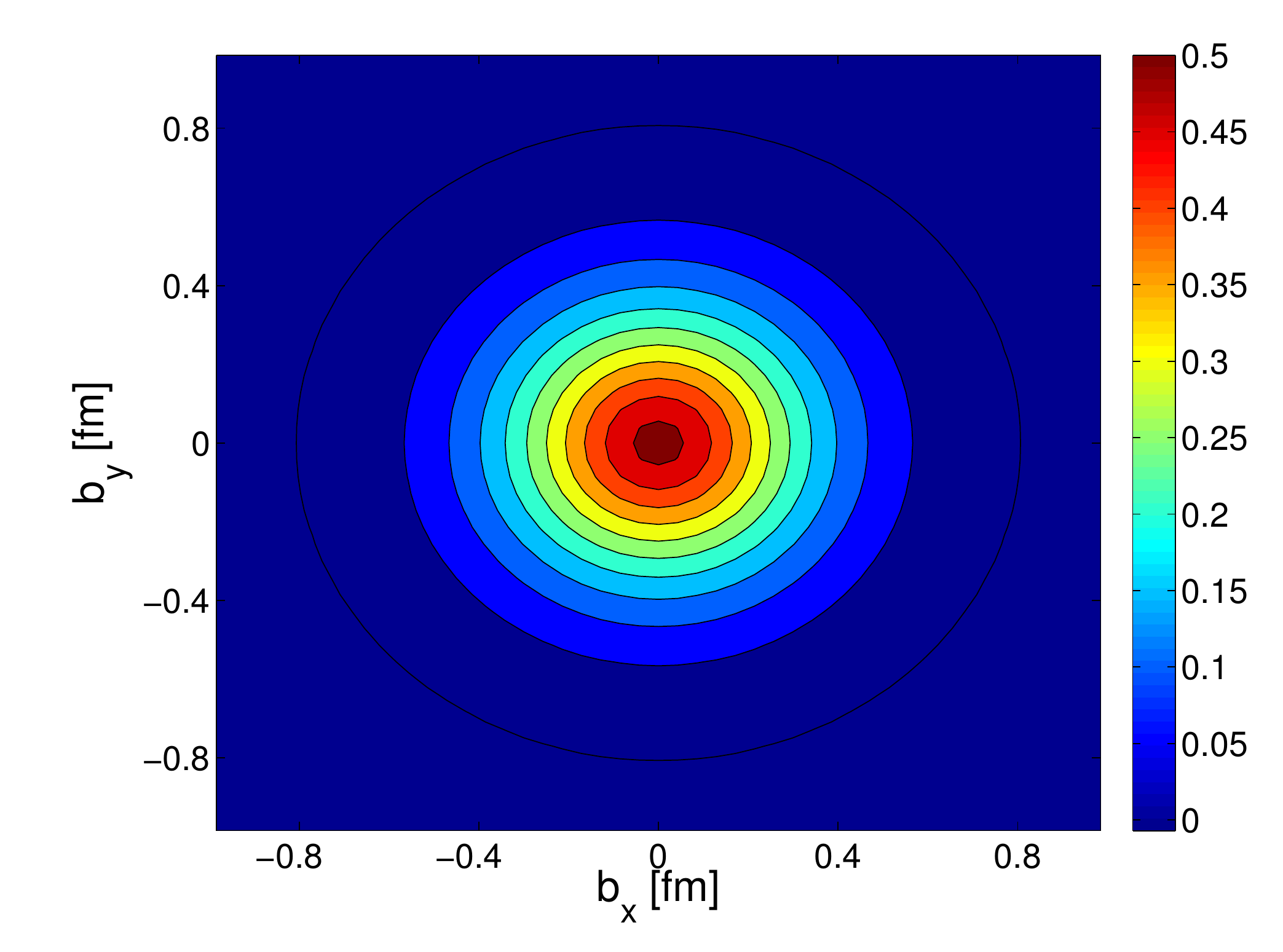}
\end{minipage}
\begin{minipage}[c]{0.98\textwidth}
\includegraphics[width=7.2cm,height=5.8cm,clip]{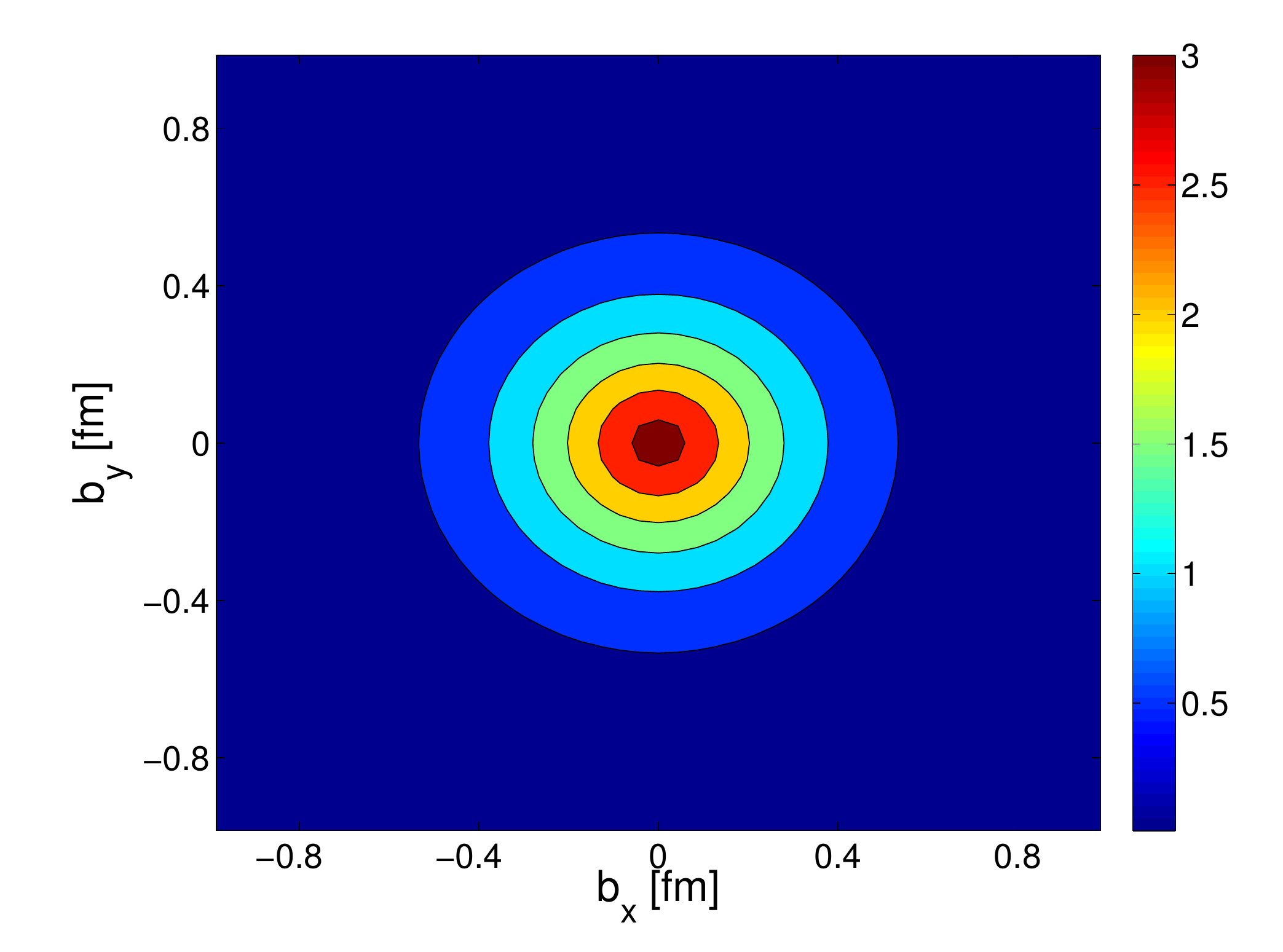}
\includegraphics[width=7.2cm,height=5.8cm,clip]{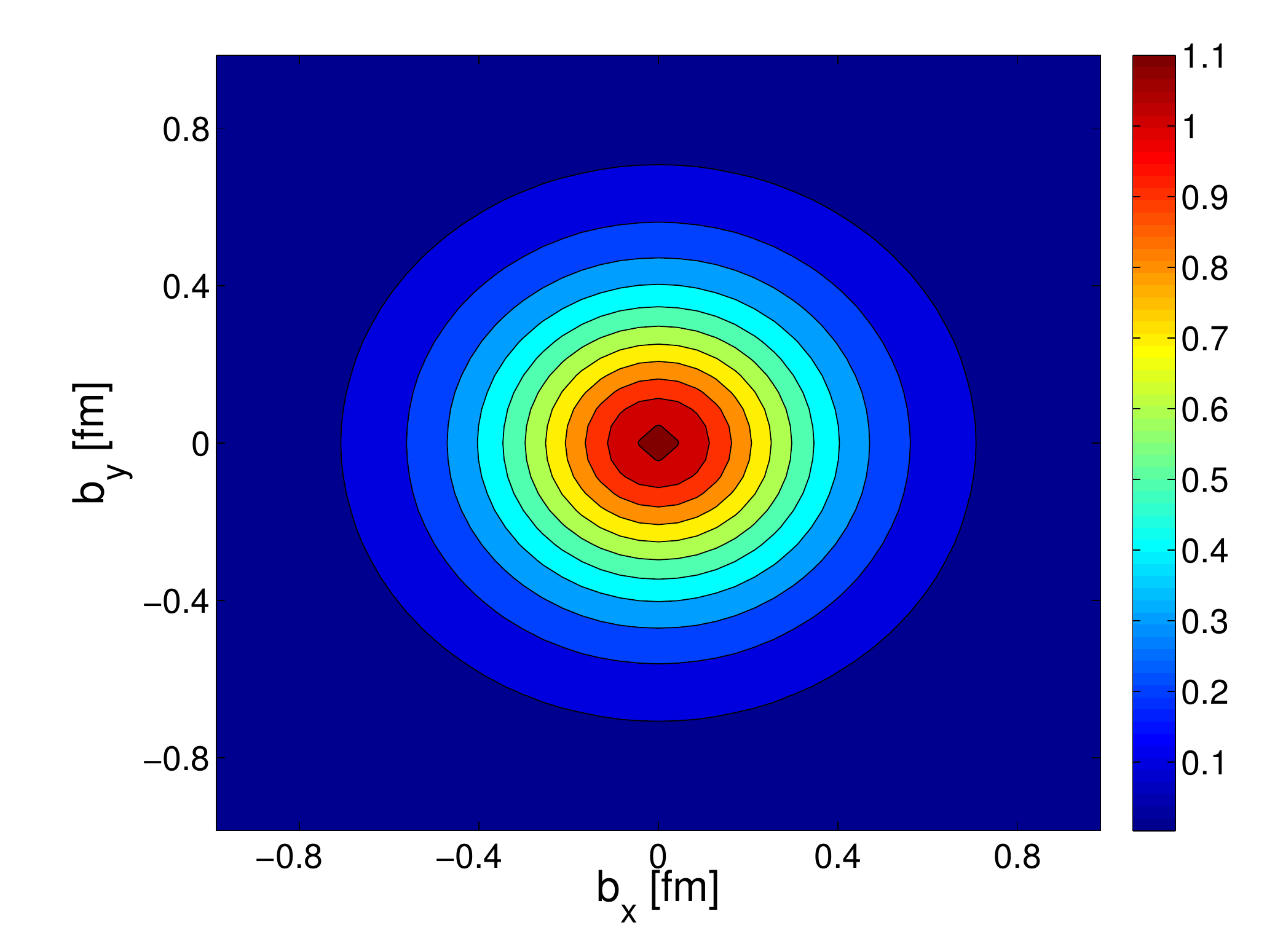}
\end{minipage}
\caption{\label{Fig7}(Color online) The monopole contribution $\frac{1}{2}H$ (top) and $\frac{1}{2}\tilde H$ (middle), and their sum (lower) corresponding to the spin distribution when both the quarks and proton are polarized along longitudinal direction; left (right) panel for $u$ ($d$) quarks.} 
\end{figure}

In Figs. \ref{Fig1} and \ref{Fig2}, we show the first $x$-moment of the spin distributions for transversely polarized quark in an unpolarized proton and for the unpolarized quark in a transversely polarized proton, respectively. The distorting effect of the dipole 
terms appears on the monopole terms which correspond to spin densities for unpolarized quarks in an unpolarized 
target. As a result the distributions get shifted toward the $\hat y$ direction for the quark or proton transversely polarized along the $\hat x$ axis. The dipolar distribution in Fig.~\ref{Fig1} has the same polarity for both the $u$ and $d$ quarks, whereas the polarity of the dipole distribution in Fig.~\ref{Fig2} is the opposite for the $u$ and $d$ quarks. The dipole distribution in Fig.~\ref{Fig2} arises from  the term $-\frac{1}{2} S_x b_y E^{q'}/M$, which gives the anomalous magnetic moment $\kappa_q$ of quark $q$ (for the $u$ quark $\kappa_q=1.673$; for the $d$ quark $\kappa_d=-2.033$). This effect provides a dynamical explanation of a non-vanishing Sivers function~\cite{Sivers:1989cc} $f_{1T}^\perp$ which describes the correlation between the intrinsic quark transverse momentum and the transverse nucleon spin. For transversely polarized quarks in an unpolarized proton, the dipolar contribution $-\frac{1}{2} s_x b_y (E'_T+2\widetilde H'_T)/M$ introduces a large distortion, however, the distortion is larger for the $d$ quark than the $u$ quark. 
This effect is related~\cite{Burk3,Burkardt:2007xm} to the Boer-Mulders function~\cite{BM98} $h_1^\perp$ describing the correlation between intrinsic transverse momentum and transverse spin of quarks. The distortion 
produces
 the anomalous tensor magnetic moment $\kappa_T$ which is positive for  both the $u$ and $d$ quarks. The distributions for the $d$ quark are little wider compared to the $u$ quark distributions. The qualitative behaviors of these spin distributions evaluated in this AdS/QCD inspired quark-diquark model are in agreement with the observations of a phenomenological model  \cite{Pasquini2} and lattice calculation\cite{Gockeler:2006zu}.

The spin distributions when both the proton and the quark are transversely polarized along $\hat x$ are shown in Fig.~\ref{Fig3}. 
 The quadrupole and the monopole terms  are of opposite signs for $u$ and $d$ quarks as a consequence of the opposite signs of the GPDs $H_T$ and $\tilde H_T$ (Fig.\ref{odd_impact_b}) for the $u$ and $d$ quarks. These results are consistent with the observations of a phenomenological model  \cite{Pasquini2}. However, the spread of quadrupole distribution in the quark-diquark model is more or less the same for the $u$ and $d$ quarks, whereas in the phenomenological model \cite{Pasquini2}, the distribution for the $u$ quark is more spread than for the $d$ quark.

The total spin distributions (Eq. \ref{eq:tr}) when the quark has the same transverse polarization as the  transversely polarized proton are shown  in Fig.~\ref{Fig4}. The polarization direction for both proton and the quark is taken to be $\hat x$.  Fig.~\ref{Fig4} shows the results for each quark after summing the two monopole contributions in the top panels of Fig.~\ref{Fig1} [$\frac{H}{2}$] and Fig.~\ref{Fig3} [$\frac{1}{2}(H_T-\Delta_b\widetilde{H}_T/4M^2)$], the two dipole contributions on the middle panels of Fig.~\ref{Fig1} [$-\frac{1}{2}(E'_T+2\widetilde H'_T)/M$] and Fig.~\ref{Fig2} [$-\frac{1}{2}E'/M$] and the quadrupole contribution of the lower panel in Fig.~\ref{Fig3} [$\frac{1}{2}(b_x^2-b_y^2)\widetilde H''_T/M^2$]. For the $u$ quark, the monopole contribution is much higher compared to the  dipole and quadrupole contributions, thus, the deformation of the spin density is small and slightly shifted in the $\hat y$ direction. In contrast the $d$ quark spin distribution has a much smaller monopole contribution and effectively shows a strong and symmetric deformation about the $\hat y$ axis stretching along the direction of the quark and proton polarization, i.e., $\hat x$ direction. 

In Fig.~\ref{Fig5}, we show the distorting effect of the dipole and quadrupole terms considering the $\hat x$-polarized quarks in a proton polarized along $\hat y$. In this case, the dipole contribution appears from $\frac{1}{2} S_yb_xE'$. The dipole distribution (Fig.~\ref{Fig5}) is rotated with respect to the case shown in Fig.~\ref{Fig1} and due to the opposite signs of the anomalous magnetic moments $\kappa_{u,d}$, the polarity for the $u$ quark remains the opposite of that for the $d$ quark.  Taking into account  the second dipole term $-\frac{1}{2} s_xb_y(E'_T+2\tilde H'_T)/M$ which is shown in Fig.~\ref{Fig1}, the total dipole distortions are displayed in the middle panel of Fig.~\ref{Fig5}. The quadrupole distortion comes from $s_xS_yb_{x}b_{y}\tilde H''_T/M^2$ and the contribution is quite small compared to the dipole distortion. The total resulting spin density which is the sum of the monopole ($\frac{1}{2}H$) and the distortions due to the dipole and  quadrupole terms is shown in Fig.~\ref{Fig6}. Due to the large monopole contribution  the distorting effect is quite small in the $u$ quark spin density, whereas due to the strong distortion, the total spin density for the $d$ quark effectively exhibits a dipolar pattern. 

\begin{table}[ht]
\centering 
\begin{tabular}{|c|c|c|c|c|c|c|}
 \hline
 & $g^u_A$& $g^d_A$&~$g_A=g_A^u-g_A^d$ & $g^{u(1)}_A$ & $g^{d(1)}_A$ & $g^{(1)}_A$\\ \hline
This model \cite{MC} & $0.71\pm0.09$ & $-0.54^{+0.19}_{-0.13}$ & $1.25^{+0.28}_{-0.22}$ & $0.18\pm0.15$ & $-0.052^{+0.003}_{-0.007}$ & $0.23^{+0.15}_{-0.16}$ \\ 
Fit to data \cite{Lead10} & $0.82\pm 0.07$ & $-0.45\pm 0.07$ & $1.27\pm0.14$ & $0.19\pm0.07$ &  $-0.06\pm0.07$ & $0.25\pm0.14$ \\
\hline
 \end{tabular} 
\caption{Axial charge and second moment of helicity distribution at the scale $\mu^2=1~ GeV^2$ and compared with LSS fit to experimental data\cite{Lead10}.} 
\label{tab_gA} 
\end{table}
\begin{figure}[htp]
\begin{minipage}[c]{0.98\textwidth}
\includegraphics[width=7.2cm,height=5.8cm,clip]{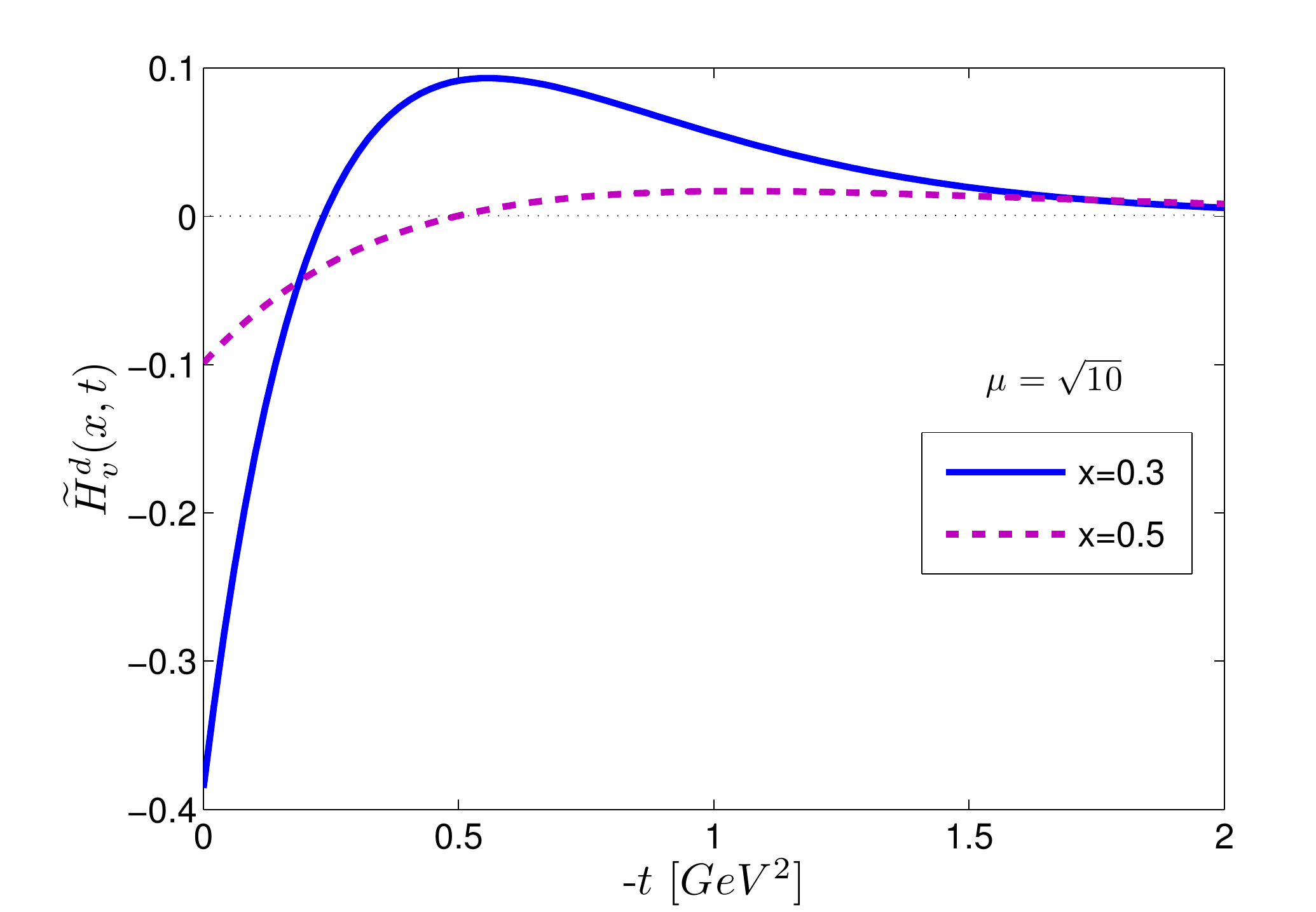}
\end{minipage}
\caption{\label{H-tilde}(Color online) Plot of $\widetilde{H}(x,t)$ GPD vs $-t$ for $d$ quark at scale $\mu=\sqrt{10}~ GeV$.} 
\end{figure}
In Fig.~\ref{Fig7}, the results are given for the quark polarization parallel to the proton helicity. Here we find only 
the monopole distribution occurring from $\frac{1}{2}H$ and $\frac{1}{2}\lambda\Lambda\widetilde{H}$. In contrast with the
other phenomenological model \cite{Pasquini2} where  $\frac{1}{2} \widetilde H^u$ has opposite sign of  $\frac{1}{2} \widetilde H^d$, 
in our model  $\frac{1}{2}\lambda\Lambda\widetilde{H}$ for both the $u$ 
and $d$
is found to be positive at the initial scale. But with increasing the scale $\mu$, the axial charges for $u$ and $d$ quarks are found to be positive and negative respectively. 
In the forward limit, the GPD $\tilde H^q$ reduces to the helicity distribution $g_1^q(x)$ which is  related to the 
axial change of the quark ($g_A^q$). Axial charges are obtained from the first moment of the helicity 
distributions $g_1^q(x)$. $g_A^q$ at $\mu^2=1$ GeV$^2$ in the quark-diquark model is given in Table.\ref{tab_gA} and compared with the
measured data \cite{Lead10}. It can be noticed that the axial charges obtained in the quark-diquark model are in more or 
less agreement with measured data. A detail comparison of axial charges in this quark-diquark model with other 
phenomenological models e.g. NQM, LFCQM, LF$\chi$QSM \cite{Lorce11} has been presented in Ref.\cite{MC}. Though the 
helicity distribution $g_1(x)$ for $d$ quark in this quark-diquark model is negative, the corresponding 
GPDs $\widetilde{H}^d(x,t)$ exhibits a positive distribution at higher values of $-t$ and it is negative at low $-t$
which can be observed in Fig.\ref{H-tilde}. This may be the reason we obtain a positive distribution for $\widetilde{H}^d$ 
at low $b_\perp$ but at higher scale get the correct sign for the axial charge $\Delta d$ which is evaluated at $t=0$. It should be 
mentioned here that $\widetilde{H}^d$ is  
also positive in the scalar diquark 
model \cite{Gutsche:2013zia} but this model is unable to reproduce the proper sign of $\Delta d$. 

\section{summary}
Using a recently proposed light-front quark-diquark model for the proton we have studied both the chiral even and 
odd leading twist GPDs.  The results of the  GPDs are presented  in the impact parameter space for zero skewness.  
Then we have studied the spin densities for different proton polarizations.  Though for longitudinally polarized proton, 
only the chiral even GPDs contribute, for transversely polarized proton both chiral even and odd GPDs and their 
derivatives are required to study the spin densities. Our study reveals how different GPDs are contributing to the proton 
spin densities for different polarizations of the quark and proton.   Monopole, dipole and quadrupole contributions to the 
spin densities are shown separately. A certain combination of $H_T$ and $\tilde{H}_T$ in impact parameter space is responsible 
for the distortion in the spin density when the active  quark and the proton, both are transversely polarized. Similarly 
for transversely polarized quarks in a unpolarized proton, the combination $(E^\prime_T +2\tilde{H}^\prime_T)$ generates 
a dipolar distortion. The anomalous tensor magnetic moment is found to be positive for both $u$ and $d$ quarks. 
For $u$ quark, monopole contribution to the spin density is large and  the dipole and quadrupole distortions are relatively small,
whereas for $d$ quark the distortions are found to be  significantly large.

 
\end{document}